\documentclass[11pt]{article}

\usepackage{amsmath,amsfonts}
\usepackage{theorem}
\usepackage{mathrsfs}

\usepackage[titletoc,toc,title]{appendix}

\usepackage{epsfig}

\usepackage{xcolor}

\usepackage{lscape}

\usepackage{multirow}
\usepackage{array}
\usepackage{rotating}

\usepackage{tikz}

\usepackage[hyphens]{url}
\usepackage[breaklinks]{hyperref}
\hypersetup{
    colorlinks=true,       
    linkcolor=red,          
    citecolor=blue,        
    filecolor=magenta,      
    urlcolor=cyan           
}

\usepackage[sort&compress]{natbib}
\bibpunct{(}{)}{,}{a}{,}{,}

\theoremstyle{plain}
\newtheorem{theorem}{Theorem}
\newtheorem{assumption}{Assumption}
\newtheorem{proposition}{Proposition}

\theoremstyle{definition}

\newtheorem{example}{Example}

\def\halmos{\mbox{\quad$\square$}}

\newcommand{\proof}[1]{\vspace{1ex}\noindent{\em #1}\hspace{0.5em}}
\newcommand{\proofend}{\vspace{1ex}}

\newcommand{\cM}{{\cal M}}

\newcommand{\prob}{\mathbb{P}}
\newcommand{\expect}{\mathbb{E}}

\newcommand{\tran}{'}

\newcommand{\add}{\color{black}}

\usepackage{setspace}
\begin{document}
\title{Recursive Utility with Investment Gains and Losses: Existence, Uniqueness, and Convergence\thanks{We are grateful to participants at the Fifth Asian Quantitative Finance Conference in Seoul, the Second Paris-Asia Conference in Quantitative Finance in Suzhou, and the 10th World Congress of The Bachelier Finance Society in Dublin. Xue Dong He acknowledges financial support from the General Research Fund of the Research Grants Council of Hong Kong SAR (Project No. 14225916). 
}
	}

\author{Jing Guo\thanks{Department of Industrial Engineering and
Operations Research, Columbia University in the City of New
York, New York, NY, 10027, US. Email: \texttt{jg3222@columbia.edu}.}
\and Xue Dong He\thanks{Room 609, William M.W. Mong Engineering Building, Department of Systems Engineering and Engineering Management, The Chinese University of Hong Kong, Shatin, N.T., Hong Kong. Email: \texttt{xdhe@se.cuhk.edu.hk}.}
}

\date{First version: June 6, 2016; This version: June 22, 2019}

\maketitle

\begin{abstract}
We consider a generalization of the recursive utility model by adding a new component that represents utility of investment gains and losses. We also study the utility process in this generalized model with constant elasticity of intertemporal substitution and relative risk aversion degree, and with infinite time horizon. In a specific, finite-state Markovian setting, we prove that the utility process uniquely exists when the agent derives nonnegative gain-loss utility, and that it can be non-existent or non-unique otherwise. Moreover, we prove that the utility process, when it uniquely exists, can be computed by starting from any initial guess and applying the recursive equation that defines the utility process repeatedly. We then consider a portfolio selection problem with gain-loss utility and solve it by proving that the corresponding dynamic programming equation has a unique solution. Finally, we extend certain previous results to the case in which the state space is infinite.

\medskip




\end{abstract}

\section{Introduction}
\citet{BarberisHuang:2009PreferenceWithFrames,BarberisHuang2008:LossAversionNarrowFraming} and \citet{BarberisEtal2006:StockMarketParticipationNarrowFraming} propose a utility specification that allows for narrow framing in a discrete-time, multiple-period setting in which an agent derives utility not only from her consumption stream but also from the investment gain and loss incurred by holding certain risky assets. The former is referred to as consumption utility, the latter as gain-loss utility. The total utility of the agent is computed based on the classical recursive utility model \citep{KrepsPorteus1978:TemporalResolutionOfUncertainty,EpsteinLZinS:89rut}: The total utility for the agent's consumption and investment starting from time $t$ is the aggregation of 1) her consumption at time $t$, 2) her gain-loss utility in the period from $t$ to $t+1$, and 3) the time-$t$ certainty equivalent of her total utility for consumption and investment starting from time $t+1$. In particular, when the gain-loss utility is set at zero, the model of narrow framing degenerates into the classical recursive utility model.

Just as in the classical recursive utility model, the aggregation of different components of utility in the model of narrow framing is achieved by a so-called aggregator function and the certainty equivalent is computed under the expected utility theory. The aggregator thus measures the elasticity of intertemporal substitution (EIS) and the certainty equivalent measures the relative risk aversion degree (RRAD) of the agent. As in many applications of the classical recursive utility model to portfolio selection and asset pricing, in their model of narrow framing, \citet{BarberisHuang:2009PreferenceWithFrames,BarberisHuang2008:LossAversionNarrowFraming} and \citet{BarberisEtal2006:StockMarketParticipationNarrowFraming} select a specific aggregator in which the EIS is constant and a specific certainty equivalent in which the RRAD is constant; see the exact forms in \eqref{eq:PowerAggregator} and \eqref{eq:PowerCE}. Furthermore, the authors adopt an infinite-horizon setting. Both the specific choice of the aggregator and certainty equivalent and the infinite-horizon setting are known to be simple and helpful in obtaining closed-form solutions to a variety of problems.

The model of narrow framing is successful in explaining some empirical findings, such as why people are averse to small, independent gambles, even when the odds are actuarially favorable; see for instance \citet{BarberisEtal2006:StockMarketParticipationNarrowFraming}. This model is further extended by \citet{DeGiorgiLegg2011:NarrowFramingProbabilityWeighting} and \citet{HeZhou2013:ReferencePoint} with various applications, and these authors also assume constant EIS and RRAD and adopt the infinite-horizon setting. Even with many successful applications, however, the existence and uniqueness of the agent's (total) utility process in the model of narrow framing have not been established. Indeed, in the infinite-horizon setting, the agent's utility is defined recursively without an end date, so its existence and uniqueness cannot be taken for granted. Surprisingly, even for the classical recursive utility, its existence and uniqueness have not been completely established; see Section \ref{se:Literature} below.

In the present paper, we consider a generalization of the recursive utility model that adds a component of gain-loss utility and thus accommodates various models of narrow framing in the literature; {\add see the recursive equation \eqref{eq:RecursiveEquation} in the following, which defines the agent's total utility per unit of her wealth.} Assuming constant EIS and RRAD, we study the existence and uniqueness of the agent's utility process in this generalized recursive utility model in a {\add specific} Markovian setting. More precisely, we assume a Markov process $\{X_t\}$ and a process $\{Y_t\}$ that is an independent sequence conditional on $\{X_t\}$. Thus, $\{X_t\}$ models the dynamics of market states and $\{Y_t\}$ can be interpreted as random noise. The asset returns in the period from $t$ to $t+1$ are assumed to be functions of $X_t$, $X_{t+1}$, and $Y_{t+1}$, {\add so the agent's consumption propensity, percentage investment in the assets, and utility of gains and losses per unit of investment} in that period are functions of $X_t$. We further assume that $\{X_t\}$ is irreducible and focus mainly on the case in which the state space $\{X_t\}$ is finite. {\add See Section \ref{subse:Model} for details of the model setting and Section \ref{subse:ModelExamples} for the relevance of the setting in portfolio selection problems.}

{\add The Markovain setting here is the same as the one assumed in \citet{HansenScheinkman2012:RecursiveUtility} and in a recent, independent work by \citet{BorovivckaStachurski2017:SpectralConditions}, except that we make different assumptions regarding the state space of $\{X_t\}$. Both of these two works study the existence and uniqueness of the classical recursive utility with non-unitary EIS and RRAD. Compared to their works, we consider additional utility of investment gains and losses, which is motivated by the aforementioned models of narrow framing. In addition, we consider unitary EIS and RRAD as well, and also study portfolio selection problems for agents with preferences as specified by \citet{BarberisHuang2008:LossAversionNarrowFraming}. On the other hand, we focus mainly on the case of a finite state space for $\{X_t\}$, while \citet{HansenScheinkman2012:RecursiveUtility} consider a general state space and \citet{BorovivckaStachurski2017:SpectralConditions} consider a compact one. See Section \ref{se:Literature} for a detailed comparison of our results with theirs and with other related works. The finite-state setting helps us to obtain more complete results than those in \citet{HansenScheinkman2012:RecursiveUtility} and \citet{BorovivckaStachurski2017:SpectralConditions}, and also makes it possible to tackle the difficulties in our analysis arising from the gain-loss utility. In addition, although the finite-state setting does not hold in many theoretical models in finance and economics,\footnote{\add See for instance \citet{BansalYaron2004:Risks}, \citet{HansenEtal2008:Consumption}, and \citet{SchorfheideEtal2018:Identifying} for such models.} it is still sufficient for many financial applications. First, finite-state Markov processes can be sufficiently flexible to describe financial data. Second, we do not impose any assumption on $\{Y_t\}$; in particular, $Y_t$ can be unbounded, so our framework accommodates the model of \citet{BarberisHuang2008:LossAversionNarrowFraming}, in which the state space for $\{X_t\}$ is a singleton and $Y_{t+1}$ follows a two-dimensional normal distribution. Third, in many numerical experiments, the state space, even when assumed to be infinite in a theoretical model, is discretized to a set of finite elements; see e.g., \citet{CampbellEtal2001:StockMarket}.}

{\add Because of the Markovian setting, identifying the agent's utility process is equivalent to solving the fixed point for an operator as defined by \eqref{eq:RecursiveUtilityMarkov} in the following, and this fixed point represents the agent's total utility divided by her consumption in the current period as a function of the market state.} We prove that when a growth condition holds, for any values of the EIS and RRAD, {\add the fixed point of the operator---or, equivalently, the agent's utility process---}uniquely exists when her gain-loss utility in each period is nonnegative. When the gain-loss utility can be negative in some states, however, the utility process can be non-existent or non-unique even in some simple settings, such as in the setting in which the EIS is less than or equal to one and the state space of $\{X_t\}$ is a singleton. In this case, we propose a sufficient condition under which the utility process uniquely exists, and this condition is nearly necessary.

We also prove that if the utility process uniquely exists, it can be obtained by starting from any positive utility as an initial guess and applying the recursive equation that defines the utility process repeatedly. This result is not only computationally useful but also economically important: it shows that as the number of periods in a finite-horizon model goes to infinity, the agent's utility in that model, for any specification of the terminal utility, converges to the one in the corresponding infinite-horizon model.

We then consider a portfolio selection problem involving an agent whose preferences are represented by the model of narrow framing proposed in \citet{BarberisHuang2008:LossAversionNarrowFraming}. We prove that a consumption-investment plan is optimal if and only if it, together with the value function of the portfolio selection problem, satisfies a dynamic programming equation. Moreover, we prove that the solution to the dynamic programming equation uniquely exists and can be computed by solving the equation recursively with any initial guess. As a result, the portfolio selection problem in a finite-horizon setting approaches that in the infinite horizon setting as the number of periods in the former goes to infinity.

We also extend {\add some of} our results to the case of a non-finite state space. Assuming nonnegative gain-loss utility, we prove the existence of the utility process for non-unitary EIS and uniqueness with further conditions on the EIS and RRAD, {\add and our results generalize those in \citet{HansenScheinkman2012:RecursiveUtility}.}

Technically, with nonnegative gain-loss utility, the proof of existence of the utility process in the present paper follows closely the approach taken by \citet{HansenScheinkman2012:RecursiveUtility} and is based on the classical Perron-Frobenius theory, {\add although some adaption is needed due to the gain-loss utility}. The proof of existence in the case of negative gain-loss utility and the proof of uniqueness in general, however, cannot follow the same approach, so we develop new methods to accomplish the proof. In addition to proving existence and uniqueness of the agent's utility process with gain-loss utility for the first time in the literature, our results, when confined to the recursive utility model, also improve upon the existing results; see the detailed literature review provided in Section \ref{se:Literature}. {\add Finally, the study of the portfolio selection problem and the techniques used therein are completely new.}

It is not only mathematically interesting but also economically important to study the issue of the existence and uniqueness of the utility process in the generalized recursive utility model. Our results show that with negative gain-loss utility, the utility process in the model of narrow framing is nonexistent or non-unique and thus is not well defined if the agent's EIS is less than or equal to one. Note that in many applications of the model of narrow framing, the gain-loss utility is indeed negative and the EIS is indeed less than one; see e.g., \citet{BarberisHuang:2009PreferenceWithFrames}, \citet{DeGiorgiLegg2011:NarrowFramingProbabilityWeighting}, and \citet{EasleyYang2014:LoassAversionSurvival}. Thus, our results suggest that one should use the model of narrow framing cautiously. Inspired by this observation, \citet{GuoHe2017:NewModelNarrowFraming} propose a new preference model that allows for narrow framing, and this new model is able to accommodate negative gain-loss utility while implying a uniquely defined utility process; see the detailed discussion therein.

The remainder of the paper is organized as follows: In Section \ref{se:Literature} we review and compare our results to the literature. In Section \ref{se:Model} we introduce the generalized recursive utility model and in Section \ref{se:ExistenceUniquenessFinite} we prove the existence and uniqueness of the utility process in a finite-state Markovian setting. In Section \ref{se:PortfolioSelection}, we consider a portfolio selection problem with narrow framing and prove the existence and uniqueness of the solution to the corresponding dynamic programming equaiton. In Section \ref{se:GeneralState}, we provide some extensions of the existence and uniqueness results to the non-finite-state Markovian setting. Section \ref{se:Conclusion} concludes. Proofs are presented in the Appendix.

\section{Literature Review}\label{se:Literature}
Recursive utility is a classical model for individual's preferences with respect to discrete-time consumption streams; see  \citet{KrepsPorteus1978:TemporalResolutionOfUncertainty} and \citet{EpsteinLZinS:89rut}. In an infinite-horizon setting, the recursive utility of consumption stream $C_t$, $t=0,1,\dots$ that is derived by an agent is represented by $U_{t}$, $t=0,1,\dots$, where $U_t$ stands for the utility of the consumption stream starting from time $t$, i.e., $C_s$, $s\ge t$. The recursive utility process $\{U_t\}$ is defined recursively by
\begin{align}\label{eq:RecursiveEquationClassical}
 U_t = H(C_t,\cM_t(U_{t+1})),\; t=0,1,\dots,
\end{align}
where $\cM_t(X)$ stands for the {\em certainty equivalent} of random quantity $X$ conditional on the information at time $t$ and $H(c,z)$ is an {\em aggregator}. There are various choices of the certainty equivalent and aggregator, but the following one, which was first proposed by \citet{KrepsPorteus1978:TemporalResolutionOfUncertainty}, is popular due to its tractability in deriving asset pricing results \citep[see e.g.,][]{EpsteinZin:1990FirstOrderRiskAverionEquityPremiumPuzzle,EpsteinLZinS:91rue}:
\begin{align}
H(c,z):&=
\begin{cases}
[(1-\beta)c^{1-\rho}+\beta z^{1-\rho}]^{\frac{1}{1-\rho}}, & 0<\rho\neq1,\\
e^{(1-\beta)\ln c+\beta\ln z},&\rho=1,
\end{cases}\label{eq:PowerAggregator}
\\
\cM_t(X):&= u^{-1}\big(\expect_t[u(X)]\big),\quad u(x):=
\begin{cases}
x^{1-\gamma},& 0<\gamma\neq1,\\
\ln(x),& \gamma=1,
\end{cases}\label{eq:PowerCE}
\end{align}
where $\expect_t$ stands for the expectation operator conditional on the information at time $t$. In addition, $\beta\in (0,1)$ is a discount rate, $\gamma$ stands for the relative risk aversion degree (RRAD),\footnote{Note that any affine transformation of $u$ does not affect the certainty equivalent $\cM_t$. In particular, in some literature $u$ takes the form $u(x)=x^{1-\gamma}/(1-\gamma)$ so that it is increasing and thus can be directly interpreted as an utility function. The form of $u$ used in the present paper is notational simpler to use in the following analysis.} and $1/\rho$ is the elasticity of intertemporal substitution (EIS); see e.g., \citet{KrepsPorteus1978:TemporalResolutionOfUncertainty} and \citet{EpsteinLZinS:89rut}.

In the following, when $\rho\ge 1$, we set {\add $H(c,0):= \lim_{z\downarrow 0}H(c,z) = 0$ for $c>0$, $H(0,z):= \lim_{c\downarrow 0}H(c,z) = 0$ for $z>0$, and $H(0,0):=\lim_{c\downarrow 0,z\downarrow 0}H(c,z)=0$.} As a result, $H(c,z)$ is well defined, takes real values, and continuous in $(c,z)\in[0,\infty)^2$. Similarly, when $\gamma= 1$, we define $u(0):=-\infty$ and $u^{-1}(-\infty):=0$; when $\gamma>1$, we define $u(0):=+\infty$ and $u^{-1}(+\infty):=0$. As a result, $\cM_t(X)$ is well defined for any nonnegative random variable $X$ {\add and increasing in $X$}. Moreover, when $\gamma\ge 1$ and $X=0$ with a positive probability, $\cM_t(X)=0$.

Note that in the infinite-horizon setting the recursive utility process is defined recursively {\em without} a terminal condition, so the existence and uniqueness of this process is not automatically guaranteed. In the following, we review the relevant literature. Note that when $\rho\ge 1$, $U_t\equiv 0$ is a trivial solution to \eqref{eq:RecursiveEquationClassical}, so in this case a non-trivial solution is referred to in the following discussion.

\citet{EpsteinLZinS:89rut} prove the existence of the recursive utility process when the aggregator is given by \eqref{eq:PowerAggregator} with $\rho<1$, assuming that consumption processes essentially have bounded growth rates.\footnote{\citet{EpsteinLZinS:89rut} use a different set of notations from ours: $\rho$ and $\alpha$ therein correspond to $1-\rho$ and $1-\gamma$, respectively, in the present paper. In the following discussion, we follow the notation used throughout the present paper. \citet{EpsteinLZinS:89rut} prove the existence of the recursive utility process when $\rho<1$, assuming that consumption processes essentially have bounded growth rates; see Theorem 3.1 and the definition of $D(b)$ therein. Although the authors also construct a solution to the recursive equation when $\rho>1$, this solution can be trivial; see the proof of Theorem 3.1 on pp. 964--965.}  \citet{Ma1993:MarketEquilibrium,Ma1996:Corrigendum,Ma1998:Attitudes} prove the existence and uniqueness of the recursive utility process by assuming that $H_z(c,z)$, the derivative of the aggregator $H(c,z)$ with respect to $z$, is bounded uniformly in $c$ and $z$ by a number that is strictly less than one.\footnote{See Assumption W4 in \citet[p. 246]{Ma1993:MarketEquilibrium} and \citet[p. 568]{Ma1996:Corrigendum}. In \citet{Ma1998:Attitudes}, the author assumes that the recursive utility for deterministic consumption flows is well defined, but this requires $H_z(c,z)$ to be bounded by a number strictly less than one as well; see Footnote 5 of \citet{Ma1998:Attitudes} and Assumption W5 in \citet{LucasJrStokey1984:OptimalGrowth}.} However, this assumption does not hold for $H$ as defined in \eqref{eq:PowerAggregator} for any $\rho>0$. \citet{Balbus2016:NonNegative} assumes that there exists $r\in(0,1)$ such that $H(c,tz)\ge t^rH(c,z)$ for any $c\ge 0$, $z>0$, and $t\in(0,1)$, which cannot hold for $H$ as defined in \eqref{eq:PowerAggregator} for any $\rho\neq 1$.\footnote{See Assumption 3 therein. Note that a similar assumption is made by \citet{LeVan_etal2008:Monotone} in their study of monotone, concave operators, so their results cannot be applied here either; see condition (P1) therein.} \citet{OzakiStreufert1996:DynamicProgramming} prove the existence and uniqueness of the recursive utility process by assuming $H_z(c,z)$ to be uniformly bounded in $c$ and $z$ and a set of conditions to hold.\footnote{In \citet[Theorem D]{OzakiStreufert1996:DynamicProgramming}, the authors assume that $\bar \delta$ and $\delta$ therein are finite, which is equivalent to assuming that $H_z(c,z)$ is bounded; see pages 403--406 therein.} However, these conditions are difficult to verify; see conditions N1--N12 in \citet[pp. 404--405]{OzakiStreufert1996:DynamicProgramming}; in addition, for $H$ as defined in \eqref{eq:PowerAggregator}, $H_z(c,z)$ is not bounded when $\rho\le 1$.

\citet{MarinacciMontrucchio2010:UniqueSolutions} consider Thompson and Blackwell aggregators and study the existence and uniqueness of the recursive utility process with these two types of aggregator.\footnote{In a recent work, \citet{Becker2017:RecursiveUtility} derive some results that are essentially the same as those in \citet{MarinacciMontrucchio2010:UniqueSolutions}. } One can check that $H$ as defined in \eqref{eq:PowerAggregator} satisfies properties (W-i), (W-ii), and (W-iii) in \citet[p. 1783]{MarinacciMontrucchio2010:UniqueSolutions}, satisfies property (W-iv) therein if and only if $\rho<1$, and does not satisfy property (W-v) therein for any $\rho>0$. Thus, $H$ as defined in \eqref{eq:PowerAggregator} with $\rho<1$ is a Thompson aggregator, but the case in which $\rho\ge 1$ is neither Thompson nor Blackwell. Moreover, $u$ as defined in \eqref{eq:PowerCE} is constant relative risk averse (CRRA), i.e., $-xu''(x)/u'(x)$ is constant in $x$, so Theorem 3-(ii) of \citet{MarinacciMontrucchio2010:UniqueSolutions} applies, showing that (i) the recursive utility process exists if consumption is bounded at each time (but the bound can be dependent on time) and (ii) uniqueness follows if the consumption growth rate satisfies a restrictive assumption.\footnote{The aggregator $H$ is $\gamma$-subhomogeneous, as defined in \citet[p. 1784]{MarinacciMontrucchio2010:UniqueSolutions}, for any $\gamma\in(0,1]$. Thus, Theorem 3-(ii) in \citet{MarinacciMontrucchio2010:UniqueSolutions} implies the existence of the recursive utility process when consumption processes belong to $L_+(w^{1/\gamma})$ for some weight function $w$, which essentially means that consumption is bounded at each time; see Section 2.2 therein. To have uniqueness, one needs to further assume that $X_t:=H(C_t,0) = (1-b)^{1/(1-\rho)}C_t, t=0,1,\dots$ satisfies $[X]_{w^{1/\gamma}}>0$. This condition, together with the condition that the consumption process is in $L_+(w^{1/\gamma})$, nearly implies that the consumption growth rate is a constant as $t$ goes to infinity; see Section 2.2 therein.} The case in which $\rho\ge 1$, however, is not studied by \citet{MarinacciMontrucchio2010:UniqueSolutions}.\footnote{Alternatively, one can consider the following transformation: $\tilde V_t:= f(V_t)$, where $f(x):=x^{1-\rho}$ when $\rho\neq 1$ and $f(x):=\ln(x)$ when $\rho=1$. Then, we have $\tilde V_t = \tilde H(C_t,\tilde \cM_t(\tilde V_{t+1}))$ for a new aggregator $\tilde H$. However, $\tilde H(0,z)$ is finite only if $\rho<1$, and the aggregators considered in \citet{MarinacciMontrucchio2010:UniqueSolutions} are assumed to take real values for any $c,z\ge 0$, so the results in \citet{MarinacciMontrucchio2010:UniqueSolutions} do not apply to the case $\rho\ge 1$ either, even if we perform the transformation.}

\citet{HansenScheinkman2012:RecursiveUtility} assume that the consumption growth rate $\frac{C_{t+1}}{C_{t}}=e^{\kappa(X_t,X_{t+1}, Y_{t+1})}$ for some function $\kappa$, where $\{X_t\}$ is a Markov process and the joint distribution of $(X_{t+1},Y_{t+1})$ conditional on $(X_t,Y_t)$ depends only on $X_t$. They show that for $H$ and $\cM_t$ as defined, respectively, in \eqref{eq:PowerAggregator} and \eqref{eq:PowerCE} with $\rho\neq 1$ and $\gamma\neq 1$, if a growth condition on the consumption process holds, the recursive utility process exists. They also show the uniqueness of the recursive utility process when $(1-\gamma)/(1-\rho)\ge 1$.

In the present paper, we consider a generalization of the recursive utility by adding to the recursive equation \eqref{eq:RecursiveEquationClassical} a component that represents utility of investment gains and losses, and this generalization allows us to accommodate a variety of utility models with narrow framing; see Section \ref{se:Model} below. We then prove that the utility process in our model (i) uniquely exists and (ii) is globally attracting in that it can be obtained by starting from any initial guess and applying the recursive equation that defines the utility process repeatedly.

Our results, when refined to the case of recursive utility, generalize the above literature as well. First, in the finite-state Markovian setting, we obtain the existence and uniqueness of the utility process for any values of $\rho$ and $\gamma$ under a mild growth condition on the consumption process, although no complete results have yet been obtained in the literature. Second, we also prove that the utility process is globally attracting, whereas in the aforementioned works, without uniqueness, the authors can only prove that the utility process is locally attracting in that it can be computed by starting from certain specific initial guesses only. Third, we also consider a portfolio selection problem, leading to a dynamic programming equation, and show that the solution to this equation is existent, unique, and globally attracting. \citet{EpsteinLZinS:89rut} consider a portfolio selection problem and show that the corresponding dynamic programming equation admits a solution, assuming certain conditions on asset returns and consumption growth rates; see Theorem 5.1 therein. After studying the existence and uniqueness of the recursive utility process, \citet{OzakiStreufert1996:DynamicProgramming} consider a portfolio selection problem and obtain the existence of the solution to the corresponding dynamic programming equation. Neither of these works, however, proves the uniqueness of the solution. Fourth, in a general Markovian setting, we generalize the results in \citet{HansenScheinkman2012:RecursiveUtility} by proving the uniqueness of the recursive utility process in the case $(1-\gamma)/(1-\rho)\in(0,1)$ and in the case $\gamma=\rho=1$.

Finally, we would like to mention a recent work by \citet{BorovivckaStachurski2017:SpectralConditions} that was carried out independently of and simultaneously with ours. These authors study the existence and uniqueness of the recursive utility process with $H$ and $\cM_t$ as defined, respectively, in \eqref{eq:PowerAggregator} and \eqref{eq:PowerCE}. Using the same Markovian setting as the one in \citet{HansenScheinkman2012:RecursiveUtility} and assuming the state process $\{X_t\}$ to be compact, the authors prove, for the case of $\rho\neq 1$ and $\gamma\neq 1$, that the existence, uniqueness, and global attractingness of the recursive utility process are all equivalent to a simple condition on the spectral radius of a certain operator that is associated with the recursive equation \eqref{eq:RecursiveEquationClassical}. We would like to emphasize that \citet{BorovivckaStachurski2017:SpectralConditions} and the present paper have different focuses, and the results in these two papers are largely different. First, \citet{BorovivckaStachurski2017:SpectralConditions} derive a sufficient and necessary condition for the existence and uniqueness of the recursive utility process, whereas the literature, including the present paper, is as yet unable to prove the necessity. Furthermore, their assumption on the state space of $\{X_t\}$ is weaker than ours: A finite state space is always compact. We, however, prove the existence and uniqueness (i) in the case in which the state space is finite and $\rho=1$ or $\gamma=1$ and (ii) in the case in which the state space can be noncompact and $(1-\gamma)/(1-\rho)\in (0,1)$, and these two cases are not covered by \citet{BorovivckaStachurski2017:SpectralConditions}. In addition, our approach to proving existence and uniqueness is different from the one employed in \citet{BorovivckaStachurski2017:SpectralConditions}. Second, we consider gain-loss utility, which is largely motivated by a set of models of narrow framing in the literature, whereas \citet{BorovivckaStachurski2017:SpectralConditions} focuses on the classical recursive utility model. Third, we also study a portfolio selection problem and the associated dynamic programming equation.

Table \ref{ta:Literature} summarizes the comparison of our results to the literature.

\begin{landscape}

\begin{table}
\centering
\caption{\small Comparison to the literature. We compare the results obtained regarding recursive utility in the present paper to those obtained by \citet{EpsteinLZinS:89rut} (EZ89), \citet{OzakiStreufert1996:DynamicProgramming} (OS96), \citet{MarinacciMontrucchio2010:UniqueSolutions} (MM10), \citet{HansenScheinkman2012:RecursiveUtility} (HS12), and a recent work, \citet{BorovivckaStachurski2017:SpectralConditions} (BS19), that was conducted independently of and simultaneously with ours. Here, $1/\rho$ refers to the EIS and $\gamma$ refers to the RRAD. The second column describes the assumptions made on the consumption process $\{C_t\}$. The third, fourth, and fifth columns show the conditions under which the recursive utility process is existent, unique, and (globally or locally) attracting, respectively. The sixth column shows whether gain-loss utility is considered. The last column shows the existence, uniqueness, and attractingness of the solution to the dynamic programming equation in a portfolio selection problem. A blank cell denotes a case in which the corresponding component/problem is not considered.}\label{ta:Literature}

\medskip

\small
    \begin{tabular}{  p{0.06\textwidth} | p{0.35\textwidth} | p{0.1\textwidth} | p{0.12\textwidth} | p{0.22\textwidth} | p{0.1\textwidth} | p{0.23\textwidth}|}
\noalign{\hrule height 1.3pt}
    & \centering Assumption & Existence & Uniqueness & \centering Attractingness &  Gain-loss & \multicolumn{1}{c|}{DP equation} \\
    \noalign{\hrule height 0.8pt}
   EZ89 & non-Markovian; bounded $C_{t+1}/C_t$& $\rho< 1$ & & local &  &existence and local attractingness for $\rho<1$\\ \hline
   OS96 & non-Markovian; 12 conditions & $\rho>1$ & $\rho>1$ &  & &existence and local attractingness for $\rho>1$\\ \hline
    MM10 & non-Markovian; bounded $C_t$ for existence and restrictive assumptions on $C_{t+1}/C_t$ for uniqueness and attractingness & $\rho<1$ & $\rho<1$ & global &  &\\ \hline
    HS12 & General Markovian; a certain  growth condition  & $\rho\neq 1$ and $\gamma\neq 1$ & $\frac{1-\gamma}{1-\rho}\ge 1$ & global when $\frac{1-\gamma}{1-\rho}\ge 1$  and local otherwise &  &\\
    \hline
    BS19 & compact-state-space Markovian; a sufficient and necessary condition & $\rho\neq 1$ and $\gamma\neq 1$ & $\rho\neq 1$ and $\gamma\neq 1$ & global &  &
    \\
      \noalign{\hrule height 0.8pt}
    & finite-state-space Markovian; a certain growth condition & any $\rho$, $\gamma$ & any $\rho$, $\gamma$  & global & any & existence, uniqueness, and global attractingness for any $\rho$, $\gamma$\\
    \cline{2-7}
   This paper & General Markovian; a certain growth condition & $\rho\neq 1$ or $\rho=\gamma=1$ & $\frac{1-\gamma}{1-\rho}\ge 1$ or $\rho=\gamma=1$ & global when $\frac{1-\gamma}{1-\rho}\ge 1$ or $\rho=\gamma=1$ and local otherwise & non-negative & \\
     \cline{2-7}
       & General Markovian; certain growth condition & $\rho\neq 1$ or $\rho=\gamma=1$ & $\frac{1-\gamma}{1-\rho}>0$ or $\rho=\gamma=1$ & global when $\frac{1-\gamma}{1-\rho}>0$ or $\rho=\gamma=1$ and local otherwise &  & \\
       \noalign{\hrule height 1.5pt}
    \end{tabular}
   \end{table}

\end{landscape}

\section{Model and Examples}\label{se:Model}
\subsection{Model}\label{subse:Model}
Consider the following equation
\begin{align}\label{eq:RecursiveEquation}
  V_t = H\left(c_t,\cM_t(A_{t+1}V_{t+1})+B_t\right),\; t=0,1,\dots,
\end{align}
where the aggregator $H$ and certainty equivalent $\cM_t$ are given by \eqref{eq:PowerAggregator} and \eqref{eq:PowerCE}, respectively. Here, $\{c_t\}$ stands for a {\em consumption propensity} process (i.e., $c_t$ stands for the percentage of wealth that is used for consumption at time $t$), $\{A_t\}$ is a process that is used to model portfolio returns, and $\{B_t\}$ is used to model the {\add utility of investment gains and losses per unit of wealth}. Our goal is to establish the existence and uniqueness of the solution $\{V_t\}$ to this equation, {\add which represents the agent's total utility per unit wealth.}

 Following \citet{HansenScheinkman2012:RecursiveUtility}, we consider equation \eqref{eq:RecursiveEquation} in a Markovian environment. More precisely, we consider a Markov process $\{(X_t,Y_t)\}$ and assume the following:
 \begin{assumption}\label{as:MarkovProperty}
 \begin{enumerate}
   \item[(i)]  $\{(X_t,Y_t)\}$ is a Markov process and the joint distribution of $(X_{t+1},Y_{t+1})$ conditional on $(X_t,Y_t)$ depends only on $X_t$.
   \item[(ii)] Consumption propensity and portfolio return dynamics evolve according to
   \begin{align*}
     \log(c_{t+1})-\log(c_t) + \log A_{t+1} = \kappa (X_t, X_{t+1},Y_{t+1}),\; t=0,1,\dots
   \end{align*}
   for some real-valued measurable function $\kappa$.
   \item[(iii)] $B_t/c_t = \varpi(X_t),t=0,1,\dots$ for some real-valued measurable function $\varpi$.
   \item[(iv)] For any state $x$, $\expect_t\left[u\left(e^{\kappa(X_t,X_{t+1},Y_{t+1})}\right)|X_t=x\right]$ exists.
 \end{enumerate}
 \end{assumption}

Assumption \ref{as:MarkovProperty}-(i) is the same as Assumption 1-a) in \citet{HansenScheinkman2012:RecursiveUtility}; {\add it implies that $\{X_t\}$ is a Markov process, and we denote its state space as $\mathbb{X}$. On the other hand, this assumption holds if $\{X_t\}$ is Markovian and $\{Y_t\}$, conditional on $\{X_t\}$, is an independent time series.}
 Assumptions \ref{as:MarkovProperty}-(ii) and -(iii) are parallel to Assumption 1-b) in \citet{HansenScheinkman2012:RecursiveUtility}, which ensure a Markovian structure in equation \eqref{eq:RecursiveEquation}. {\add Compared to the setting in \citet{HansenScheinkman2012:RecursiveUtility}, we have two additional terms, $A_t$ and $B_t$; the relevance of adding them to the model and the above Markovian assumption will become clear in Section \ref{subse:ModelExamples}.}
 We assume the state space $\mathbb{X}$ to be a metric space, so the measurability in Assumption \ref{as:MarkovProperty} is with respect to Borel $\sigma$-algebra of $\mathbb{X}$.

Dividing \eqref{eq:RecursiveEquation} by $c_t$ on both sides and using the homogeneity of $H$, we obtain
\begin{align}\label{eq:RecursiveEquationNormalized}
  V_t/c_t = H\Big(1,\cM_t\big(A_{t+1}(c_{t+1}/c_t)(V_{t+1}/c_{t+1})\big)+B_t/c_t\Big),\; t=0,1,\dots.
\end{align}
Thus, to solve equation \eqref{eq:RecursiveEquation}, we only need to solve $\{V_t/c_t\}$ from \eqref{eq:RecursiveEquationNormalized}. Moreover, because of Assumption \ref{as:MarkovProperty}, we restrict ourselves to Markovian solutions to \eqref{eq:RecursiveEquationNormalized}, i.e., $V_t/c_t = f(X_t),t=0,1,\dots$ for some function $f$. Then, the solution to equation \eqref{eq:RecursiveEquationNormalized} becomes the fixed point of operator $\mathbb T$, defined as
\begin{align}\label{eq:RecursiveUtilityMarkov}
  \mathbb T f(x):=H\left(1,u^{-1}\left(\expect_t\left[u\left(e^{\kappa(X_t,X_{t+1},Y_{t+1})}f(X_{t+1})\right)|X_t=x\right]\right)+\varpi(x)\right),\; x\in \mathbb X.
\end{align}
{\add Note that $f$ represents the agent's total utility divided by her consumption in the current period.}

Denote by ${\cal X}$ the space of measurable functions on $\mathbb{X}$, ${\cal X}_+$ the space of nonnegative measurable functions on $\mathbb{X}$, i.e., ${\cal X}_+:=\{f\in{\cal X}|f(x)\ge 0,x\in \mathbb{X}\}$, ${\cal X}_+^o$ the space of nonnegative functions on $\mathbb{X}$ that are not zero, i.e., ${\cal X}_+^o:=\{f\in{\cal X}_+|f\neq 0\}$, and ${\cal X}_{++}$ the space of positive functions on $\mathbb{X}$, i.e., ${\cal X}_{++}:=\{f\in{\cal X}|f(x)>0,x\in \mathbb{X}\}$. Recalling the definitions of $H$, $u$, and $\mathbb{T}$, we can see that the domain of $\mathbb{T}$ is contained in ${\cal X}_+$.

In the following, {\add denote by $\mathbb{R}$ the set of real numbers. For $a\in\mathbb{R}$, we denote $a^+:=\max(a,0)$ and $a^-:=\max(-a,0)$. For} any $f\in{\cal X}$, we denote its positive part as $f^+$, i.e., $f^+(x):=\max(f(x),0)$. For any $f_1,f_2\in {\cal X}$, $f_1\ge f_2$ means $f_1(x)\ge f_2(x),x\in\mathbb{X}$ and $f_1>f_2$ means $f_1(x)>f_2(x),x\in\mathbb{X}$. Any $a\in \mathbb{R}$ also denotes the function on $\mathbb{X}$ that takes value $a$ in all states.

\subsection{Examples}\label{subse:ModelExamples}
\subsubsection{Recursive Utility Model}\label{subsubse:RecursiveUPortfolio}
Recall the recursive utility model \eqref{eq:RecursiveEquationClassical}. Denote by $\{W_t\}$ the agent's wealth process corresponding to a consumption strategy, i.e., a consumption process $\{C_t\}$, and an investment strategy, i.e., the process of the dollar amount invested in asset $i$, $\{\Theta_{i,t}\}$, $i=1,\dots,n$. Then, the wealth dynamics evolve according to
\begin{align*}
  W_{t+1} = \left(W_t-C_t-\sum_{i=1}^n\Theta_{i,t}\right)R_{f,t+1}+\sum_{i=1}^n\Theta_{i,t}R_{i,t+1},\; t=0,1,\dots,
\end{align*}
where $R_{i,t+1}$ and $R_{f,t+1}$ are the gross returns of asset $i$ and the risk-free asset, respectively, in period $t$ to $t+1$.
Because of the homogeneity of $H$, $\cM_t$, and $G_{i,t}$, the agent's {\em utility per unit wealth}, $U_t/W_t$, satisfies
\begin{align*}
  U_t/W_t = H\left(c_t,\cM_t\Big((1-c_t)R_{p,t+1}(U_{t+1}/W_{t+1})\Big)\right)
\end{align*}
where $c_t:=C_t/W_t$ is the consumption propensity at time $t$, $\theta_{i,t}:=\Theta_{i,t}/(W_t-C_t)$ is the percentage allocation to risky asset $i$ at time $t$, $i=1,\dots,n$, and
\begin{align}\label{eq:PortfolioReturn}
  R_{p,t+1}:=R_{f,t+1}+\sum_{i=1}^n\theta_{i,t}(R_{i,t+1}-R_{f,t+1})
\end{align}
is the portfolio return in period $t$ to $t+1$. If we denote $V_t:=U_t/W_t$, then $\{V_t\}$ solves \eqref{eq:RecursiveEquation} with $A_{t+1}  = (1-c_t)R_{p,t+1}$ and $B_t=0$.

\subsubsection{Models that Allow for Narrow Framing}\label{subsubse:NarrowFramingPortfolio}
\citet{BarberisHuang:2009PreferenceWithFrames,BarberisHuang2008:LossAversionNarrowFraming} and \citet{BarberisEtal2006:StockMarketParticipationNarrowFraming} consider a model of narrow framing: in addition to consumption utility, the agent evaluates each risky asset in a separate mental account and derives utility from the investment gain and loss in the asset. Thus, the agent's utility process $\{U_t\}$ is defined recursively as follows:
\begin{align}\label{eq:NarrowFraming}
  U_t =  H\left(C_t,\cM_t(U_{t+1})+\sum_{i=1}^nb_{i}G_{i,t}\right),\; t=0,1,\dots,
\end{align}
where $b_i\ge 0$ is a constant and $G_{i,t}$ stands for the utility of the gain and loss experienced by the agent for her investment in asset $i$.

In \citet{BarberisHuang:2009PreferenceWithFrames,BarberisHuang2008:LossAversionNarrowFraming} and \citet{BarberisEtal2006:StockMarketParticipationNarrowFraming},
\begin{align}\label{eq:UtilityGainLossDollar}
  G_{i,t} = \expect_t\left[\max\big(\Theta_{i,t}(R_{i,t+1}-R_{f,t+1}),0\big)+ k \min\big(\Theta_{i,t}(R_{i,t+1}-R_{f,t+1}),0\big)\right]
\end{align}
for some $k\ge 1$. Indeed, $G_{i,t}$ represents the preference value of the agent's position in asset $i$ under prospect theory \citep{KahnemanDTverskyA:79pt,TverskyKahneman1992:CPT} with the reference point as the risk-free return, utility function as a piece-wise linear function in which parameter $k$ measures the loss aversion degree of the agent, and no probability weighting. Thus, $G_{i,t}$ captures the agent's utility of the gain and loss for her investment in asset $i$ due to narrow framing. Later, \citet{DeGiorgiLegg2011:NarrowFramingProbabilityWeighting} generalize \eqref{eq:UtilityGainLossDollar} by considering a piece-wise power utility function and nonlinear probability weighting functions. \citet{HeZhou2013:ReferencePoint} consider the case in which there is only one risky asset, but the reference point therein can be different from the risk-free return. All of the above variants of the model of narrow framing can be written in the form \eqref{eq:NarrowFraming}.\footnote{In \citet{BarberisNHuangM:01ma}, \citet{BarberisNHuangMSantosT:01ptap}, and \citet{LiYang2013:AssetPricing} the utility of gains and losses is scaled by a power transformation of the aggregate consumption in the market, and their models take the form of \eqref{eq:NarrowFraming} with a time-varying, random $b_i$.}

Now, define $V_t:=U_t/W_t$ to be the agent's utility per unit wealth in the model of narrow framing and $c_t:=C_t/W_t$ as the consumption propensity at time $t$. Then, $\{V_t\}$ solves equation \eqref{eq:RecursiveEquation} with $A_{t+1}  = (1-c_t)R_{p,t+1}$ and $B_t=\big(\sum_{i=1}^nb_{i}G_{i,t}\big)/W_t$.

\subsubsection{Markovian Assumption}\label{subsubse:MarkovianSetting}
In the following, we show that Assumption \ref{as:MarkovProperty} {\add is appropriate for the above  examples.} To this end, we consider the model of narrow framing \eqref{eq:NarrowFraming} with gain-loss utility $G_{i,t}$ specified in \eqref{eq:UtilityGainLossDollar} and assume for simplicity that $\theta_{i,t}\ge 0$. Denote
\begin{align}\label{eq:UtilityGainLoss}
  g_{i,t} = \expect_t\left[(R_{i,t+1}-R_{f,t+1})\mathbf 1_{R_{i,t+1}>R_{f,t+1}}+ k (R_{i,t+1}-R_{f,t+1})\mathbf 1_{R_{i,t+1}<R_{f,t+1}}\right],
\end{align}
which {\add stands for the utility of gains and losses per unit of investment in asset $i$.}
Then, $V_t:=U_t/W_t$ solves equation \eqref{eq:RecursiveEquation} with
\begin{align}\label{eq:ABinNarrowFraming}
  A_{t+1} = (1-c_t)R_{p,t+1},\quad B_t = (1-c_t)\sum_{i=1}^nb_i \theta_{i,t}g_{i,t}.
\end{align}
Note {\add that $A_{t+1}$ stands for the growth of the agent's wealth in period $t$ to $t+1$ and $B_t$ stands for the agent's utility of gains and losses per unit of wealth.}

Suppose that the gross return rate of risky asset $i$ in period $t$ to $t+1$ is $R_{i,t+1} = r_i(X_t,X_{t+1},Y_{t+1})$, $t=0,1,\dots$, for some function $r_i$ and that the gross return rate of the risk-free asset in period $t$ to $t+1$ is $R_{f,t+1} = r_0(X_t)$, $t=0,1,\dots$, for some function $r_0$. Because, conditional on $(X_t,Y_t)$, the joint distribution of $(X_{t+1},Y_{t+1})$ depends only on $X_t$, it is natural for the agent to consider Markovian strategies only, i.e., to consider $c_t = c(X_t)$, $\theta_{i,t} = \theta_{i}(X_t)$, $t=0,1,\dots$, $i=1,\dots,n$, for some functions $c$ and $\theta_i$'s. {\add Consequently, Assumption \ref{as:MarkovProperty}-(ii) holds. On the other hand, one can see that $g_{i,t}$ depends on $X_t$ only, so the utility of gains and losses per unit of wealth $B_t$ is a function of $X_t$. Because the consumption propensity $c_t=c(X_t)$, we conclude that Assumption \ref{as:MarkovProperty}-(iii) holds as well.}

Finally, we have
\begin{align*}
   \frac{A_{t+1}c_{t+1}}{c_t} = \frac{(1-c_t)R_{p,t+1}c_{t+1}}{c_t}  = \frac{\big(1-C_t/W_t\big) \big(W_{t+1}/(W_t-C_t)\big)\big(C_{t+1}/W_{t+1}\big)}{C_t/W_t}=\frac{C_{t+1}}{C_t}.
\end{align*}
Thus, $\exp\big(\kappa(X_t,X_{t+1},Y_{t+1})\big)$ stands for consumption growth rate $C_{t+1}/C_t$.

\section{Existence, Uniqueness, and Convergence}\label{se:ExistenceUniquenessFinite}

In this section, we study the existence and uniqueness of the solution to \eqref{eq:RecursiveEquation}---that is, of the fixed point of \eqref{eq:RecursiveUtilityMarkov}---when the state space of $\{X_t\}$ is finite. Thus, we impose
\begin{assumption}\label{as:FiniteState}
     The state space for $\{X_t\}$ is finite and $\{X_t\}$ is irreducible.
\end{assumption}

\citet{HansenScheinkman2012:RecursiveUtility} consider a general Markov process when studying the solution to \eqref{eq:RecursiveEquationClassical}, and they {\add implicitly }assume the existence of the Perron-Frobenius eigenvalue and eigenvector of a linear operator and the stochastic stability of $\{X_t\}$ after a change of measure; {\add see equation [4] and Assumption 2 therein. These assumptions hold automatically when Assumption \ref{as:FiniteState} is in place; see Proposition \ref{prop:Perron} below. Their results, however, cannot be applied here because we consider utility of gains and losses in our model, and even for the case of recursive utility, they do not obtain the uniqueness for a large range of parameter values; see the detailed discussion following Theorem \ref{th:ExistUniqueFiniteStatePositive} below.}

Note that we assume $\{X_t\}$ to be irreducible. This assumption is necessary for the existence of the stationary distribution of $\{X_t\}$, which will be used in the following. Note also that we do not impose any assumptions on $\{Y_t\}$; in particular, $Y_t$ can be unbounded.

{\add When $\mathbb{X}$ is finite, $\mathbb{M}f$ defined by $\mathbb{M}f(x):=\mathbb{E}_t\left[u\left(e^{\kappa(X_t,X_{t+1},Y_{t+1})}f(X_{t+1})\right)|X_t=x \right],x\in\mathbb{X}$ is continuous in $f\in{\cal X}_{+}$. Indeed, fix any $f\in {\cal X}_+$ and consider a sequence $\{f_n\}$ that converges to $f$. Denote by $\mathbb{X}_1$ the set of $x$ such that $f(x)=0$ and $\mathbb{X}_2$ the set of $x$ such that $f(x)>0$. Define $\epsilon_n:=2^{-n} + \sup_{m\ge n}\max_{x\in\mathbb{X}}|f_m(x)-f(x)|$. Then $\epsilon_n$ decreasingly converges to 0 as $n\rightarrow +\infty$, and $f_n\le f +\epsilon_n$. Consequently, for any fixed $x\in\mathbb{X}$, we have
\begin{align*}
  \mathbb{M}f_n(x)\le u^{-1}\left(\mathbb{E}_t\big[u(e^{\kappa(X_t,X_{t+1},Y_{t+1})}(f(X_{t+1}) + \epsilon_n))|X_t=x\big]\right).
\end{align*}
If $\gamma<1$, or if $\mathbb{X}_1=\emptyset$, or if $\prob(X_{t+1}\in\mathbb{X}_1|X_t=x)=0$, the dominated convergence theorem shows that
\begin{align*}
  \mathbb{E}_t\big[u(e^{\kappa(X_t,X_{t+1},Y_{t+1})}(f(X_{t+1}) + \epsilon_n))|X_t=x\big]\rightarrow \mathbb{E}_t\big[u(e^{\kappa(X_t,X_{t+1},Y_{t+1})}f(X_{t+1})))|X_t=x\big]
\end{align*}
as $n\rightarrow +\infty$, so we conclude that
\begin{align*}
  \limsup_{n\rightarrow +\infty}\mathbb{M} f_n(x) \le \limsup_{n\rightarrow +\infty}u^{-1}\left(\mathbb{E}_t\big[u(e^{\kappa(X_t,X_{t+1},Y_{t+1})}(f(X_{t+1}) + \epsilon_n))|X_t=x\big]\right) = \mathbb{M}f(x).
\end{align*}
If $\gamma >1$ and $\prob(X_{t+1}\in\mathbb{X}_1|X_t=x)>0$, in which case $\mathbb{M}f(x)=0$, we have
\begin{align*}
  \mathbb{E}_t\big[u(e^{\kappa(X_t,X_{t+1},Y_{t+1})}(f(X_{t+1}) + \epsilon_n))|X_t=x\big]\ge \epsilon_n^{1-\gamma}\mathbb{E}_t\big[u(e^{\kappa(X_t,X_{t+1},Y_{t+1})})\mathbf 1_{X_{t+1}\in\mathbb{X}_1}|X_t=x\big]\rightarrow +\infty
\end{align*}
as $n\rightarrow +\infty$, so
\begin{align*}
  &\limsup_{n\rightarrow +\infty}\mathbb{M} f_n(x) \le \limsup_{n\rightarrow 0}u^{-1}\left(\mathbb{E}_t\big[u(e^{\kappa(X_t,X_{t+1},Y_{t+1})}(f(X_{t+1}) + \epsilon_n))|X_t=x\big]\right) \\
  &\le u^{-1}\left(\liminf_{n\rightarrow +\infty}\mathbb{E}_t\big[u(e^{\kappa(X_t,X_{t+1},Y_{t+1})}(f(X_{t+1}) + \epsilon_n))|X_t=x\big]\right) = 0 = \mathbb{M}f(x).
\end{align*}
Similarly, when $\gamma=1$ and $\prob(X_{t+1}\in\mathbb{X}_1|X_t=x)>0$, we also have $\limsup_{n\rightarrow 0}\mathbb{M} f_n(x)\le \mathbb{M}f(x)$.

On the other hand, we need to show $\liminf_{n\rightarrow +\infty}\mathbb{M}f_n(x)\ge \mathbb{M}f(x)$. This is trivially true when $\gamma \ge 1$ and $\prob(X_{t+1}\in\mathbb{X}_1|X_t=x)>0$ because in this case $\mathbb{M}f(x)=0$. In the remaining cases in which $\gamma<1$, or $\mathbb{X}_1=\emptyset$, or $\prob(X_{t+1}\in\mathbb{X}_1|X_t=x)=0$, using the dominated convergence theorem, we can easily show $\liminf_{n\rightarrow +\infty}\mathbb{M}f_n(x)\ge \mathbb{M}f(x)$. Thus, we conclude that $\lim_{n\rightarrow +\infty} \mathbb{M}f_n=f$.

Now, recalling that $H(c,z)$ is continuous in $(c,z)\in[0,+\infty)^2$, we conclude that $\mathbb{T}$ is continuous.} However, we cannot apply the classical Brouwer fixed point theorem to prove the existence and uniqueness of the fixed point of $\mathbb{T}$. First, the domain of $\mathbb{T}$ under consideration in the following, namely, ${\cal X}_{++}$, is not compact. Second, the Brouwer theorem does not imply uniqueness of the fixed point. Third, the Brouwer theorem does not show how to compute the fixed point; however, we will provide an easy algorithm to compute the fixed point.

\subsection{Changing the Probability Measure}
We follow \citet{HansenScheinkman2012:RecursiveUtility} in performing a change of probability measure based on the classical Perron-Frobenius theory. To this end, consider the following operator
\begin{align*}
  \mathbb{U}h (x):=\expect_t\left[u\left(e^{\kappa(X_t,X_{t+1},Y_{t+1})}\right) h(X_{t+1})|X_t=x\right],\; x\in \mathbb{X}.
\end{align*}
With Assumptions \ref{as:MarkovProperty} and \ref{as:FiniteState}, this operator is well defined. Denote by $\mathbf P$ the transition matrix of $\{X_t\}$,
 i.e., $\mathbf P_{x,y} = \prob(X_{t+1}=y|X_t=x)$, $x,y\in\mathbb{X}$. Define matrix $\tilde {\mathbf P}$ by $\tilde {\mathbf P}_{x,y}:=\mathbf P_{x,y}\expect_t\big[u\big(e^{\kappa(X_t,X_{t+1},Y_{t+1})}\big)|X_t=x,X_{t+1}=y\big]$, $x,y\in\mathbb{X}$.

\begin{proposition}\label{prop:Perron}
Suppose Assumptions \ref{as:MarkovProperty} and \ref{as:FiniteState}
 hold.
\begin{enumerate}
  \item[(i)] Suppose $\gamma \neq 1$. Then, there exist $\eta>0$ and $v\in {\cal X}_{++}$ such that
  \begin{align}\label{eq:PerronCase1}
    \mathbb{U}v (x) = \eta v(x),\; x\in \mathbb{X}.
  \end{align}
  Moreover, $\eta$ and $v$ are the Perron-Frobenius eigenvalue and eigenvector of $\tilde {\mathbf P}$, respectively.
  \item[(ii)] Suppose $\gamma=1$. Then, there exist $\eta \in\mathbb{R}$ and $v\in {\cal X}$ such that
  \begin{align}\label{eq:PerronCase2}
    \expect_t[\kappa(X_t,X_{t+1},Y_{t+1})|X_t=x] = -\expect_t[v(X_{t+1})|X_{t}=x] + v(x) + \eta ,\; x\in \mathbb{X}.
  \end{align}
  In addition,
  \begin{align*}
    \eta = \sum_{x\in\mathbb{X}}\pi_x\expect_t[\kappa(X_t,X_{t+1},Y_{t+1})|X_t=x],
  \end{align*}
  where vector $(\pi_x)_{x\in\mathbb X}$ is the stationary distribution of $\{X_t\}$.
  \item[(iii)] Define $\delta:=u^{-1}(\eta)$. Then,
      \begin{align*}
        \delta=\max_{f\in{\cal X}_{++}} \min_{x\in\mathbb{X}} \frac{u^{-1}\Big(\expect_t\left[u\left(e^{\kappa(X_t,X_{t+1},Y_{t+1})}f(X_{t+1})\right)|X_t=x\right]\Big)}{f(x)}.
      \end{align*}
\end{enumerate}
\end{proposition}

Proposition \ref{prop:Perron}-(i) is the same as equation [4] in \citet{HansenScheinkman2012:RecursiveUtility}, but Proposition \ref{prop:Perron}-(ii) is new, {\add as these authors do not consider the case $\gamma=1$.}\footnote{Note that the notations in \citet{HansenScheinkman2012:RecursiveUtility} are different from ours: $\eta$ therein corresponds to $\ln \eta$ in the present paper.} Proposition \ref{prop:Perron}-(iii) transforms $\eta$ obtained in Proposition \ref{prop:Perron}-(i) and -(ii) into $\delta$ that is easy to use in the following. {\add More importantly, it provides an economic interpretation for $\delta$ by representing it as a special form of the certainty equivalent of the consumption growth rate $e^{\kappa(X_t,X_{t+1},Y_{t+1})}$. This interpretation is not available in \citet{HansenScheinkman2012:RecursiveUtility}. }

As we will see, $\delta$ is critical in proving the existence and uniqueness of the fixed point of $\mathbb{T}$. Thus, it is important to compute $\delta$, i.e., to compute $\eta$. When $\gamma\neq 1$, $\eta$ is the Perron-Frobenius eigenvalue of $\tilde {\mathbf P}$, so its computation has been studied extensively in the literature; see e.g., \citet{Chanchana2007:Algorithm}. When $\gamma =1$, $\eta$ is actually the expectation of $\kappa(X_t,X_{t+1},Y_{t+1})$ under the stationary distribution of $\{X_t\}$, which is also easy to compute.

\subsection{Case of Nonnegative Gain-Loss Utility}
\begin{theorem}\label{th:ExistUniqueFiniteStatePositive}
  Suppose Assumptions \ref{as:MarkovProperty} and \ref{as:FiniteState} hold. Assume $\varpi(x)\ge 0$, $x\in \mathbb{X}$. Recall $\delta$ as defined in Proposition \ref{prop:Perron} and assume $\beta \delta^{1-\rho}<1$.
  Then, the fixed point of $\mathbb{T}$ in ${\cal X}_{++}$ uniquely exists. Moreover, for any $f\in {\cal X}_{++}$, $\{\mathbb{T}^nf\}_{n\ge 0}$ converges to the fixed point.
\end{theorem}

Theorem \ref{th:ExistUniqueFiniteStatePositive} shows that when the state space of $\{X_t\}$ is finite and $\varpi$ is nonnegative, the fixed point of $\mathbb{T}$ in ${\cal X}_{++}$ and, thus, the utility process defined by \eqref{eq:RecursiveEquation} uniquely exist provided that $\beta \delta^{1-\rho}<1$. Condition $\beta \delta^{1-\rho}<1$ is the same as the one in \citet[Proposition 6]{HansenScheinkman2012:RecursiveUtility}, where the authors study the existence and uniqueness of the classical recursive utility (without gain-loss utility). {\add Thus, in a finite-state setting and in the case of recursive utility (by setting $\varpi\equiv 0$), Theorem \ref{th:ExistUniqueFiniteStatePositive} generalizes the results in \citet{HansenScheinkman2012:RecursiveUtility} because in the latter the authors do not consider the case of unitary EIS and RRAD nor prove the uniqueness when $(1-\gamma)/(1-\rho)<1$.}

Note that we restrict the domain of $\mathbb{T}$ to ${\cal X}_{++}$ although $\mathbb{T}$ is well defined on ${\cal X}_+$. This is because $\mathbb{T}$ can have nonpositive fixed points. For example, when $\rho\ge 1$ and $\varpi\equiv0$, $0$ is a fixed point of $\mathbb{T}$. When $\rho\ge 1$, $\gamma\ge 1$, the transition matrix of $\{X_t\}$ is positive, and $\varpi(x)=0$ for some $x\in\mathbb{X}$, we can verify that $H(1,\varpi(x)),x\in\mathbb{X}$ is a fixed point of $\mathbb{T}$ but is not in ${\cal X}_{++}$. The fixed points in these two examples, however, are not economically meaningful in representing the total utility of the agent's consumption and investment: given a positive consumption stream and nonnegative gain-loss utility, we expect the agent's total utility to be positive. Thus, we need to exclude such fixed points by restricting the domain of $\mathbb{T}$ to ${\cal X}_{++}$ and, by doing so, we obtain the uniqueness of the fixed point.

Theorem \ref{th:ExistUniqueFiniteStatePositive} also provides a simple algorithm to compute the fixed point: one can start from any positive function, e.g., a positive constant function, to do iteration and then obtain a sequence that eventually converges to the fixed point. This result provides another reason why nonpositive fixed points of $\mathbb{T}$, if they exist, are not desirable: These fixed points cannot be obtained by a recursive algorithm with any positive starting point.

In the above algorithm, one can also choose a nonnegative function, i.e., $f\in{\cal X}_+$, as the initial guess, provided that $\mathbb{T}^mf\in {\cal X}_{++}$ for some $m$. Such $m$ exists (i) for any $f\in{\cal X}_+$ if $\rho<1$ because $H(1,0)=(1-\beta)^{1/(1-\rho)}>0$ and (ii) for any $f\in {\cal X}_+^o$ if $\gamma<1$ because $\{X_t\}$ is irreducible and $u^{-1}\left(\expect[u(Z)]\right)>0$ for any nonnegative, nonzero random variable $Z$ when $\gamma<1$. If $\rho\ge 1$ and $\gamma\ge 1$, however, $\{\mathbb{T}^nf\}_{n\ge 0}$ may not converge to the fixed point of $\mathbb{T}$ in ${\cal X}_{++}$. For instance, suppose $\mathbb{X}$ contains two elements, e.g., $x_1$ and $x_2$, the transition matrix of $\{X_t\}$ is positive, and $\varpi\equiv 0$. Consider $f\in {\cal X}_+^o$ such that $f(x_1)=0$ and $f(x_2)>0$. Note that $H(1,0)=0$ because $\rho\ge 1$ and that $u^{-1}(\expect[u(Z)])=0$ for any nonnegative random variable taking zero with a positive probability because $\gamma \ge 1$. We then immediately obtain that $\mathbb{T}f=0$ and thus the limit $\{\mathbb{T}^nf\}_{n\ge 0}$ is 0; in other words, this sequence does not converge to the fixed point of $\mathbb{T}$ in ${\cal X}_{++}$.

The convergence of $\{\mathbb{T}^nf\}_{n\ge 0}$ to the fixed point of $\mathbb{T}$ for any positive $f$ is economically important: it shows that a finite-horizon model, in which the utility at the terminal time is positive, converges to the infinite-horizon model when the number of periods in the former model goes to infinity. Moreover, the utility at the terminal time in the former model is irrelevant, provided that it is positive.

\subsection{Case of Negative Gain-Loss Utility}\label{subse:NegativeGainLoss}
We first illustrate that when $\varpi(x)<0$ for some $x\in\mathbb{X}$, $\mathbb{T}$ can have zero, one, or multiple fixed points, depending on the parameter values.

\begin{example}\label{ex:FixedPointNegative}
  Suppose $\mathbb{X}$ is a singleton. Then, operator $\mathbb{T}$ becomes a function on $[0,+\infty)$, and we denote this function as $T(f)$. In this case, $\delta$ defined in Proposition \ref{prop:Perron} becomes $u^{-1}\big(\expect\big[u(e^{\kappa(X_t,X_{t+1},Y_{t+1})})\big]\big)$. Then, function $T(f)$ can be written as $T(f) = H(1,\delta f + \varpi)$. We assume $\beta \delta^{1-\rho}<1$, and Theorem \ref{th:ExistUniqueFiniteStatePositive} shows that the fixed point of $T$ in $(0,+\infty)$ uniquely exists when $\varpi\ge 0$. Next, we consider the case in which $\varpi<0$.

  It is obvious that the domain of $T$ is $[-\varpi/\delta,+\infty)$. Straightforward computation yields
  \begin{align*}
    \lim_{f\downarrow -\varpi/\delta}T(f) = \begin{cases}
    (1-\beta)^{1/(1-\rho)}, & \rho<1,\\
    0, & \rho\ge 1,
    \end{cases}\quad  \lim_{f\uparrow +\infty}T(f) = \begin{cases}
    +\infty, & \rho\le 1,\\
    (1-\beta)^{1/(1-\rho)}, & \rho> 1,
    \end{cases}\\
        \lim_{f\downarrow -\varpi/\delta}T'(f) = \begin{cases}
    +\infty, & \rho\le 1,\\
     \left(\beta \delta^{1-\rho}\right)^{1/(1-\rho)}, & \rho> 1,
    \end{cases}\quad  \lim_{f\uparrow +\infty}T'(f) = \begin{cases}
   \left(\beta \delta^{1-\rho}\right)^{1/(1-\rho)}, & \rho<1,\\
    0, & \rho\ge 1.
    \end{cases}
  \end{align*}
  Moreover, $T$ is strictly increasing and concave.

  We first consider the case in which $\rho\ge 1$. Note that in this case $T(-\varpi/\delta)=0<-\varpi/\delta$. Because $T'(-\varpi/\delta)>1$ and $T'(+\infty)<1$, we conclude that except in a very special case in which the identity line is tangent to $T$, it is either the case in which $T$ has no fixed point or the case in which $T$ has two fixed points; see Figure \ref{fi:rho>1}.

  Next, consider the case in which $\rho<1$. If $T(-\varpi/\delta) = (1-\beta)^{1/(1-\rho)}\le -\varpi/\delta$, we conclude, as in the case in which $\rho\ge 1$, that except in a very special case in which the identity line is tangent to $T$, it is either the case in which $T$ has no fixed point or the case in which $T$ has two fixed points. If $(1-\beta)^{1/(1-\rho)}>-\varpi/\delta$, then the fixed point exists and is unique; see Figure \ref{fi:rho<1}.

  Note that $\exp[\kappa(X_t,X_{t+1},Y_{t+1})]$ stands for the consumption growth rate in the model of narrow framing in Section \ref{subsubse:NarrowFramingPortfolio}, so $\delta$ stands for the certainty equivalent of the consumption growth rate and thus is decreasing with respect to the RRAD. On the other hand, $-\varpi$ stands for the disutility of loss. We can see that with $\rho<1$, inequality $(1-\beta)^{1/(1-\rho)}>-\varpi/\delta$ holds if $\beta$ is small, $\delta$ is large, and $-\varpi $ is small. Thus, we can conclude that the agent's total utility is well defined when her EIS is strictly larger than one, her time discounting is large, her consumption growth rate is high, her RRAD is low, and her disutility of loss is small.

\begin{figure}
\centering\begin{minipage}[t]{0.48\textwidth}
\scalebox{0.4}[0.4]{\includegraphics{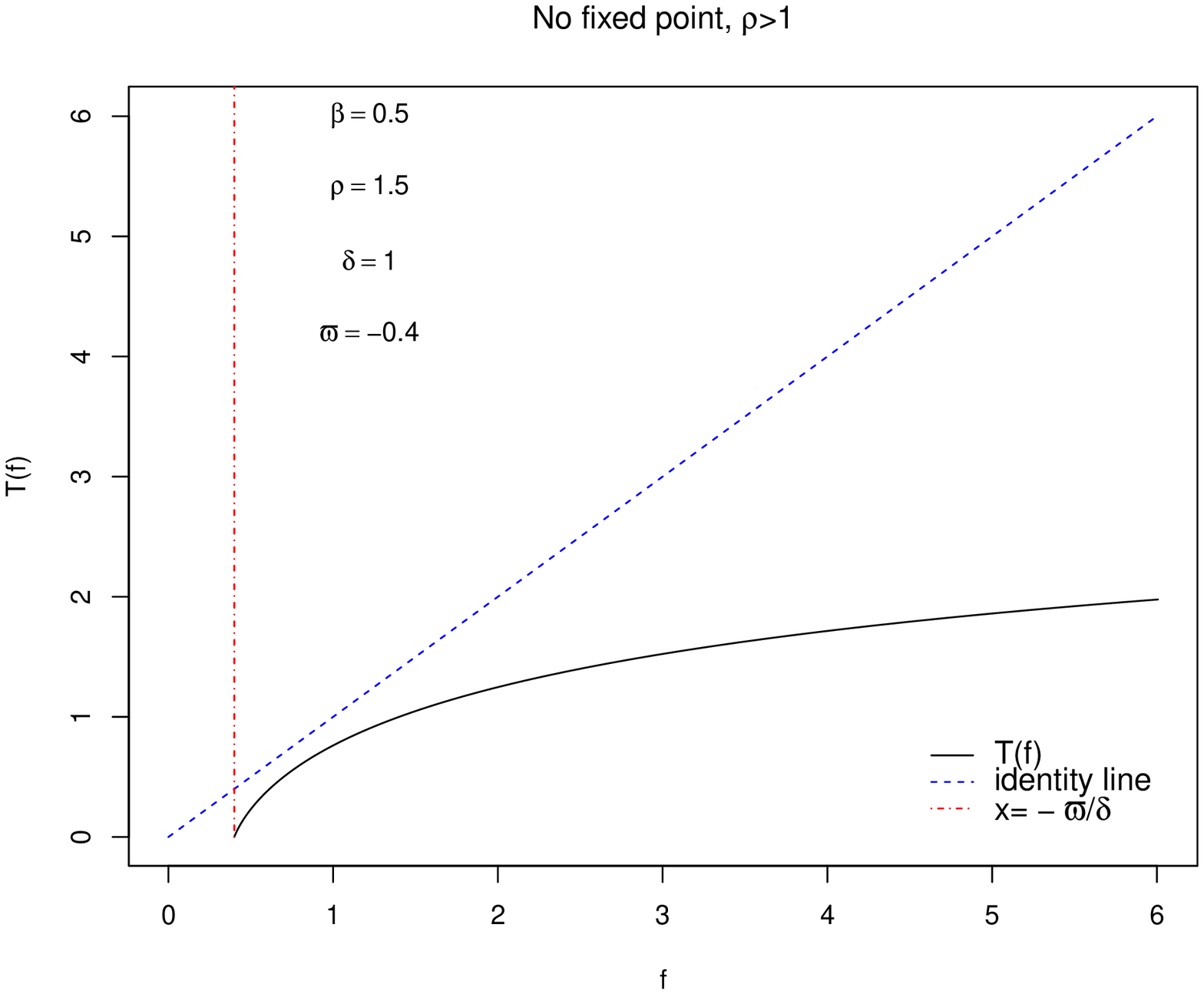}}
\end{minipage}
\hfill
\begin{minipage}[t]{0.48\textwidth}
\scalebox{0.4}[0.4]{\includegraphics{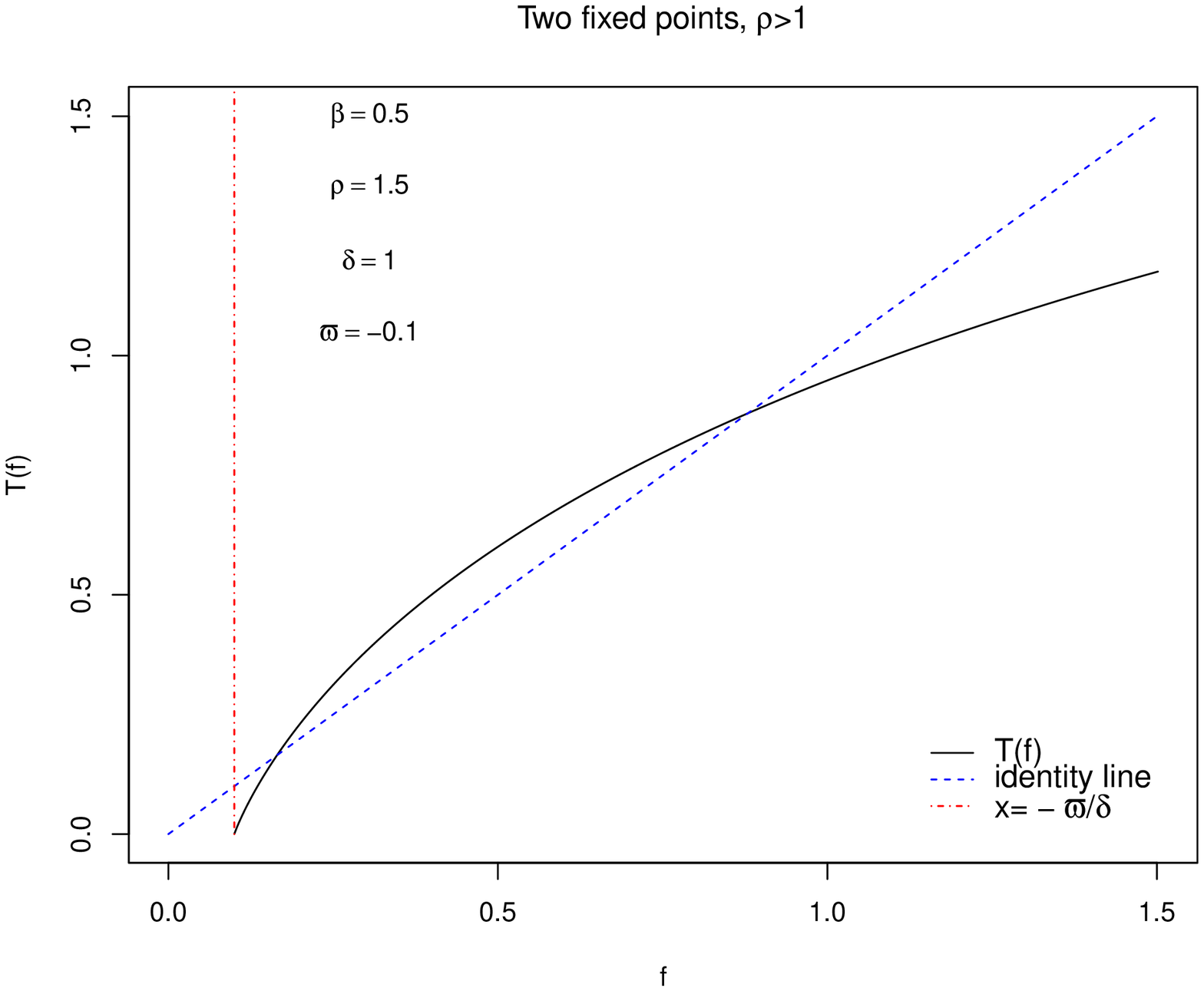}}
\end{minipage}
\caption{$T(f)$ in Example \ref{ex:FixedPointNegative} without a fixed point (left-hand panel) and with two fixed points (right-hand panel) when $\rho>1$. The solid lines in the two panels stand for $T(f)$ and the dashed lines stand for the identity function. Note that the domain of $T$ is $[-\varpi/\delta,+\infty)$, as indicated by the dash-dotted lines.}\label{fi:rho>1}
\end{figure}

\begin{figure}
\centering\begin{minipage}[t]{0.325\textwidth}
\scalebox{0.27}[0.27]{\includegraphics{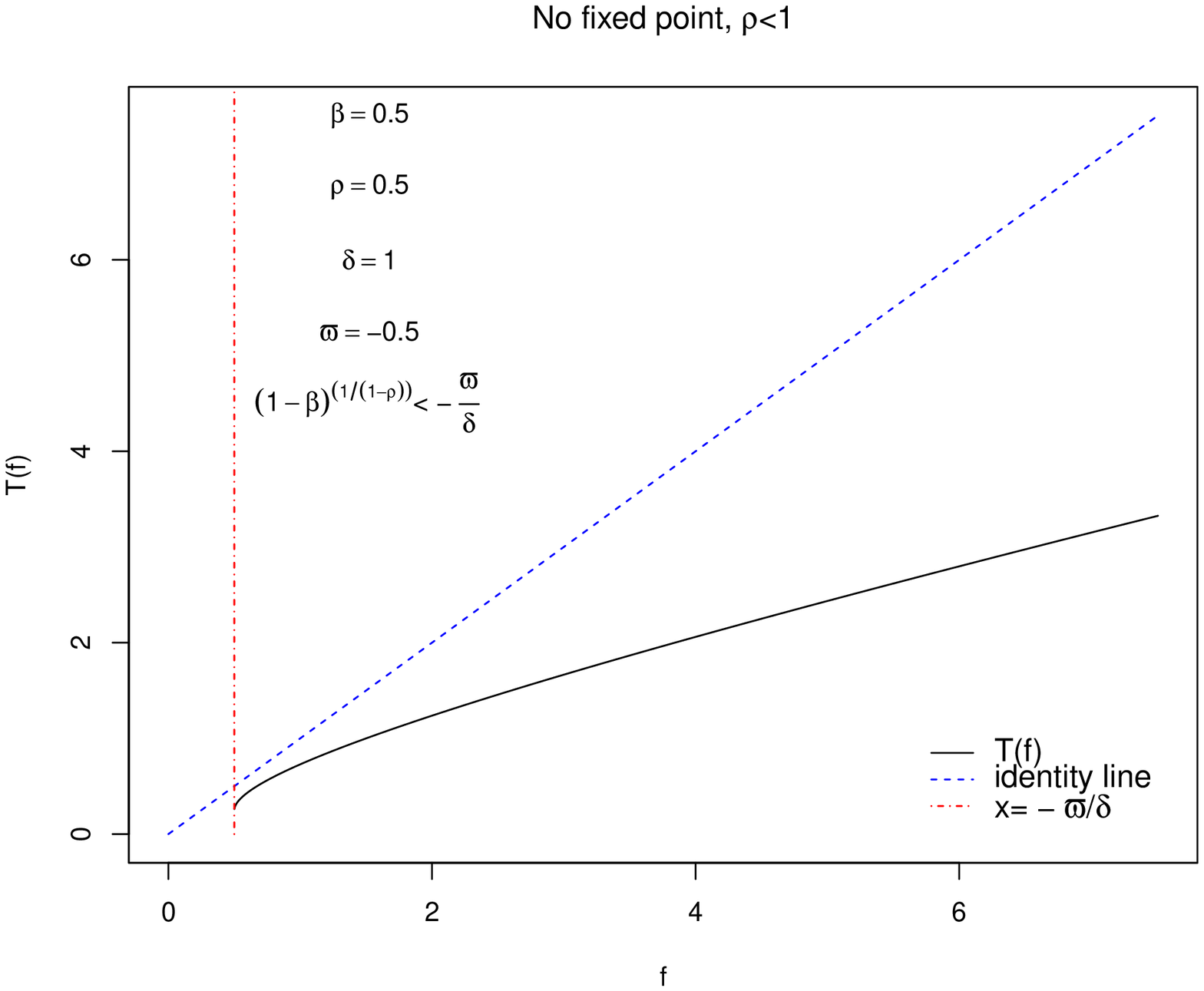}}
\end{minipage}
\hfill
\begin{minipage}[t]{0.325\textwidth}
\scalebox{0.27}[0.27]{\includegraphics{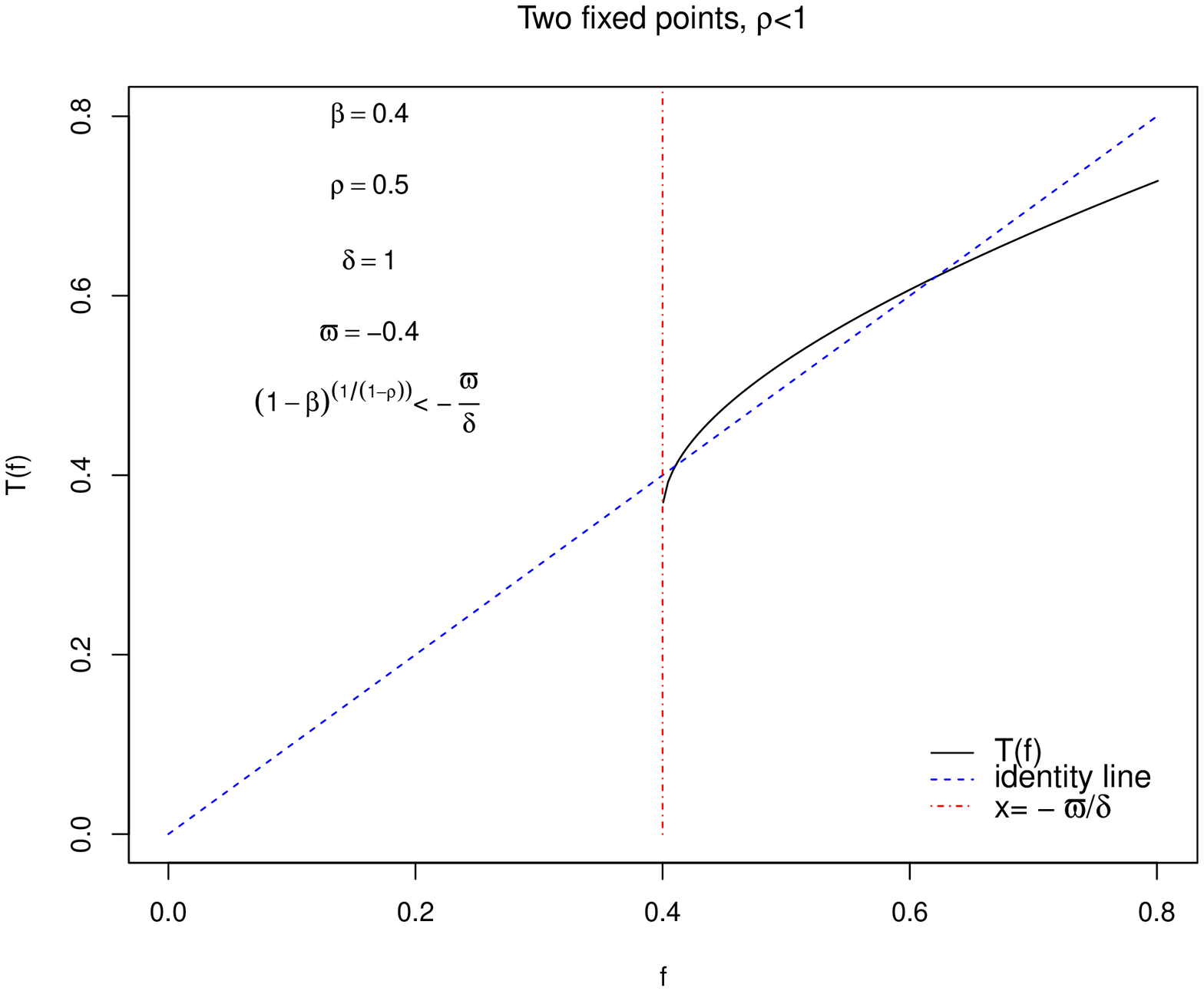}}
\end{minipage}
\hfill
\begin{minipage}[t]{0.325\textwidth}
\scalebox{0.27}[0.27]{\includegraphics{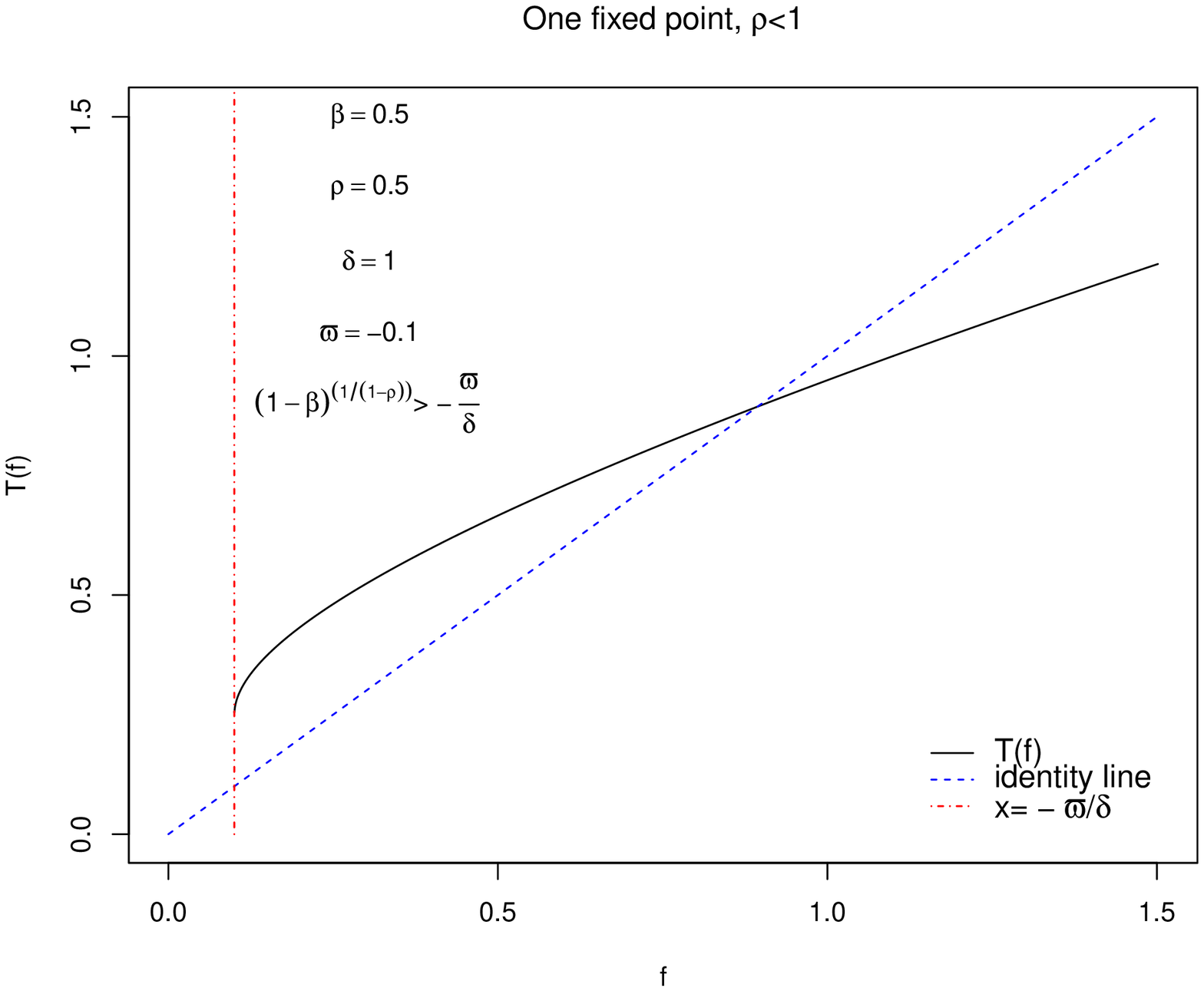}}
\end{minipage}
\caption{$T(f)$ in Example \ref{ex:FixedPointNegative} without a fixed point (left-hand panel), with two fixed points (middle pane), and with one fixed point (right-hand panel) when $\rho<1$. The solid lines in the three panels stand for $T(f)$ and the dashed lines stand for the identity function. Note that the domain of $T$ is $[-\varpi/\delta,+\infty)$, as indicated by the dash-dotted lines. In the left-hand and middle panels, $(1-\beta)^{1/(1-\rho)}<-\varpi/\delta$, and in the right-hand panel, $(1-\beta)^{1/(1-\rho)}>-\varpi/\delta$.}\label{fi:rho<1}
\end{figure}

  \end{example}

Example \ref{ex:FixedPointNegative} shows that we need some conditions on model parameters in order to establish the existence and uniqueness of the fixed point of $\mathbb{T}$ when $\varpi$ is negative in some states.

\begin{assumption}\label{as:BoundedDisutilityLoss}
  Denote
  \begin{align}
    f_0(x):=H(1,\varpi^+(x)),\; x\in\mathbb{X}.
  \end{align}
  Assume $\mathbb{T}f_0$ is well defined, i.e.,
  \begin{align*}
    u^{-1}\left[\expect_t\left(u\left(e^{\kappa(X_t,X_{t+1},Y_{t+1})}f_0(X_{t+1})\right)|X_t=x\right)\right] + \varpi(x)\ge 0,\; x\in\mathbb{X},
  \end{align*}
  and $\mathbb{T}^mf_0>f_0$ for some $m\ge 1$.
\end{assumption}

\begin{theorem}\label{th:ExistUniqueFiniteStateNegative}
  Suppose Assumptions \ref{as:MarkovProperty}--\ref{as:BoundedDisutilityLoss} hold. Assume $\beta\delta^{1-\rho}<1$, where $\delta$ is defined as in Proposition \ref{prop:Perron}.
   Then, the fixed point of $\mathbb{T}$ in its domain uniquely exists and is strictly larger than $f_0$ point-wisely. Moreover, for any $f$ such that $\mathbb{T}f$ is well defined, sequence $\{\mathbb T^{n} f\}_{n\ge 0}$ converges to the fixed point of $\mathbb{T}$.
\end{theorem}

Theorem \ref{th:ExistUniqueFiniteStateNegative} shows the existence and uniqueness of the fixed point of $\mathbb{T}$ when $\varpi$ can go negative. Moreover, the calculation of the fixed point is easy: Start from any $f$ such that $\mathbb{T}f$ is well defined and apply $\mathbb{T}$ repeatedly. Then, the resulting sequence converges to the fixed point. As discussed in the case of nonnegative $\varpi$, this algorithm implies that a finite-horizon model converges to the infinite-horizon model when the number of periods in the former goes to infinity.

Assumption \ref{as:BoundedDisutilityLoss} is crucial in order to obtain the existence and uniqueness of the fixed point of $\mathbb{T}$, so we discuss it in detail in the following:

(i) Note that if $\mathbb{T}f_0$ is well defined, we must have $\mathbb{T}f_0\ge f_0$. However, this is insufficient to guarantee the uniqueness of the fixed point of $\mathbb{T}$. Indeed, in the setting of Example \ref{ex:FixedPointNegative}, if $(1-\beta)^{1/(1-\rho)}=-\varpi/\delta$, $\mathbb{T}f_0$ is well defined and, actually, $\mathbb{T}f_0=f_0$. We already showed in that example that $\mathbb{T}$ has two fixed points, that one of them is $f_0$, and that both can represent the utility process. Thus, to guarantee the uniqueness, we need further conditions, and Assumption \ref{as:BoundedDisutilityLoss} serves the purpose.

(ii) Assumption \ref{as:BoundedDisutilityLoss} implies that $\mathbb{T}f_0(x)>f_0(x)$ for some $x\in\mathbb{X}$. The reverse is also true when $\gamma<1$ or $f_0\in{\cal X}_{++}$. Indeed, suppose $\mathbb{T}f_0(x_0)>f_0(x_0)$ for some $x_0\in\mathbb{X}$. Then, for any $y\in\mathbb{X}$ such that the transition probability from $y$ to $x_0$ is positive, either of the conditions that $\gamma<1$ and that $f_0\in {\cal X}_{++}$ implies
\begin{align*}
  u^{-1}\left(\expect_t\left[u\left(e^{\kappa(X_t,X_{t+1},Y_{t+1})}\mathbb{T}f_0(X_{t+1})\right)|X_t=y\right]\right)
  >u^{-1}\left(\expect_t\left[u\left(e^{\kappa(X_t,X_{t+1},Y_{t+1})}f_0(X_{t+1})\right)|X_t=y\right]\right).
\end{align*}
As a result, $\mathbb{T}^2f_0(y)=\mathbb{T}(\mathbb{T}f_0)(y)>\mathbb{T}f_0(y)$, and because of the irreducibility of $\{X_t\}$, we conclude that $\mathbb{T}^mf_0>f_0$ for some $m\ge 1$.

(iii) When $\gamma \ge 1$ and $f_0(x_0)=0$ for some $x_0\in\mathbb{X}$, which can be the case if and only if $\varpi(x_0)\le 0$ and $\rho\ge 1$, it is possible that $\mathbb{T}f_0$ is well defined, $\mathbb{T}f_0(x)>f_0(x)$ for {\em some} $x\in\mathbb{X}$, and the fixed point of $\mathbb{T}$ is {\em not} unique. For instance, consider $\{X_t\}$ with state space $\mathbb{X}=\{x_1,x_2,x_3\}$ such that $\prob(X_{t+1}=x_3|X_{t}=x_1)=1$, $\prob(X_{t+1}=x_3|X_t=x_2)=1$, $p_1:=\prob(X_{t+1}=x_1|X_t=x_3)>0$, and $p_2:=\prob(X_{t+1}=x_2|X_t=x_3)>0$. Then, $\{X_t\}$ is irreducible. Suppose $\kappa\equiv 0$. Suppose $\varpi(x_2)>0$, $\varpi(x_3)>0$, and $\varpi(x_1):=-H(1,\varpi(x_3))<0$. Then, one can verify that $\mathbb{T}f_0(x_1)=H(1,0)=f_0(x_1)=0$, $\mathbb{T}f_0(x_2) = H(1,f_0(x_3)+\varpi(x_2)) >f_0(x_2)>0$, and $\mathbb{T}f_0(x_3)=H(1,\varpi(x_3))=f_0(x_3)>0$. Moreover, it is straightforward to see that $\mathbb{T}f_0$ is a fixed point of $\mathbb{T}$. On the other hand, suppose $\rho>1$ and $\gamma>1$. Then, straightforward calculation shows that
\begin{align*}
  \frac{d\mathbb{T}(f_0+\epsilon\mathbf{1} )(x_1)}{d \epsilon}\Big |_{\epsilon=0} &= H_z(1,0)=\beta^{1/(1-\rho)},\\
  \frac{d\mathbb{T}(f_0+\epsilon\mathbf{1} )(x_2)}{d \epsilon}\Big |_{\epsilon=0} &= H_z(1,f_0(x_3)+\varpi(x_2)),\\
   \frac{d\mathbb{T}(f_0+\epsilon\mathbf{1} )(x_3)}{d \epsilon}\Big |_{\epsilon=0} & =H_z(1,\varpi(x_3))p_1^{1/(1-\gamma)},
\end{align*}
where $\mathbf{1}$ stands for the constant function taking value 1 and $H_z$ is the partial derivative of $H(c,z)$ with respect to $z$. Because $\rho>1$, $\gamma>1$, and $\beta<1$, with sufficiently small (but positive) $p_1$, $\varpi(x_2)$, and $\varpi(x_3)$, we have  $\frac{d\mathbb{T}(f_0+\epsilon\mathbf{1} )(x_i)}{d \epsilon}\big |_{\epsilon=0}>1$, $i=1,2,3$. As a result, there exists $\epsilon>0$ such that $\mathbb{T}(f_0+\epsilon\mathbf{1} )\ge f_0+\epsilon\mathbf{1}$. Consequently, $\{\mathbb{T}^n(f_0+\epsilon\mathbf{1})\}_{n\ge 0}$ is increasing and converges because $\mathbb{T}f\le (1-\beta)^{1/(1-\rho)}$ for any $f$. It is obvious that the convergent point is a fixed point of $\mathbb{T}$ and is different from $\mathbb{T}f_0$ because $f_0(x_1)+\epsilon>0=\mathbb{T}f_0(x_1)$.

(iv) When $\rho\ge 1$, $\gamma\ge 1$, and the transition matrix of $\{X_t\}$ is positive, Assumption \ref{as:BoundedDisutilityLoss} does not hold and thus Theorem \ref{th:ExistUniqueFiniteStateNegative} cannot apply if $\varpi(x)<0$ for some $x\in\mathbb{X}$. Indeed, in this case, we have
\begin{align*}
  u^{-1}\left(\expect_t\left[u\left(e^{\kappa(X_t,X_{t+1},Y_{t+1})}f_0(X_{t+1})\right)|X_t=x\right]\right)=0
\end{align*}
because $f_0(x)=0$ and $\prob(X_{t+1}=x|X_t=x)>0$, and $\gamma\ge 1$. Consequently, $\mathbb{T}f_0$ is not well defined because $\varpi(x)<0$.

(v) Theorem \ref{th:ExistUniqueFiniteStateNegative} cannot cover Theorem \ref{th:ExistUniqueFiniteStatePositive}. Indeed, suppose $\varpi\equiv 0$ and $\rho\ge 1$. Then, Assumption \ref{as:BoundedDisutilityLoss} does not hold, so Theorem \ref{th:ExistUniqueFiniteStateNegative} cannot apply. Theorem \ref{th:ExistUniqueFiniteStatePositive}, however, can still apply. Therefore, Theorem \ref{th:ExistUniqueFiniteStatePositive} is more comprehensive than Theorem \ref{th:ExistUniqueFiniteStateNegative} when $\varpi$ is nonnegative, and Theorem \ref{th:ExistUniqueFiniteStateNegative} is useful when $\varpi$ goes negative.

\section{Portfolio Selection and Dynamic Programming Equation}\label{se:PortfolioSelection}
\subsection{Model}\label{subse:PortfolioSelectionModel}
Consider the portfolio selection problem with narrow framing as discussed in Section \ref{subsubse:NarrowFramingPortfolio}. The agent's total utility $U_t$ is given by \eqref{eq:NarrowFraming}, and thus her total utility per unit wealth $U_t/W_t$ satisfies \eqref{eq:RecursiveEquation} with $A_{t+1}$ and $B_t$ as given by \eqref{eq:ABinNarrowFraming}. Suppose the gross return rate of risky asset $i$ in period $t$ to $t+1$ is $R_{i,t+1} = r_i(X_t,X_{t+1},Y_{t+1})$ for some function $r_i$ and the gross return rate of the risk-free asset in period $t$ to $t+1$ is $R_{f,t+1} = r_0(X_t)$ for some function $r_0$. Suppose the agent chooses consumption propensity $c_t= c(X_t)$ and portfolio $\theta_t=(\theta_1(X_t),\dots,\theta_n(X_t))\tran$ at time $t$ for some functions $c$ and $\theta_i$'s. For simplicity, we assume $\theta_{i}\ge 0$, and the following analysis can be performed without any additional difficulty for the case of negative $\theta_{i}$. Then, the agent's total utility per unit wealth $U_t/W_t=F_{c,\theta}(X_t)$, where $F_{c,\theta}$ is a fixed point of
\begin{align*}
  \mathbb{V}_{c,\theta}F(x):=H\Big(c(x), u^{-1}\left(\expect_t\left[u\big((1-c(x))R_\theta(x,X_{t+1},Y_{t+1})F(X_{t+1})\big)|X_t=x\right]\right)\\
  +(1-c(x))\sum_{i=1}^nb_i\theta_i(x)g_i(x)\Big),\; x\in\mathbb{X}
\end{align*}
with $R_\theta(X_t,X_{t+1},Y_{t+1}):=r_0(X_t)+\sum_{i=1}^n\theta_i(X_t)(r_i(X_t,X_{t+1},Y_{t+1})-r_0(X_t))$ and
\begin{align*}
  g_i(x):=\expect_t\left[(R_{i,t+1}-R_{f,t+1})\mathbf 1_{R_{i,t+1}>R_{f,t+1}}+ k (R_{i,t+1}-R_{f,t+1})\mathbf 1_{R_{i,t+1}<R_{f,t+1}}|X_t=x\right], x\in\mathbb{X}.
\end{align*}

For any $c$ and $\theta$ such that $0<c(X_t)< 1$ and $R_\theta(X_t,X_{t+1},Y_{t+1})>0$, $F$ is a fixed point of $\mathbb{V}_{c,\theta}$ if and only if $f(x):=F(x)/c(x)$ is a fixed point of $\mathbb{T}_{c,\theta}$, where
  \begin{align}
  \mathbb T_{c,\theta} f(x):=H\left(1,u^{-1}\left(\expect_t\left[u\left(e^{\kappa_{c,\theta}(X_t,X_{t+1},Y_{t+1})}f(X_{t+1})\right)|X_t=x\right]\right)+\varpi_{c,\theta}(x)\right),\; x\in \mathbb X
\end{align}
and
\begin{align}
  \kappa_{c,\theta}(X_t,X_{t+1},Y_{t+1}):&=\ln c(X_{t+1})-\ln c(X_t) + \ln (1-c(X_t)) + \ln R_p(X_t,X_{t+1},Y_{t+1}),\label{eq:kappa}\\
  \varpi_{c,\theta}(X_t):&= \frac{1-c(X_t)}{c(X_t)}\sum_{i=1}^nb_i\theta_i(X_t)g_i(X_t).\label{eq:varpi}
\end{align}
Denote $\delta$ in Proposition \ref{prop:Perron} as $\delta_{c,\theta}$ when $\kappa$ and $\varpi$ therein are set to be $\kappa_{c,\theta}$ and $\varpi_{c,\theta}$, respectively.

For each $x\in\mathbb{X}$, consider a set $I_x\subseteq(0,1)$ and a set $J_x\subseteq\mathbb{R}^n$. Define
\begin{align*}
  {\cal A}:=\{(c,\theta)|c(x)\in I_x,\theta(x)\in J_x,x\in\mathbb{X}\}.
\end{align*}
{\add In view of the results obtained in Section \ref{se:ExistenceUniquenessFinite}, we need the following assumption:}
\begin{assumption}\label{as:Feasibility}
  For each $(c,\theta)\in{\cal A}$, $R_\theta(X_t,X_{t+1},Y_{t+1})>0$, $\beta \delta_{c,\theta}^{1-\rho}<1$, and it is either the case in which $\varpi_{c,\theta}(x)\ge 0,x\in\mathbb{X}$ or the case in which $\mathbb{T}_{c,\theta}f_{0,c,\theta}$ with $f_{0,c,\theta}(x):=H(1,\varpi_{c,\theta}^+(x)), x\in\mathbb{X}$ is well defined, and $\mathbb{T}_{c,\theta}^mf_{0,c,\theta}>f_{0,c,\theta}$ for some $m\ge 1$.

\end{assumption}

With Assumptions \ref{as:FiniteState} and \ref{as:Feasibility} in place, Theorems \ref{th:ExistUniqueFiniteStatePositive} and \ref{th:ExistUniqueFiniteStateNegative} show that the fixed point of $\mathbb{T}_{c,\theta}$ in ${\cal X}_{++}$ uniquely exists for any $(c,\theta)\in{\cal A}$. Thus, if the agent consumes $C_s=c(X_s)W_s$ and invests $\Theta_{i,s}=\theta_i(X_s)(W_s-C_s)$ dollars in risky asset $i$, $i=1,\dots,n$ at time $s\ge t$, her utility $U_t$ is well defined. As a result, the following portfolio selection problem
\begin{align}\label{eq:PortSelectionNarrowFraming}
  \max_{(\{C_s\}_{s\ge t},\{\Theta_s\}_{s\ge t})\in {\cal B}_t} \; U_t
\end{align}
is well defined, where
\begin{align*}
  {\cal B}_t:=\{(\{C_s\}_{s\ge t},\{\Theta_s\}_{s\ge t})| C_s = c(X_s)W_s,\Theta_s=\theta(X_s)(W_s-C_s)\text{ for some }(c,\theta)\in {\cal A}\}.
\end{align*}
Note that for each $X_t=x$, problem \eqref{eq:PortSelectionNarrowFraming} is equivalent to
\begin{align}\label{eq:PortfolioSelectionMarkovian}
  \max_{(c,\theta)\in {\cal A}} \; F_{c,\theta}(x).
\end{align}

\subsection{Dynamic Programming}
The dynamic programming equation associated with the portfolio selection problem \eqref{eq:PortfolioSelectionMarkovian} can be derived heuristically as
\begin{align}\label{eq:DynamicPrograming}
\Phi(x)=\mathbb{W}\Phi(x),\; x\in \mathbb{X},
\end{align}
where
\begin{align}
&\mathbb{W}\Phi(x):=\max_{\bar c\in I_x} H\left(\bar c,(1-\bar c)\max_{\bar \theta \in J_x}D_{\Phi}(x,\bar \theta)\right),\; x\in\mathbb{X} \text{ and }\label{eq:WOperator}\\
&D_{\Phi}(x, \bar \theta):=\sum_{i=1}^n\bar \theta_ib_ig_i(x)\notag \\
&\quad  +u^{-1}\bigg(\expect_t\Big[u\Big(\big(r_0(x)+\sum_{i=1}^n\bar \theta_{i}
(r_i(x,X_{t+1},Y_{t+1})-r_0(x))\big)\Phi(X_{t+1})\Big)|X_t=x\Big]\bigg).\label{eq:DPhi}
\end{align}
{\add Note that the domain of $\mathbb{W}$ is the set of $\Phi$ in ${\cal X}_+$ such that $\max_{\bar \theta \in J_x}D_{\Phi}(x,\bar \theta)\ge 0$ for any $x\in \mathbb{X}$.}

\begin{proposition}\label{prop:DynamicProgramming}
  Suppose Assumptions \ref{as:FiniteState} and \ref{as:Feasibility} hold. Suppose $\Phi\in{\cal X}_{++}$ is a solution to \eqref{eq:DynamicPrograming}. Then, $\Phi(x)\ge F_{c,\theta}(x),x\in\mathbb{X}$ for any $(c,\theta)\in {\cal A}$. Moreover, if there exists $(c^*,\theta^*)\in {\cal A}$ such that $(c^*(x),\theta^*(x))$ is a maximizer of \eqref{eq:WOperator} for each $x\in\mathbb{X}$, then $(c^*,\theta^*)$ and $\Phi$ are a maximizer and the optimal value, respectively, of \eqref{eq:PortfolioSelectionMarkovian} for all $x\in \mathbb{X}$.
\end{proposition}

Proposition \ref{prop:DynamicProgramming} shows that the solution to the dynamic programming equation, if it exists, must be the solution to \eqref{eq:PortfolioSelectionMarkovian}.

\begin{theorem}\label{prop:DynamicProgramingSolution}
  Suppose that Assumptions \ref{as:FiniteState} and \ref{as:Feasibility} hold, {\add $\sup_{(c,\theta)\in{\cal A}} \beta\delta_{c,\theta}^{1-\rho}<1$, and that for each $x\in\mathbb{X}$, $J_x$ is compact and $I_x$ is closed relative to $(0,1)$ (i.e., $I_x=(0,1)\cap \tilde I_x$ for some closed set $\tilde I_x\subseteq \mathbb{R}$).
  \begin{enumerate}
    \item[(i)] Suppose $\varpi_{c,\theta}\ge 0$ for any $(c,\theta)\in{\cal A}$. Then, the fixed point of $\mathbb{W}$ in ${\cal X}_{++}$ uniquely exists, $\{\mathbb{W}^n\Phi\}_{n\ge 0}$ converges to the fixed point of $\mathbb{W}$ in ${\cal X}_{++}$ for any $\Phi\in{\cal X}_{++}$, and there exists $(c^*,\theta^*)\in {\cal A}$ such that $(c^*(x),\theta^*(x))$ is a maximizer of \eqref{eq:WOperator} for each $x\in\mathbb{X}$.
    \item[(ii)] Suppose $\varpi_{c,\theta}(x)< 0$ for some $x\in\mathbb{X}$ and some $(c,\theta)\in{\cal A}$, and define
          \begin{align}\label{eq:Phi0}
   \Phi_0(x):=\max_{\bar c\in I_x}H\left(\bar c,(1-\bar c)\max_{\bar \theta \in J_x}\big(\sum_{i=1}^n\bar \theta_ib_ig_i(x)\big)^+\right).
  \end{align}
  Then, $\Phi_0$ is in the domain of $\mathbb{W}$. Assume that there exists $m\ge 0$ such that $\max_{\bar \theta\in J_x}D_{\mathbb{W}\Phi_0^m}(x,\bar \theta)>0,\forall x\in\mathbb{X}$. Then, the fixed point of $\mathbb{W}$ in ${\cal X}_{++}$ uniquely exists, $\{\mathbb{W}^n\Phi\}_{n\ge 0}$ converges to the fixed point of $\mathbb{W}$ in ${\cal X}_{++}$ for any $\Phi$ in the domain of $\mathbb{W}$, and there exists $(c^*,\theta^*)\in {\cal A}$ such that $(c^*(x),\theta^*(x))$ is a maximizer of \eqref{eq:WOperator} for each $x\in\mathbb{X}$.
  \end{enumerate}
  }
\end{theorem}

Theorem \ref{prop:DynamicProgramingSolution} shows the existence and uniqueness of the solution to the dynamic programming equation {\add and the existence of corresponding maximizer $(c^*,\theta^*)$. Note that we assume $J_x$, the feasible set of percentage investment in the risky assets, to be compact and $I_x$, the feasible set of percentage consumption, to be a closed set relative to $(0,1)$; in particular, $I_x$ can be $(0,1)$. Note also that $\sup_{(c,\theta)\in{\cal A}} \beta\delta_{c,\theta}^{1-\rho}<1$ is implied by Assumption \ref{as:Feasibility} when $I_x$ and $J_x$ are compact for all $x\in\mathbb{X}$.\footnote{\add Indeed, in this case, ${\cal A}$ is compact. Because the eigenvalue of a matrix $A$ is continuous in $A$ and because $\expect[\kappa_{c,\theta}(X_t,X_{t+1},Y_{t+1})]$ under the stationary distribution of $\{X_t\}$ is continuous in $(c,\theta)$, $\beta \delta_{c,\theta}^{1-\rho}<1$ for any $(c,\theta)\in {\cal A}$ implies that $\sup_{(c,\theta)\in{\cal A}} \beta\delta_{c,\theta}^{1-\rho}<1$.}
When $\varpi_{c,\theta}(x)< 0$ for some $x\in\mathbb{X}$ and some $(c,\theta)\in{\cal A}$, we impose an additional assumption: there exists $m\ge 0$ such that $\max_{\bar \theta\in J_x}D_{\mathbb{W}\Phi_0^m}(x,\bar \theta)>0,\forall x\in\mathbb{X}$, and the following Proposition shows that this assumption can be easily satisfied.\footnote{\add The condition in Proposition \ref{prop:SufficientCond}-(iii) stipulates that the gain-loss utility is nonnegative for certain investment strategies $\theta$ and in certain states $x$. It holds particularly when zero investment in the risky assets (i.e., $\theta=0$) is allowed.}

\begin{proposition}\label{prop:SufficientCond}
  Let Assumptions \ref{as:FiniteState} and \ref{as:Feasibility} hold, and for each $x\in\mathbb{X}$, $J_x$ is compact and $I_x$ is closed relative to $(0,1)$. Suppose $\varpi_{c,\theta}(x)< 0$ for some $x\in\mathbb{X}$ and some $(c,\theta)\in{\cal A}$ and recall $\Phi_0$ as defined in
  \eqref{eq:Phi0}. Then, there exists $m\ge 0$ such that $\max_{\bar \theta\in J_x}D_{\mathbb{W}\Phi_0^m}(x,\bar \theta)>0,\forall x\in\mathbb{X}$ if one of the following conditions holds:
  \begin{enumerate}
  \item[(i)] $\rho\ge 1$.
    \item[(ii)] There exists $a\in(0,1)$ such that $J_x\subset (0,a]$ for any $x\in \mathbb{X}$.
    \item[(iii)] There exists $x\in\mathbb{X}$ such that $\max_{\bar \theta \in J_x}\left(\sum_{i=1}^n\bar \theta_ib_ig_i(x)\right)\ge 0$.
  \end{enumerate}
\end{proposition}

}

{\add Theorem \ref{prop:DynamicProgramingSolution} also shows that starting from any $\Phi$ that is positive when $\varpi_{c,\theta}\ge 0$ for any $(c,\theta)\in{\cal A}$ or any $\Phi$ in the domain of $\mathbb{W}$ in other cases,} by applying the dynamic programming equation repeatedly, one eventually obtains the solution to the equation. This result shows that the optimal consumption and portfolio in a finite-horizon model converges to those in the infinite-horizon model when the number of periods in the former goes to infinity.

Note that $D_\Phi(x,\bar \theta)$ is strictly concave in $\bar \theta$ for each $x$ and $\Phi$ and that $H(\bar c,(1-\bar c)z)$ is strictly concave in $\bar c$ for any given $z> 0$. Thus, for each $x$ and $\Phi$, the maximization problem in the right-hand side of the dynamic programming equation \eqref{eq:DynamicPrograming}, i.e., the problem in \eqref{eq:WOperator}, can be solved easily. As a result, $\mathbb{W}\Phi$ can be easily computed, and once we find the fixed point of $\mathbb{W}$, the optimal control $(c^*,\theta^*)$ can also be solved easily.

Finally, when $\varpi_{c,\theta}(x)< 0$ for some $x\in\mathbb{X}$ and some $(c,\theta)\in{\cal A}$, equation \eqref{eq:Phi0} provides a simple choice of $\Phi$ {\add in the domain of $\mathbb{W}$, which can be easily computed}. Note that $\big(\sum_{i=1}^n\bar \theta_ib_ig_i(x)\big)^+$ is convex in $\theta_i$ and $H\left(\bar c,(1-\bar c)z\right)$ is concave in $\bar c$ for any $z>0$, so the maximization in $\bar \theta_i$ and $\bar c$ can be computed easily.

{\add

\subsection {Verification of Assumptions}
There are two crucial assumptions to verify in order to apply Proposition \ref{prop:DynamicProgramming} and Theorem \ref{prop:DynamicProgramingSolution}: (i) $\sup_{(c,\theta)\in{\cal A}} \beta\delta_{c,\theta}^{1-\rho}<1$ and (ii) for any $(c,\theta)\in {\cal A}$ such that $\varpi_{c,\theta}(x)<0$ for some $x\in\mathbb{X}$, $\mathbb{T}_{c,\theta}^mf_{0,c,\theta}>f_{0,c,\theta}$ for some $m\ge 1$. In this subsection, we provide sufficient conditions for these two assumptions.

\begin{proposition}\label{prop:VerifyCon}
  Let Assumption \ref{as:FiniteState} hold. Then, $\sup_{(c,\theta)\in{\cal A}} \beta\delta_{c,\theta}^{1-\rho}<1$ if one of the following holds: (i) $\rho=1$, (ii) $\rho<1$ and
  \begin{align}
    \max_{x\in\mathbb{X}}\max_{(c,\theta)\in {\cal A}}\left\{ (1-c(x))u^{-1}\left(\mathbb{E}_t\left[u\big(R_{\theta}(X_t,X_{t+1},Y_{t+1})\big)\mid X_t=x\right]\right)\right\}< \beta^{-1/(1-\rho)}, \label{eq:VerifyConRho<1}
  \end{align}
  and (iii) $\rho>1$ and
  \begin{align}
    \min_{x,\in\mathbb{X}}\min_{(c,\theta)\in {\cal A}}\left\{ (1-c(x))u^{-1}\left(\mathbb{E}_t\left[u\big(R_{\theta}(X_t,X_{t+1},Y_{t+1})\big)\mid X_t=x\right]\right)\right\}> \beta^{-1/(1-\rho)}.\label{eq:VerifyConRho>1}
  \end{align}
\end{proposition}

Because $u^{-1}\left(\mathbb{E}_t\left[u\big(R_{\theta}(X_t,X_{t+1},Y_{t+1})\big)\mid X_t=x\right]\right)$ is concave in $\theta$, its maximization and minimization with respect to $\theta$ can be computed easily, and thus conditions \eqref{eq:VerifyConRho<1} and \eqref{eq:VerifyConRho>1} can be verified.


\begin{proposition}\label{prop:VerifyCon2}
  Let Assumption \ref{as:FiniteState} hold, denote ${\cal A}^-$ as the set of $(c,\theta)\in {\cal A}$ such that $\varpi_{c,\theta}(x)<0$ for some $x\in\mathbb{X}$, and denote ${\cal A}_2^-$ as the set of $\theta$ such that $\theta(x)\in J_x$ for any $x\in\mathbb{X}$ and $\sum_{i=1}^nb_i\theta_i(x)g_i(x)<0$ for certain $x\in\mathbb{X}$. Denote $\underline{i}:=\min_{x\in\mathbb{X}}\inf_{\bar c \in I_x}\bar c$ and $\bar{i}:=\max_{x\in\mathbb{X}}\max_{\bar c \in I_x}\bar c$. If
  \begin{align}
    &u^{-1}\left(\mathbb{E}_t\left[u\left(R_\theta(X_t,X_{t+1},Y_{t+1})\min_{\bar c\in\{\underline{i},\bar{i}\}}H\left(\bar c,(1-\bar c)\left(\sum_{i=1}^nb_i\theta_i(X_{t+1})g_i(X_{t+1})\right)^+\right)\right)\mid X_t=x\right]\right)\notag\\
    &>\left(\sum_{i=1}^nb_i\theta_i(x)g_i(x)\right)^-,\; \forall x\in\mathbb{X},\theta\in {\cal A}^-_2,\label{eq:VerifyCon2}
  \end{align}
  then $\mathbb{T}_{c,\theta}f_{0,c,\theta}>f_{0,c,\theta}$ for any $(c,\theta)\in {\cal A}^-$. When $\rho<1$, a sufficient condition for \eqref{eq:VerifyCon2} to hold is the following:
  \begin{align}
    H(\underline{i},0)u^{-1}\left(\mathbb{E}_t\left[u\left(R_\theta(X_t,X_{t+1},Y_{t+1})\right)\mid X_t=x\right]\right) - \left(\sum_{i=1}^nb_i\theta_i(x)g_i(x)\right)^->0,\; \forall x\in\mathbb{X},\theta\in {\cal A}^-_2.\label{eq:VerifyCon2Suff}
  \end{align}
\end{proposition}

Suppose $\rho\ge 1$, $\gamma \ge 1$, and the transition matrix of $\{X_t\}$ is positive. Then, the left-hand side of \eqref{eq:VerifyCon2} is 0 for any $\theta$ that makes $\sum_{i=1}^nb_i\theta_i(x)g_i(x)=0$ for certain $x$, so \eqref{eq:VerifyCon2} does not hold. This is not surprising because the analysis in Section \ref{subse:NegativeGainLoss} shows that it is difficult to define the agent's total utility when the utility for investment gain and loss is negative. Condition \eqref{eq:VerifyCon2}, however, is useful and can be verified when $\rho<1$. Indeed, a sufficient condition is given by \eqref{eq:VerifyCon2Suff}, where the left-hand side of the inequality therein is concave in $\theta$ and thus its minimum with respect to $\theta$ can be computed easily.

Note that \eqref{eq:VerifyCon2Suff} cannot be satisfied if $\underline{i}=0$---that is, when the agent chooses to consume very little.\footnote{\add Intuitively, suppose $\gamma\ge 1$, $\prob(X_{t+1}=x|X_t=x)>0$, and in a market state $X_t=x$, the agent derives negative gain-loss utility, i.e., $\varpi_{c,\theta}^-(x) = c(x)^{-1}(1-c(x))(\sum_{i=1}^n b_i\theta_i(x)g_i(x))^->0$. Suppose the agent consumes little in this state, i.e., $c(x)\approx0$, . Then, one can compute that $c(1-c)f_{c,\theta}\approx 0$ in state $x$ and, consequently, the certainty equivalent of $c(X_{t+1})(1-c(X_{t+1}))^{-1}f_{c,\theta}(X_{t+1})$ is nearly 0. As a result, it is impossible that this certainty equivalent plus $c^{-1}(1-c)\varpi_{c,\theta}$ is larger than 0 in state $x$, so $\mathbb{T}_{c,\theta}f_{0,c,\theta}$ is not even well defined.} This, however, does not undermine our portfolio selection results. First, when the gain-loss utility is always nonnegative, in particular when the agent's preferences are modeled by recursive utility, we only need to verify $\sup_{(c,\theta)\in{\cal A}} \beta\delta_{c,\theta}^{1-\rho}<1$, so the conditions in Proposition \ref{prop:VerifyCon2} are irrelevant. Second, it is reasonable for us to focus on strategies that satisfy $\mathbb{T}_{c,\theta}f_{0,c,\theta}>f_{0,c,\theta}$ because other strategies generate too much disutility of losses and thus should not be preferred.

}

\subsection{A Numerical Example}
We consider a market with a risky stock and a risk-free asset, and we can regard the stock as the market portfolio. Set the length of each period to be one year. To construct the return of the stock, we assume that the stock pays a dividend every year and that the dividend growth rates are i.i.d. and follow the distribution as given in Table \ref{ta:Consumption}.

\begin{table}[ht]
\caption{Distribution of the dividend growth rate. The distribution is assumed to be the same as the one in Table I of \citet{ChapmanPolkovnichenko2009:FirstOrderRiskAversion}, which is obtained by using the historical gross consumption growth from 1949 to 2006.}\label{ta:Consumption}

\medskip
\centering 
\begin{tabular}{c |c c c c c c c c c}
  \noalign{\hrule height 1.5pt}
State & 1 & 2 & 3 & 4 & 5 & 6 & 7 & 8 & 9\\
\hline
 Outcome & 0.976 & 0.993 & 1.002 & 1.011 & 1.019 & 1.028 & 1.037 & 1.045 & 1.054 \\
Probability & 0.03 & 0.03 & 0.10 & 0.16 & 0.24 & 0.19 & 0.13 &
0.09 & 0.03
\\
  \noalign{\hrule height 1.5pt}
\end{tabular}
\label{table:GrowthRateStates}
\end{table}

We assume that the market is governed by a two-state Markovian process $\{X_t\}$ that takes values in $\mathbb{X}=\{0,1\}$. We assume the price-dividend ratio at time $t$ to be $\varphi(X_t)$ and the risk-free gross return rate in period $t$ to $t+1$ to be $r_0(X_t)$; i.e., both are functions of $X_t$. As a result, the gross return rate of the stock in period $t$ to $t+1$ is
\begin{align*}
  r(X_t,X_{t+1},Y_{t+1}) = Y_{t+1}(\varphi(X_{t+1})+1)/\varphi(X_t),
\end{align*}
where $Y_{t+1}$ refers to the dividend growth rate in period $t$ to $t+1$.

We assume the transition matrix of $\{X_t\}$ to be
\begin{align*}
\left( \begin{array}{cc} 0.6&0.4\\
0.2&0.8\end{array} \right).
\end{align*}
We also set the risk-free total return rate and the price-dividend ratio to be
\begin{align*}
r_0(X_t)&=\left( \begin{array}{cc} 1.03\\
1.03\end{array} \right),\quad
\varphi(X_t)=\left( \begin{array}{cc} 30.25\\
39.75\end{array} \right),
\end{align*}
respectively,
so that the mean and {\add volatility} of the stock return under the stationary distribution of $\{X_t\}$ are 6\% and $15\%$, respectively, and, consequently, the equity premium is 3\%.


We set the loss aversion degree $k=1.5$, so $g(X_t)$, which measures the gain-loss utility, is
\begin{align*}
g(X_t)&=\expect_t\big[\nu\left(r(X_t,X_{t+1},Y_{t+1})-r_0(X_t)\right)|X_t\big]=\left( \begin{array}{cc} 0.1532\\
-0.0551\end{array} \right),
\end{align*}
where $\nu(x):=x\mathbf1_{x>0}+kx\mathbf1_{x\le 0}$.
Finally, we set {\add$\beta=0.937$, $b=0.00065$,} $\rho=0.5$, and $\gamma=8$.

Consider the feasible set {\add$I_x=[0.45\%,100\%)$} and $J_x=[0,100\%]$, $x=0,1$. {\add With the help of Propositions \ref{prop:VerifyCon} and \ref{prop:VerifyCon2}, we can verify that all assumptions in Theorem \ref{prop:DynamicProgramingSolution} hold.\footnote{\add Here, we set the lower bound of $I_x$ to be a positive number in order to have $\mathbb{T}_{c,\theta}>f_{0,c,\theta}$ for any $(c,\theta)\in{ \cal A}$. } Then, we apply this theorem} to calculate the optimal consumption and portfolio and the value function, and the results are as follows:
{\add
\begin{align*}
	c^*(X_t)=\left( \begin{array}{cc} 5.85\%\\
		7.30\%\end{array} \right),\quad \theta^*(X_t)=\left( \begin{array}{cc} 100\%\\
		15.0\%\end{array} \right),\quad \Phi(X_t)
	=\left( \begin{array}{cc} 0.0679\\
		0.0544\end{array} \right).
\end{align*}
}

\section{When the State Space is Not Finite}\label{se:GeneralState}
In this section, we study the existence and uniqueness of the solution to \eqref{eq:RecursiveEquation}, i.e., the fixed point of $\mathbb{T}$, when the state space of $\{X_t\}$ is not finite. We consider only the case in which $\varpi$ is nonnegative for two reasons. First, when $\varpi$ is negative, by imposing a similar condition to Assumption \ref{as:BoundedDisutilityLoss}, we can prove that the fixed point of $\mathbb{T}$ exists, but we do not have uniqueness, so we chose not to present the results here. Second, a new model of narrow framing that is proposed by \citet{GuoHe2017:NewModelNarrowFraming} represents a special case of \eqref{eq:RecursiveEquation} with $\varpi\equiv 0$.


\begin{proposition}\label{th:ExistenceGeneral}
  Suppose Assumption \ref{as:MarkovProperty} holds, $\rho\neq 1$, and $\varpi(x)\ge 0,x\in\mathbb{X}$. Suppose that the results in Proposition \ref{prop:Perron}-(i) and -(ii) hold and recall $v$ defined therein. When $\gamma\neq 1$, denote as $\tilde \prob$ the probability measure that is obtained by a change of measure using the Radon-Nikodym density $M_{t+1}:= \eta^{-1}e^{(1-\gamma)\kappa(X_t,X_{t+1},Y_{t+1})} v(X_{t+1})/v(X_t)$, and, when $\gamma =1$, $\tilde \prob$ simply refers to the original probability measure. Denote as $\tilde \expect$ the expectation operator corresponding to $\tilde \prob$. For each $p\in[1,+\infty)$, denote space ${\cal X}_+$ when equipped with the $L^p$ norm under the stationary distribution of $\{X_t\}$ under $\tilde \prob$ as $\tilde L^p_+(\mathbb X)$. Define operator $\mathbb{S}$ on ${\cal X}_{+}$ by
    \begin{align*}
\begin{split}
  &\mathbb{S}g(x):=\frac{1-\beta}{\tilde u^{-1}(v(x))} + \beta \delta^{1-\rho}\left\{\left[\tilde u^{-1}\left(\tilde \expect_t\left[\tilde u\big(g(X_{t+1})\big) |X_t=x\right]\right)\right]^{\frac{1}{1-\rho}}+
\frac{\delta^{-1}\varpi(x)}{ u^{-1}\big(v(x)\big)}\right\}^{1-\rho},
\end{split}
\end{align*}
where $\tilde u(x):=u(x^{1/(1-\rho)}),x\ge 0$, and define $\alpha:=(1-\gamma)/(1-\rho)$. Then, $f$ is a fixed point of $\mathbb{T}$ in ${\cal X}_{++}$ if and only if $g(x):=\big(f(x)/u^{-1}(v(x))\big)^{1-\rho},x\in\mathbb{X}$ is a fixed point of $\mathbb{S}$ in the same space. Moreover, with the assumptions that $\beta \delta^{1-\rho}<1$ and that the stationary distribution of $\{X_t\}$ under $\tilde \prob$ exists, the following results hold:
  \begin{enumerate}
    \item[(i)] If $\alpha\ge 1$ and $\mathbb{S} 0\in \tilde L^\alpha_+(\mathbb X)$, then $\mathbb{S}$ is a contraction mapping in $\tilde L^\alpha_+(\mathbb X)$ and its unique fixed point is positive.
    \item[(ii)] If $\alpha<1$ and $\mathbb{S} 0\in \tilde L^{\alpha'}_+(\mathbb X)$ for some $\alpha'\ge 1$, then the limit of $\{\mathbb{S}^n0\}$ exists, belongs to $\tilde L^{\alpha'}_+(\mathbb X)$, and is the minimum fixed point of $\mathbb{S}$ in ${\cal X}_{++}$.
    \item[(iii)] If $\alpha \in(0,1)$ and $\varpi\equiv 0$, then $f$ is a fixed point of $\mathbb{T}$ in ${\cal X}_{++}$ if and only if $h(x):=u(f(x))/v(x),x\in\mathbb{X}$ is a fixed point of the following operator in the same space:
    \begin{align*}
      \bar{\mathbb{S}} h(x):=\tilde u\left(\frac{1-\beta}{\tilde u^{-1}(v(x))} + \beta \delta^{1-\rho}\tilde u^{-1}\left(\tilde \expect_t\left[h(X_{t+1}) |X_t=x\right]\right)\right).
    \end{align*}
    Moreover, if $\bar{\mathbb{S}}0\in \tilde L^1_+(\mathbb X)$, then $\bar{\mathbb{S}}$ is a contraction mapping in $\tilde L^1_+(\mathbb X)$, and its unique fixed point is positive.
  \end{enumerate}
\end{proposition}

Propositions \ref{th:ExistenceGeneral}-(i) and -(ii) are completely parallel to Proposition 6 in \citet{HansenScheinkman2012:RecursiveUtility}: When $\rho\neq 1$ and $\varpi$ is nonnegative, if (i) $\eta$, $v$, and $\delta$ in Proposition \ref{prop:Perron} are well defined and $\beta \delta^{1-\rho}<1$ and (ii) the stationary distribution of $\{X_t\}$ exists after a specific change of measure, then the fixed point of $\mathbb{T}$ exists. Moreover, when $\alpha=(1-\gamma)/(1-\rho)\ge 1$, the fixed point is unique. Our proof is also analogous to that in \citet{HansenScheinkman2012:RecursiveUtility}. Note that just as in \citet{HansenScheinkman2012:RecursiveUtility}, in general we are unable to prove uniqueness when $\alpha<1$. {\add Following \citet{HansenScheinkman2012:RecursiveUtility}, we do not discuss here the issue of when the results in Proposition \ref{prop:Perron} hold and when the stationary distribution of $\{X_t\}$ under $\tilde \prob$ exists. For sufficient conditions, one can refer to \add{Assumption 7.2} in \citet{HansenScheinkman2009:LongTermRisk} and Proposition 9.2 in \citet{EthierKurtz1986:MarkovProcesses}.}

When $\alpha\in(0,1)$, we also prove uniqueness in the recursive utility model, namely, in the case $\varpi\equiv 0$; see Proposition \ref{th:ExistenceGeneral}-(iii). This result generalizes those in \citet[Proposition 6]{HansenScheinkman2012:RecursiveUtility} nontrivially.

\begin{proposition}\label{prop:ExistUniqueGeneralStateMyopic}
	Suppose Assumption \ref{as:MarkovProperty} holds and the stationary distribution of $\{X_t\}$ exists. Suppose $\rho=\gamma=1$ and $\varpi(x)\ge 0,x\in\mathbb{X}$. Then, $f$ is a fixed point of $\mathbb{T}$ in ${\cal X}_{++}$ if and only if $g:=\ln f$ is a fixed point of $\mathbb{S}$ in ${\cal X}$, where
	\begin{align}\label{eq:SGeneralMyopic}
	\begin{split}
	\mathbb{S}g(x):= & \beta\ln \left[e^{\expect_t(\kappa(X_t,X_{t+1},Y_{t+1})|X_t=x)}e^{\expect_t\left(g(X_{t+1})|X_t=x\right)} + \varpi(x)\right],\;  x\in\mathbb X.
	\end{split}
	\end{align}
	Denote ${\cal X}$ equipped with the $L^1$ norm under the stationary distribution of $\{X_t\}$ as $L^1(\mathbb{X})$, and assume there exists $g\in L^1(\mathbb{X})$ such that $\mathbb{S}g\in L^1(\mathbb{X})$. Then, $\mathbb{S}$ is a contraction mapping on $L^1(\mathbb{X})$ and thus the fixed point of $\mathbb{S}$ uniquely exists in $L^1(\mathbb{X})$.
\end{proposition}

Proposition \ref{prop:ExistUniqueGeneralStateMyopic} shows the existence and uniqueness of the fixed point of $\mathbb{T}$ when $\rho=\gamma=1$, provided that $\{X_t\}$ has a stationary distribution and $\varpi$ is nonnegative. 

Finally, we show that the solution to \eqref{eq:RecursiveEquation} uniquely exists when $\gamma=\rho=1$ even in a non-Markovian setting.
\begin{proposition}\label{prop:ExistUniqueNonMarkovMyopic}
	Suppose $\rho=\gamma=1$, $A_t> 0$, and $B_t\ge 0$. Then, $\{V_t\}$ is a positive solution to \eqref{eq:RecursiveEquation} if and only if $\{\ln(V_t/c_t)\}$ is a fixed point of
	\begin{align}
	({\cal S}Z)_t:=\beta\ln \left[e^{\expect_t(\ln c_{t+1} - \ln c_t+ \ln A_{t+1})}e^{\expect_t\left(Z_{t+1}\right)} + (B_t/c_t)\right],\; t=0,1,\dots.
	\end{align}
	Moreover, if there exist $\alpha\in(\beta, 1)$ and $\{Z_t\}\in {\cal L}^{1,\alpha}$, the space of $\{{\cal F}_t\}$-adapted processes with norm $||Z||:=\sum_{t=0}^{\infty}\alpha^t\expect(|Z_t|)$, such that ${\cal S}Z\in {\cal L}^{1,\alpha}$, then ${\cal S}$ is a contraction mapping on ${\cal L}^{1,\alpha}$ and thus the fixed point of ${\cal S}$ on this space uniquely exists.
\end{proposition}

\section{Conclusion}\label{se:Conclusion}
We considered a generalization of the recursive utility model that adds a component of gain-loss utility and thus accommodates a variety of models of narrow framing encountered in the literature. Assuming constant EIS and RRAD, we studied the existence and uniqueness of the agent's utility process in this generalized model.

We assumed a Markovian setting: the asset returns in the period from $t$ to $t+1$ are assumed to be functions of $X_t$, $X_{t+1}$, and $Y_{t+1}$, so the agent's consumption propensity, percentage investment, {\add and utility of gains and losses per unit of investment} for the assets in that period are functions of $X_t$, where $\{X_t\}$ is a Markov process that represents market states and $\{Y_t\}$ is an independent sequence conditional on $\{X_t\}$ and thus represents random noise. We further assumed that $\{X_t\}$ is irreducible and that its state space is finite.

We proved that the utility process uniquely exists for any values of the EIS and RRAD when the gain-loss utility is nonnegative. We then illustrated by an example that when the state space of $\{X_t\}$ is a singleton and the EIS is less than or equal to one, the utility process is either non-existent or non-unique if the gain-loss utility is negative. We then proposed a sufficient condition under which the utility process uniquely exists when the gain-loss utility is negative, and this condition is nearly necessary. We also proved that if the utility process uniquely exists, it can be computed by starting from any initial guess and applying the recursive equation that defines the utility process repeatedly.

We then considered a portfolio selection problem with narrow framing and proved that a consumption and portfolio plan is optimal if and only if it, together with the value function of the portfolio selection problem, satisfies a dynamic programming equation. Moreover, we proved that the solution to the dynamic programming equation uniquely exists and can be computed by solving the equation recursively with any starting point.

Finally, we extended some of the previous results to the setting of non-finite state spaces.

\appendix
\section{Proofs}\label{se:Proofs}
\proof{Proof of Proposition \ref{prop:Perron}}
We first consider (i). Because of Assumption \ref{as:MarkovProperty}-(i), we have
\begin{align*}
  \mathbb{U}h (x):=\expect_t\left[ h(X_{t+1})\expect_t\left[u\left(e^{\kappa(X_t,X_{t+1},Y_{t+1})}\right)|X_t,X_{t+1}\right]|X_t=x\right]=\sum_{y\in\mathbb{X}}\tilde {\mathbf{P}}_{x,y}h(y).
\end{align*}
Because $\mathbf{P}$ is irreducible and $e^\kappa$ is positive, we conclude that $\tilde {\mathbf{P}}$ is also irreducible. Thus, we have \eqref{eq:PerronCase1},
where $\eta$ and $v$ are the Perron-Frobenius eigenvalue and eigenvector of $\tilde {\mathbf P}$, respectively, and $\eta>0$, $v\in{\cal X}_{++}$; see e.g., \citet[p. 673]{Meyer2000:MatrixAnalysis}.

Next, we consider (ii). It is straightforward to see that \eqref{eq:PerronCase2} is equivalent to
\begin{align}\label{eq:PerronCase2Vector}
  \mathbf P v = v + \eta \mathbf 1- w,
\end{align}
where $w$ denotes the vector of $\expect_t[\kappa(X_t,X_{t+1},Y_{t+1})|X_t=x], x\in\mathbb{X}$ and $\mathbf 1$ denotes the vector of all ones. Because $\mathbf P$ is an irreducible stochastic matrix, the kernel of $\mathbf{I}-\mathbf{P}^\top$, where $\mathbf{I}$ is the identity mapping, is the linear space spanned by the left-Perron-Frobenius eigenvector of $\mathbf P$, namely, by the stationary distribution $\pi$ of $\{X_t\}$. As a result, the range of $\mathbf{I}-\mathbf{P}$ is the space of all vectors that are orthogonal to $\pi$. By the definition of $\eta$, $\eta \mathbf 1- w$ is orthogonal to $\pi$ and thus is in the range of $\mathbf{I}-\mathbf{P}$. As a result, there exists $v$ such that \eqref{eq:PerronCase2Vector} holds. Moreover, by multiplying the stationary distribution $\pi$ on both sides of \eqref{eq:PerronCase2Vector}, we can see that $\eta$ is uniquely determined.

Finally, we prove (iii). We first consider the case in which $\gamma\neq 1$. Because $\eta$ is the Perron-Frobenius eigenvalue of $\tilde {\mathbf P}$, according to the max-min version of the Collatz-Wielandt formula \citep[p. 673]{Meyer2000:MatrixAnalysis}, we have
\begin{align*}
  \eta &= \max_{g\in{\cal X}_+^o} \min_{x\in\mathbb{X},g(x)\neq 0}\frac{\sum_{y\in\mathbb{X}} \tilde {\mathbf P}_{x,y}g(y)}{g(x)} \\
  &=\max_{g\in{\cal X}_+^o} \min_{x\in\mathbb{X},g(x)\neq 0} \frac{\sum_{y\in\mathbb{X}}\mathbf P_{x,y}\expect_t\left[u\left(e^{\kappa(X_t,X_{t+1},Y_{t+1})}\right)g(X_{t+1})|X_t=x,X_{t+1}=y\right]}{g(x)}\\
  & = \max_{g\in{\cal X}_+^o} \min_{x\in\mathbb{X},g(x)\neq 0} \frac{\expect_t\left[u\left(e^{\kappa(X_t,X_{t+1},Y_{t+1})}\right)g(X_{t+1})|X_t=x\right]}{g(x)}.
\end{align*}
Moreover, it is straightforward to see that the maximum in the above formula is attained when $g$ is chosen to be the Perron-Frobenius eigenvector of $\tilde {\mathbf P}$. Because the eigenvector lies in ${\cal X}_{++}$, we conclude that ${\cal X}_+^o$ can be replaced with ${\cal X}_{++}$ in the above formula. Now, recalling $\delta = \eta^{1/(1-\gamma)}$ and setting $f=g^{1/(1-\gamma)}$, we conclude that when $\gamma<1$,
\begin{align}\label{eq:Calculatedelta}
  \delta = \max_{f\in{\cal X}_{++}} \min_{x\in\mathbb{X}} \frac{u^{-1}\Big(\expect_t\left[u\left(e^{\kappa(X_t,X_{t+1},Y_{t+1})}f(X_{t+1})\right)|X_t=x\right]\Big)}{f(x)}.
\end{align}
Similarly, according to the min-max version of the Collatz-Wielandt formula \citep[p. 669]{Meyer2000:MatrixAnalysis},\footnote{The formula therein is presented for postive matrices, but it also holds for irreducible nonnegative matrices because the Perron-Frobenius eigenvectors for these matrices are positive; see for instance \citet[p. 673]{Meyer2000:MatrixAnalysis}.} we have
\begin{align*}
  \eta
  & = \min_{g\in{\cal X}_{++}} \max_{x\in\mathbb{X}} \frac{\expect_t\left[u\left(e^{\kappa(X_t,X_{t+1},Y_{t+1})}\right)g(X_{t+1})|X_t=x\right]}{g(x)}.
\end{align*}
Recalling $\delta = \eta^{1/(1-\gamma)}$ and setting $f=g^{1/(1-\gamma)}$, we conclude that \eqref{eq:Calculatedelta} also holds when $\gamma>1$.

Finally, we show that \eqref{eq:Calculatedelta} also holds when $\gamma =1$. For each $f\in{\cal X}_{++}$, denote
\begin{align*}
  \xi_f:&=\min_{x\in\mathbb{X}} \frac{u^{-1}\Big(\expect_t\left[u\left(e^{\kappa(X_t,X_{t+1},Y_{t+1})}f(X_{t+1})\right)|X_t=x\right]\Big)}{f(x)}\\
  & = \exp\left\{\min_{x\in\mathbb{X}} \Big(\expect_t\left[\kappa(X_t,X_{t+1},Y_{t+1})+\ln f(X_{t+1})|X_t=x\right] - \ln f(x)\Big)\right\}.
\end{align*}
As a result, $\ln (\xi_f) + \ln f(X_t)\le \expect_t\left[\kappa(X_t,X_{t+1},Y_{t+1})+\ln f(X_{t+1})|X_t\right]$. Taking expectation on both sides under the stationary distribution of $\{X_t\}$ and recalling $\eta$ that is derived in part (ii) of the proof, we conclude that $\ln(\xi_f) + \expect[\ln f(X_t)] \le \eta +\expect[\ln f(X_{t+1})]$, which implies $\ln(\xi_f)\le \eta$. Therefore, we conclude
\begin{align}\label{eq:Calculatedeltaproof}
  \eta\ge \max_{f\in{\cal X}_{++}} \min_{x\in\mathbb{X}} \Big(\expect_t\left[\kappa(X_t,X_{t+1},Y_{t+1})+\ln f(X_{t+1})|X_t=x\right] - \ln f(x)\Big).
\end{align}
On the other hand, recall $v$ defined in part (ii) of the proof. Then, $e^v\in {\cal X}_{++}$ and \eqref{eq:PerronCase2Vector} can be written as
\begin{align*}
  \expect_t\left[\ln e^{v(X_{t+1})}|X_t=x\right] = \ln e^{v(x)} + \eta - \expect_t\left[\kappa(X_t,X_{t+1},Y_{t+1})|X_t=x\right],\; x\in\mathbb{X}.
\end{align*}
Combining the above with \eqref{eq:Calculatedeltaproof}, we immediately conclude that
\begin{align*}
  \eta= \max_{f\in{\cal X}_{++}} \min_{x\in\mathbb{X}} \Big(\expect_t\left[\kappa(X_t,X_{t+1},Y_{t+1})+\ln f(X_{t+1})|X_t=x\right] - \ln f(x)\Big).
\end{align*}
Therefore, \eqref{eq:Calculatedelta} holds. \halmos
\proofend

\proof{Proof of Theorem \ref{th:ExistUniqueFiniteStatePositive}}
For ease of exposition, the proof is divided into two parts.

\medskip
\noindent {\em Part One: existence and uniqueness of the fixed point}
\medskip

In the first part of the proof, we show the existence and uniqueness of the fixed point of $\mathbb{T}$ in ${\cal X}_{++}$. The proof of the case in which $\alpha:=(1-\gamma)/(1-\rho)\ge 1$ follows exactly the same line as in \citet{HansenScheinkman2012:RecursiveUtility}, but some adaptation is needed to accommodate the gain-loss utility, so we sketch the proof in the following. For the case in which $\alpha \in (-\infty,0)\cup (0,1)$, the proof of the existence mimics the idea of \citet{HansenScheinkman2012:RecursiveUtility}, but the proof of the uniqueness and the global attractingness of the fixed point is completely new because they are not proved in \citet{HansenScheinkman2012:RecursiveUtility}. In addition, \citet{HansenScheinkman2012:RecursiveUtility} did not consider the case $\gamma=1$ or the case $\rho=1$ either.

Observe that $\mathbb{T}f$ is well-defined for any $f\in{\cal X}_+$ and that $\mathbb{T}$ is increasing. We first note from Proposition \ref{prop:Perron}-(i) that when $\gamma\neq 1$, with $\eta$, $\delta$, and $v$ as defined in Proposition \ref{prop:Perron}, we can define $M_{t+1}:=\eta^{-1}e^{(1-\gamma)\kappa(X_{t+1},Y_{t+1},X_t)} v(X_{t+1})/v(X_t)$ and show that $M_{t+1}>0$ and $\expect_t[M_{t+1}]=1$. As a result, we can define a new measure $\tilde \prob$ by using $M_{t+1}$ as the Radon-Nikodym density. Note that $\{X_t\}$ is still an irreducible Markov process under $\tilde \prob$. Denote the corresponding expectation as $\tilde \expect$. Then,
\begin{align*}
\begin{split}
  &\expect_t\left[u\left(e^{\kappa(X_t,X_{t+1},Y_{t+1})}f(X_{t+1})\right)|X_t=x\right]\\
  =&\expect_t\left[M_{t+1}v(X_{t+1})^{-1}v(X_t)\eta f(X_{t+1})^{1-\gamma}|X_t=x\right]\\
  =& \tilde \expect_t\left[v(X_{t+1})^{-1}v(X_t)\eta f(X_{t+1})^{1-\gamma}|X_t=x\right].
\end{split}
\end{align*}
As a result, we obtain
\begin{align}\label{eq:ChangeOfMeasure1}
\begin{split}
  &u^{-1}\left(\expect_t\left[u\left(e^{\kappa(X_t,X_{t+1},Y_{t+1})}f(X_{t+1})\right)|X_t=x\right]\right)
   \\&\quad = \delta u^{-1}\big(v(x)\big)u^{-1}\left(\tilde \expect_t\left[u\left(f(X_{t+1})/u^{-1}\big(v(X_{t+1})\big)\right)|X_t=x\right]\right).
  \end{split}
\end{align}
After careful calculation, one can conclude from Proposition \ref{prop:Perron}-(ii) that \eqref{eq:ChangeOfMeasure1} holds for the case $\gamma=1$ as well, with $\tilde \expect$ replaced by $\expect$. Therefore, in the following, we will use \eqref{eq:ChangeOfMeasure1} regardless of the value of $\gamma$, and $\tilde \expect$ stands for $\expect$ when $\gamma=1$.
Using \eqref{eq:ChangeOfMeasure1} and the homogeneity of $H(c,z)$, we obtain
\begin{align}\label{eq:ChangeOfMeasure2}
  \frac{\mathbb{T}f(x)}{u^{-1}\big(v(x)\big)} = H\left(\frac{1}{u^{-1}(v(x))},\delta \left[u^{-1}\left(\tilde \expect_t\left[u\left(\frac{f(X_{t+1})}{u^{-1}\big(v(X_{t+1})\big)}\right)|X_t=x\right]\right) + \frac{\delta^{-1}\varpi(x)}{ u^{-1}\big(v(x)\big)}\right] \right).
\end{align}

We first consider the case in which $\rho \neq 1$. In this case, denoting $g(x):=\big(f(x)/u^{-1}(v(x))\big)^{1-\rho}$, we conclude that $f$ is a fixed point of $\mathbb{T}$ in ${\cal X}_{++}$ if and only if $g$ is a fixed point of $\mathbb{S}$ in ${\cal X}_{++}$, where $\mathbb{S}$ is an operator on ${\cal X}_+$ defined as
\begin{align}\label{eq:RecursiveUtilityMarkovTransformed}
\begin{split}
  &\mathbb{S}g(x):=\frac{1-\beta}{\tilde u^{-1}(v(x))} + \beta \delta^{1-\rho}\left\{\left[\tilde u^{-1}\left(\tilde \expect_t\left[\tilde u\big(g(X_{t+1})\big) |X_t=x\right]\right)\right]^{\frac{1}{1-\rho}}+
\frac{\delta^{-1}\varpi(x)}{ u^{-1}\big(v(x)\big)}\right\}^{1-\rho}
\end{split}
\end{align}
with $\tilde u(x):=u(x^{1/(1-\rho)}),x\ge 0$.

It is easy to see that $\mathbb{S}$ is an increasing mapping from ${\cal X}_+$ into ${\cal X}_{++}$.
Consider function $\varphi(z):=\left(z^{\frac{1}{1-\rho}} + a\right)^{1-\rho},z\ge 0$ for some $a\ge 0$. It is straightforward to see that $\varphi'(z) = \left(z^{\frac{1}{1-\rho}}/(z^{\frac{1}{1-\rho}} + a)\right)^{\rho}\le 1$.
Consequently, $|\varphi(z_1)-\varphi(z_2)|\le |z_1-z_2|$ for any $z_1,z_2\ge 0$.
As a result, for any $g_1$ and $g_2$, we have
\begin{align*}
\begin{split}
  |\mathbb S g_1 (x)-\mathbb S g_2(x)|\le \beta \delta^{1-\rho}\left|\tilde u^{-1}\left(\tilde \expect_t\left[\tilde u\big(g_1(X_{t+1})\big) |X_t=x\right]\right)-\tilde u^{-1}\left(\tilde \expect_t\left[\tilde u\big(g_2(X_{t+1})\big) |X_t=x\right]\right)\right|.
\end{split}
\end{align*}

When $\alpha=(1-\gamma)/(1-\rho)\ge 1$, $\tilde u(x) = x^{\alpha}$ is a convex power function, so we conclude from the above inequality that
\begin{align*}
  |\mathbb S g_1 (x)-\mathbb S g_2(x)|\le \beta \delta^{1-\rho}\left[\tilde \expect_t\left(|g_1(X_{t+1})-g_2(X_{t+1})|^\alpha |X_t=x\right)\right]^{\frac{1}{\alpha}},\; x\in\mathbb {X},
\end{align*}
which implies
\begin{align*}
   |\mathbb S g_1 (X_t)-\mathbb S g_2(X_t)|^\alpha \le\left(\beta \delta^{1-\rho}\right)^\alpha \tilde \expect_t\left(|g_1(X_{t+1})-g_2(X_{t+1})|^\alpha|X_t\right).
\end{align*}
Recall that $\{X_t\}$ is an irreducible Markov chain under measure $\tilde \prob$, so it has a unique stationary distribution. Taking expectation on both sides of the inequality under this stationary distribution, and noting that the marginal distributions of $X_t$ and $X_{t+1}$ are the same, we conclude
\begin{align}\label{eq:ContractionMapping}
  \left[\tilde \expect\left(|\mathbb S g_1 (X_t)-\mathbb S g_2(X_t)|^\alpha\right)\right]^{1/\alpha}\le  \beta \delta^{1-\rho}\left[\tilde \expect\left(| g_1 (X_{t})- g_2(X_{t})|^\alpha\right)\right]^{1/\alpha}.
\end{align}
Because $\beta \delta^{1-\rho}<1$, $\mathbb S$ is a contraction mapping on ${\cal X}_+$ with norm $ \left[\tilde \expect |g (X_t)|^\alpha\right]^{1/\alpha}$. Consequently, for any $g\in{\cal X}_+$, the limit of $\{\mathbb{S}^ng\}_{n\ge 0}$ exists and is the unique fixed point of $\mathbb{S}$ in ${\cal X}_+$. Moreover, because $\mathbb{S}g(x)\ge (1-\beta)/\tilde u^{-1}(v(x))>0,x\in\mathbb{X}$, the fixed point must lie in ${\cal X}_{++}$. As a result, $\mathbb{T}$ has a unique fixed point in ${\cal X}_{++}$.


When $\alpha<1$, we consider the following operator:
\begin{align}\label{eq:STilde}
\begin{split}
  \tilde {\mathbb{S}}g(x):=\frac{1-\beta}{\tilde u^{-1}(v(x))} + \beta \delta^{1-\rho}\left\{\left(\tilde \expect_t\left[g(X_{t+1})|X_t=x\right]\right)^{\frac{1}{1-\rho}}+
\frac{\delta^{-1}\varpi(x)}{ u^{-1}\big(v(x)\big)}\right\}^{1-\rho}.
\end{split}
\end{align}
Because $\tilde u$ is either a concave power function or a logarithmic function when $\alpha<1$, we have $\tilde u^{-1}\left(\expect_t[\tilde u(Z)]\right)\le \expect_t[Z]$ for any nonnegative random variable $Z$. As a result, $\mathbb{S}g(x)\le \tilde {\mathbb{S}}g(x),x\in\mathbb{X}$ for any $g$, and in particular for $g_0(x):=(1-\beta)/\tilde u^{-1}(v(x))>0,x\in\mathbb{X}$. One can see that both $\{\mathbb{S}^ng_0\}_{n\ge 0}$ and $\{\tilde {\mathbb{S}}^n g_0\}_{n\ge 0}$ are increasing sequences, and that the former is dominated by the latter. On the other hand, following the same proof as in the case in which $\alpha\ge 1$, we can show that $\tilde {\mathbb{S}}$ is a contraction mapping from ${\cal X}_+$ into ${\cal X}_{++}$. As a result, $\{\tilde {\mathbb{S}}^n g_0\}_{n\ge 0}$ converges, and so does $\{\mathbb{S}^ng_0\}_{n\ge 0}$. Consequently, the limit of $\{\mathbb{S}^ng_0\}_{n\ge 0}$ is a fixed point of $\mathbb{S}$ and lies in ${\cal X}_{++}$, and thus the fixed point of $\mathbb{T}$ in ${\cal X}_{++}$ exists.
We then show the uniqueness of the fixed point of $\mathbb{T}$ in ${\cal X}_{++}$ when $\alpha<1$. For the sake of contradiction, suppose there are two distinct fixed points $f_1$ and $f_2$ in ${\cal X}_{++}$. Without loss of generality, we assume $f_1(x)<f_2(x)$ for some $x\in\mathbb{X}$. Define $x^*:=\underset{x\in \mathbb{X}}{\text{argmin}}\; f_1(x)/f_2(x)$
and denote the corresponding minimum value as $r^*$. Because $\mathbb{X}$ is finite and $f_i$'s are positive, $x^*$ is well defined and $r^*\in(0,1)$. Define $f(x):=r^*f_2(x),x\in\mathbb{X}$. Then, $f(x)\le f_1(x),x\in\mathbb{X}$ and $f(x^*)=f_1(x^*)$. Denote $\mathbb{I}$ as the identity mapping. Then, for each
%
%
%
$x\in\mathbb{X}$, we have
\begin{align*}
   (\mathbb{T}-\mathbb{I})f(x) &= H\left(1, u^{-1}\left(\expect_t\left[u\left(e^{\kappa(X_t,X_{t+1},Y_{t+1})}r^*f_2(X_{t+1})\right)|X_t=x\right]\right)+\varpi(x)\right) - r^*f_2(x)\\
   & > r^*    H\left(1, u^{-1}\left(\expect_t\left[u\left(e^{\kappa(X_t,X_{t+1},Y_{t+1})}f_2(X_{t+1})\right)|X_t=x\right]\right)+\varpi(x)\right) - r^*f_2(x)\\
   & = r^*(\mathbb{T}-\mathbb{I})f_2(x) = 0,
\end{align*}
where the inequality is the case because $\beta<1$ and $\varpi(x)\ge 0$ and the last equality is the case because $f_2$ is a fixed point of ${\mathbb{T}}$. In particular, we have $ {\mathbb{T}}f(x^*)>f(x^*)=f_1(x^*)$. On the other hand, because $  {\mathbb{T}}$ is increasing and $f\le f_1$, we have $ {\mathbb{T}} f \le  {\mathbb{T}} f_1 = f_1$, where the equality is the case because $f_1$ is a fixed point of $ {\mathbb{T}}$. Speficially, $ {\mathbb{T}} f(x^*)\le f_1(x^*)$. Thus, we have a contradiction, so the fixed point of $ {\mathbb{T}}$ must be unique.

Next, we consider the case in which $\rho=1$. In this case,
\begin{align*}
  \mathbb{T}f(x) &=  \exp\bigg\{\beta\ln\Big[u^{-1}\left(\expect_t\left(u(e^{\kappa(X_t,X_{t+1},Y_{t+1})}f(X_{t+1}))|X_t=x\right)\right)+\varpi(x)\Big]\bigg\},\; x\in\mathbb{X}.
\end{align*}
Because $\mathbb{X}$ is finite, there exists $\epsilon>0$ such that
\begin{align}\label{eq:LowerStartingPoint}
      \epsilon\le \min_{x\in\mathbb{X}} \left[u^{-1}\left(\expect_t\left(u(e^{\kappa(X_t,X_{t+1},Y_{t+1})})|X_t=x\right)\right)\right]^{\frac{\beta}{1-\beta}}.
    \end{align}
It is straightforward to verify that for such $\epsilon>0$, $\mathbb{T} \epsilon\ge \epsilon>0$. Because $\mathbb{T}$ is increasing, $\{\mathbb{T}^n \epsilon\}_{n\ge 0}$ is an increasing sequence. On the other hand, because $\beta<1$ and $\mathbb{X}$ is finite, there exists $N>\epsilon$ such that $\mathbb{T}N\le N$. Consequently, $\mathbb{T}^n \epsilon\le \mathbb{T}^n N\le N,n\ge 0$, so the limit of $\{\mathbb{T}^n \epsilon\}_{n\ge 0}$ exists and is a fixed point of $\mathbb{T}$ in ${\cal X}_{++}$. Using the same proof as the one in the case in which $\rho\neq 1$ and $\alpha<1$, we can show that the fixed point of $\mathbb{T}$ in ${\cal X}_{++}$ is unique.

\medskip
\noindent {\em Part Two: computation of the fixed point.}

\medskip

In the second part of the proof, we show that $\{\mathbb{T}^n f\}_{n\ge 0}$ converges to the fixed point of $\mathbb{T}$ for any $f\in{\cal X}_{++}$. Denote the fixed point as $f^*$.

Because $f\in{\cal X}_{++}$, $f^*\in{\cal X}_{++}$, and $\mathbb{X}$ is finite, there exists $r\ge 1$ such that $f\le rf^*$ and $f\ge (1/r)f^*$. Then,
\begin{align*}
  \mathbb{T}(rf^*)\le r\mathbb{T}(f^*)=rf^*,
\end{align*}
where the inequality is the case because $\beta \le 1$ and $\varpi\ge 0$ and the equality is the case because $f^*$ is the fixed point of $\mathbb{T}$. Consequently, $\{\mathbb{T}^n(rf^*)\}_{n\ge 0}$ is a decreasing sequence. Similarly, $\{\mathbb{T}^n((1/r)f^*)\}_{n\ge 0}$ is an increasing sequence. Moreover, $\mathbb{T}^n((1/r)f^*)\le \mathbb{T}^nf \le\mathbb{T}^n(rf^*)$ because $\mathbb{T}$ is increasing. As a result, both $\{\mathbb{T}^n(rf^*)\}_{n\ge 0}$ and $\mathbb{T}^n((1/r)f^*)$ converge in ${\cal X}_{++}$, and the convergent points are fixed points of $\mathbb{T}$. Because the fixed point of $\mathbb{T}$ is unique, both $\{\mathbb{T}^n(rf^*)\}_{n\ge 0}$ and $\mathbb{T}^n((1/r)f^*)$ converge to this fixed point, namely to $f^*$. By the squeeze theorem, $\{\mathbb{T}^nf\}_{n\ge 0}$ also converges to $f^*$.
\halmos
\proofend

\proof{Proof of Theorem \ref{th:ExistUniqueFiniteStateNegative}. }
Define operator $\mathbb{T}_+$ on ${\cal X}_+$ by
\begin{align*}
  \mathbb T_+ f(x):=H\left(1,u^{-1}\left(\expect_t\left[u\left(e^{\kappa(X_t,X_{t+1},Y_{t+1})}f(X_{t+1})\right)|X_t=x\right]\right)+\varpi^+(x)\right),\; x\in \mathbb X.
\end{align*}
It is obvious that $\mathbb{T}f\le \mathbb{T}_+f$ for any $f$. According to Assumption \ref{as:BoundedDisutilityLoss}, sequence $\{\mathbb{T}^nf_0\}_{n\ge 0}$ is increasing. Consequently, $\mathbb{T}_+f_0\ge \mathbb{T}f_0\ge f_0$, and thus $\{\mathbb{T}_+^nf_0\}_{n\ge 0}$ is also an increasing sequence and dominates $\{\mathbb{T}^nf_0\}_{n\ge 0}$. By Assumption \ref{as:BoundedDisutilityLoss}, $\mathbb{T}^mf_0(x)>f_0(x)\ge 0,x\in\mathbb{X}$ for some $m\ge 0$. As a result, $\mathbb{T}_+^nf_0\in {\cal X}_{++}$ for sufficiently large $n$, and thus $\{\mathbb{T}_+^nf_0\}_{n\ge 0}$ converges to the fixed point of $\mathbb{T}_+$ in ${\cal X}_{++}$ according to Theorem \ref{th:ExistUniqueFiniteStatePositive}. Consequently, the limit of $\{\mathbb{T}^nf_0\}_{n\ge 0}$ exists, is a fixed point of $\mathbb{T}$, and is strictly larger than $f_0$ point-wisely.

Next, we show the uniqueness of the fixed point of $\mathbb{T}$. We first note that for any fixed point $f^*$ of $\mathbb{T}$, we have $f^*= \mathbb{T}f^*\ge f_0$. Because $\mathbb{T}^mf_0>f_0$ for some $m\ge 0$, $f^*$ must be strictly larger than $f_0$ point-wisely. Now, for the sake of contradiction, suppose that we have two distinct fixed points $f_1$ and $f_2$. We already showed that $f_i(x)>f_0(x),x\in\mathbb{X}$, $i=1,2$.
 Without loss of generality, we assume $f_1(x)<f_2(x)$ for some $x\in\mathbb{X}$. Define
\begin{align*}
  x^*:=\text{argmin}_{x\in\mathbb{X}}\frac{f_1(x)-f_0(x)}{f_2(x)-f_0(x)}
\end{align*}
and denote the corresponding minimum value as $r^*$. Because $\mathbb{X}$ is finite, $x^*$ must exist and $r^*\in(0,1)$. Define $f(x):=r^*f_2(x)+(1-r^*)f_0(x),x\in\mathbb{X}$. Then, one can verify that $f_0(x)<f(x)\le f_1(x),x\in\mathbb{X}$ and $f(x^*)=f_1(x^*)$. Because $\mathbb{T}f_0$ is well defined, so is $\mathbb{T}f$.
Recall that $\mathbb{T}^mf_0(x)>f_0(x),x\in\mathbb{X}$ for some $m\ge 1$. Because $\mathbb{T}$ is increasing and concave, so is $\mathbb{T}^m$. Denote $\mathbb{I}$ as the identity mapping. Then, for any $x\in\mathbb{X}$,
\begin{align*}
  (\mathbb{T}^m-\mathbb{I})f(x) \ge r^*(\mathbb{T}^m-\mathbb{I})f_2(x)+(1-r^*)(\mathbb{T}^m-\mathbb{I})f_0(x) = (1-r^*)(\mathbb{T}^m-\mathbb{I})f_0(x)>0,
\end{align*}
where the first inequality is the case due to the concavity of $\mathbb{T}^m$ and the equality is the case because $f_2$ is a fixed point of $\mathbb{T}$. Thus, $\mathbb{T}^mf(x^*)>f(x^*)=f_1(x^*)$. On the other hand, $\mathbb{T}^mf(x)\le \mathbb{T}^mf_1(x)=f_1(x),x\in\mathbb{X}$ because $\mathbb{T}$ is increasing and $f_1$ is a fixed point of $\mathbb{T}$. In particular, $\mathbb{T}^mf(x^*)\le f_1(x^*)$, which is a contradiction.

Finally, we show that for any $f$ such that $\mathbb{T}f$ is well defined, $\{\mathbb{T}^nf\}_{n\ge 0}$ converges to the fixed point of $\mathbb{T}$. We first note that $\mathbb{T}f\ge f_0$ and that $\mathbb{T}f_0$ is well defined according to Assumption \ref{as:BoundedDisutilityLoss}. As a result, $\mathbb{T}^n f$ is well defined for any $n\ge 0$. Recall that $\mathbb{T}^mf_0(x)>f_0(x),x\in\mathbb{X}$ for some $m\ge 1$, so $\mathbb{T}^{m+1}f(x)\ge \mathbb{T}^mf_0(x)>f_0(x),x\in\mathbb{X}$. Thus, in the following, we assume $f(x)>f_0(x),x\in\mathbb{X}$ without loss of generality.

Denote $f^*$ as the unique fixed point of $\mathbb{T}$, and recall that we already showed that $f^*(x)>f_0(x),x\in\mathbb{X}$. Because $\mathbb{X}$ is finite, there must exist $r\in (0,1]$ such that $f-f_0\le (f^*-f_0)/r$, i.e., $rf+(1-r)f_0\le f^*$. Then,
\begin{align*}
  f^*=\mathbb{T}f^* \ge \mathbb{T}(r f + (1-r)f_0)\ge r\mathbb{T}f + (1-r)\mathbb{T}f_0,
\end{align*}
where the equality is the case because $f^*$ is the fixed point, the first inequality is the case because $\mathbb{T}$ is increasing, and the second inequality is the case because $\mathbb{T}$ is concave. Applying $\mathbb{T}$ on both sides of the above inequality, we conclude that $f^*\ge r\mathbb{T}^n f + (1-r)\mathbb{T}^nf_0$ for any $n\ge 0$, which implies
\begin{align*}
  \mathbb{T}^n f\le \left[f^*-(1-r)\mathbb{T}^nf_0\right]/r.
\end{align*}
On the other hand, we have $\mathbb{T}^nf \ge \mathbb{T}^nf_0$. Because $\{\mathbb{T}^nf_0\}_{n\ge 0}$ converges to $f^*$, the squeeze theorem shows that $\{\mathbb{T}^nf\}_{n\ge 0}$ converges to $f^*$ as well. \halmos
\proofend

\proof{Proof of Proposition \ref{prop:DynamicProgramming}.}
{\add One can observe from \eqref{eq:DPhi} that for any $\Phi$ in the domain of $\mathbb{W}$, we have $\big((1-c(x))/c(x)\big)\max_{\bar \theta\in J_x}D_\Phi(x,\bar \theta)\ge \varpi_{c,\theta}^+(x),x\in\mathbb{X}$ for any $(c,\theta)\in {\cal A}$.

Now, suppose $\Phi\in {\cal X}_{++}$ is a solution to \eqref{eq:DynamicPrograming} and consider any $(c,\theta)\in{\cal A}$. Then, we have $\Phi/c=\mathbb{W}\Phi/c\ge f_{0,c,\theta}$, which implies that $\mathbb{V}_{c,\theta}\Phi = c \mathbb{T}_{c,\theta}(\Phi/c)$ is well defined due to Assumption \ref{as:Feasibility}. Moreover, by \eqref{eq:DynamicPrograming} we have $\Phi=\mathbb{W}\Phi\ge \mathbb{V}_{c,\theta}\Phi$.} Consequently, $\{\mathbb{V}_{c,\theta}^n\Phi\}_{n\ge 0}$ is a decreasing sequence and so is $\{\mathbb{T}_{c,\theta}^n(\Phi/c)\}$. By Theorem \ref{th:ExistUniqueFiniteStatePositive}, its limit is the fixed point of $\mathbb{V}_{c,\theta}$, i.e., $F_{c,\theta}$. Thus, $\Phi(x)\ge F_{c,\theta}(x),x\in\mathbb{X}$.

If there exists $(c^*,\theta^*)\in {\cal A}$ such that $(c^*(x),\theta^*(x))$ is a maximizer of \eqref{eq:WOperator} for each $x\in\mathbb{X}$, then $\Phi = \mathbb{V}_{c^*,\theta^*}\Phi$. From the uniqueness of the fixed point of $\mathbb{V}_{c^*,\theta^*}$, we conclude that $\Phi=F_{c^*,\theta^*}$. As a result, $(c^*,\theta^*)$ and $\Phi$ are a maximizer and the optimal value, respectively, of \eqref{eq:PortfolioSelectionMarkovian}.\halmos
\proofend

\proof{Proof of Theorem \ref{prop:DynamicProgramingSolution}.}
{\add  We first derive some properties for $\mathbb{W}$. Because $J_x$ is compact for each $x\in\mathbb{X}$, we immediately conclude that for any fixed $\Phi\in {\cal X}_+$, there exists $\theta$ such that $\theta(x)\in J_x$ and $D_{\Phi}(x,\theta(x)) = \max_{\bar \theta\in J_x}D_\Phi(x,\bar \theta)$ for all $x\in \mathbb{X}$. Moreover, $\max_{\bar \theta\in J_x}D_\Phi(x,\bar \theta)$ is continuous and increasing in $\Phi\in {\cal X}_+$. Consequently, in its domain, $\mathbb{W}$ is increasing and satisfies
\begin{align}\label{eq:DynamicProgramingOperator}
  \mathbb{W}\Phi(x)=\max_{ (c,\theta)\in {\cal A}} H\left(c(x),(1-c(x))D_{\Phi}(x,\theta(x))\right)=\max_{ (c,\theta)\in {\cal A}} c(x){\mathbb{T}}_{c,\theta}(\Phi/c)(x),\; x\in\mathbb{X},
\end{align}
where we set $H(c,z)=-\infty$ for $z<0$ for convenience.

Define $\varphi_y(\bar c):=H(\bar c,(1-\bar c)y)$ for $y\ge 0$. Then, for each $x\in\mathbb{X}$, $\max_{\bar c\in I_x}H(\bar c,(1-\bar c)y) = \max_{\bar c\in \bar I_x}H(\bar c,(1-\bar c)y)$ is continuous in $y\ge 0$, where $\bar I_x$ is the closure of $I_x$ in $\mathbb{R}$ and thus is a compact set. We then conclude from \eqref{eq:DynamicProgramingOperator} and from the continuity of $\max_{\bar \theta\in J_x}D_\Phi(x,\bar \theta)$ in $\Phi$ that $\mathbb{W}$ is continuous in its domain.

For any $y> 0$, straightforward calculation yields
\begin{align*}
  \varphi_y'(\bar c) = H(\bar c, (1-\bar c)y)^\rho (1-\bar c)^{-\rho}(1-\beta)\left[\left(\frac{1-\bar c}{\bar c}\right)^\rho-y^{1-\rho}\frac{\beta}{1-\beta}\right].
\end{align*}
One can see that $ \varphi_y'(\bar c)$ is positive (respectively negative) when $\bar c$ is sufficiently close to 0 (respectively close to 1). For each $x\in \mathbb{X}$, because $I_x$ is close relative to $(0,1)$, $\max_{\bar c\in I_x}\varphi_y(\bar c)$ is uniquely attained by certain $c(x)\in I_x$, and
\begin{align}\label{eq:OptimalCBound}
   \sup\{\bar c\in I_x\mid \bar c\le i_y\}\le c(x)\le \inf\{\bar c\in I_x\mid \bar c\ge i_y\},
\end{align}
where $i_y:=\left(1+y^{\frac{1-\rho}{\rho}}\big(\beta/(1-\beta)\big)^{1/\rho}\right)^{-1}\in(0,1)$, $\sup\emptyset:=i_y$, and $\inf\emptyset:=i_y$. As a result, for each $\Phi$ such that $\max_{\bar \theta\in J_x}D_\Phi(x,\bar \theta)>0,\forall x\in\mathbb{X}$, there exists $(c,\theta)\in{\cal A}$ such that $\mathbb{W}\Phi = c{\mathbb{T}}_{c,\theta}(\Phi/c)$, i.e., the maximum in \eqref{eq:DynamicProgramingOperator} is attained by $(c,\theta)$.

The remaining proof of the theorem is divided into three parts.
}

\medskip
\noindent {\em Part One: existence of the fixed point}
\medskip

We prove the existence of the fixed point of $\mathbb{W}$ in this part. We first consider the case in which there exists $(c,\theta)\in{\cal A}$ such that $\varpi_{c,\theta}(x)<0$ for some $x\in\mathbb{X}$.

Recall $f_{0,c,\theta}$ as defined in Assumption \ref{as:BoundedDisutilityLoss}, i.e., $f_{0,c,\theta}=H(1,\varpi^+_{c,\theta})$ for each $(c,\theta)$, and recall
\begin{align*}
  \Phi_0(x)=\max_{\bar c\in I_x}H\left(\bar c,(1-\bar c)\max_{\bar \theta \in J_x}\big(\sum_{i=1}^n\bar \theta_ib_ig_i(x)\big)^+\right)=\max_{(c,\theta)\in{\cal A}}c(x)f_{0,c,\theta}(x),\; x\in\mathbb{X}.
\end{align*}
By Assumption \ref{as:Feasibility}, for each $(c,\theta)\in {\cal A}$, $\mathbb{T}_{c,\theta}f_{0,c,\theta}$ is well defined. Because $\Phi_0\ge cf_{0,c,\theta}$, i.e., $\Phi_0/c\ge f_{0,c,\theta}$, for each $(c,\theta)\in{\cal A}$, we conclude that $ {\mathbb{T}}_{c,\theta}(\Phi_0/c)$ is well defined for any $(c,\theta)\in{\cal A}$, so {\add $\Phi_0$ is in the domain of $\mathbb{W}$}. Moreover, because $\mathbb{T}_{c,\theta}(f)\ge f_{0,c,\theta}$ for any $(c,\theta)\in{\cal A}$ and $f$ such that $\mathbb{T}_{c,\theta}(f)$ is well defined, we conclude that $\mathbb{T}_{c,\theta}(\Phi_0/c)\ge f_{0,c,\theta}$ for any $(c,\theta)\in{\cal A}$. As a result, $\mathbb{W}\Phi_0\ge \Phi_0$. Now, define $\Phi_n:=\mathbb{W}^n\Phi_0$, $n\ge 1$. Becuase $\mathbb{W}$ is increasing, $\{\Phi_n\}_{n\ge 0}$ is an increasing sequence.

Because $\varpi_{c_0,\theta_0}(x)<0$ for some $x\in\mathbb{X}$ and some $(c_0,\theta_0)\in{\cal A}$, Assumption \ref{as:Feasibility} yields that $\mathbb{T}^m_{c_0,\theta_0}f_{c_0,\theta_0}>f_{c_0,\theta_0}$ for some $m\ge 1$. On the other hand,
\begin{align*}
  \mathbb{W}\Phi_0\ge c_0\mathbb{T}_{c_0,\theta_0}(\Phi_0/c_0)\ge c_0\mathbb{T}_{c_0,\theta_0}(f_{c_0,\theta_0}),
\end{align*}
which implies $\Phi_1/c_0=(\mathbb{W}\Phi_0)/c_0\ge \mathbb{T}_{c_0,\theta_0}(f_{c_0,\theta_0})$. Repeating the the above calculation, we conclude that $\Phi_m/c_0\ge \mathbb{T}^m_{c_0,\theta_0}(f_{c_0,\theta_0})>f_{c_0,\theta_0}$. As a result, $\Phi_m>c_0f_{c_0,\theta_0}$, showing that $\Phi_m\in{\cal X}_{++}$. {\add Also recall that it is assumed that $\max_{\bar \theta\in J_x}D_{\mathbb{W}^{m'}\Phi_{0}}(x,\bar \theta)>0,\forall x\in\mathbb{X}$ for some $m'\ge 0$.}

 Now, we show that $\{\Phi_n\}_{n\ge 0}$ is bounded from above and thus its limit exists. {\add The above shows that without loss of generality, we can assume $\Phi_n\in {\cal X}_{++}$ and $\max_{\bar \theta\in J_x}D_{\Phi_{n}}(x,\bar \theta)>0,\forall x\in\mathbb{X}$ for all $n\ge 1$. Then, }for each $n\ge 1$, there exists $(c_n,\theta_n)\in{\cal A}$ such that
\begin{align*}
  \Phi_n\le \Phi_{n+1} = \mathbb{W}\Phi_n = c_n{\mathbb{T}}_{c_n,\theta_n}(\Phi_n/c_n),
\end{align*}
which implies $\Phi_n/c_n\le {\mathbb{T}}_{c_n,\theta_n}(\Phi_n/c_n)$. Because $\Phi_n\in {\cal X}_{++}$, according to Theorems \ref{th:ExistUniqueFiniteStatePositive} and \ref{th:ExistUniqueFiniteStateNegative}, $\{{\mathbb{T}}_{c_n,\theta_n}^m(\Phi_n/c_n)\}_{m\ge 0}$ converges to $f_{c_n,\theta_n}$, the fixed point of $\mathbb{T}_{c_n,\theta_n}$, as $m$ goes to infinity. Consequently, $\Phi_n/c_n\le f_{c_n,\theta_n}$, i.e., $\Phi_n\le c_nf_{c_n,\theta_n}$. Thus, we only need to show that $\{c_nf_{c_n,\theta_n}\}_{n\ge 1}$ is bounded. {\add When $\rho>1$, we have $c_nf_{c_n,\theta_n}\le f_{c_n,\theta_n}=\mathbb{T}_{c_n,\theta_n}f_{c_n,\theta_n}\le (1-\beta)^{1/(1-\rho)}$, where the first inequality is the case because $c_n\le 1$ and the second one is the case because $H(1,z)\le (1-\beta)^{1/(1-\rho)}$.} Thus, in the following, we only need to consider the case in which $\rho \le 1$.

{\add Recall $\mathbb{U}$, $\eta$, $\delta$, and $v$ in Proposition \ref{prop:Perron}, and denote them as $\mathbb{U}_{c,\theta}$, $\eta_{c,\theta}$, $\delta_{c,\theta}$, and $v_{c,\theta}$, respectively, when $\kappa$ is replaced by $\kappa_{c,\theta}$. Define
\begin{align*}
\hat \kappa_{c,\theta}(X_t,X_{t+1},Y_{t+1}) := \ln c(X_t)-\ln c(X_{t+1}) + \kappa_{c,\theta}(X_t,X_{t+1},Y_{t+1})
\end{align*}
and $\hat{\mathbb{U}}_{c,\theta}h(x):=\mathbb{E}_t\left[u(e^{\hat \kappa_{c,\theta}(X_t,X_{t+1},Y_{t+1})})h(X_{t+1})|X_t=x\right],x\in\mathbb{X}$. Then, straightforward calculation yields that for $\gamma\neq 1$, $\eta_{c,\theta}$ and $\hat v_{c,\theta}:=c^{1-\gamma}v_{c,\theta}$ are, respectively, the Perron-Frobenius eigenvalue and eigenvector of the operator $\hat{\mathbb{U}}_{c,\theta}$. For $\gamma=1$, with $\hat v_{c,\theta}:=\ln c + v_{c,\theta}$, we have
\begin{align*}
\mathbb{E}_t\left[\hat \kappa_{c,\theta}(X_t,X_{t+1},Y_{t+1})|X_t=x\right] = -\mathbb{E}\left[\hat v_{c,\theta}(X_{t+1})|X_t=x\right] + \hat v_{c,\theta}(x) + \eta_{c,\theta},\; \forall x\in\mathbb{X}.
\end{align*}
In the following, when $\gamma\neq 1$, we always identify $\hat v_{c,\theta}$, Perron-Frobenius eigenvector of $\hat{\mathbb{U}}_{c,\theta}$, to the one that has unitary $L^1$ norm. When $\gamma =1$, by the proof of Proposition \ref{prop:Perron}, $\hat v_{c,\theta}$ is the solution to $(\mathbf{I}-\mathbf{P})\hat v_{c,\theta} = \hat w_{c,\theta}-\eta_{c,\theta}\mathbf 1$, where $\hat w_{c,\theta}(x):=\mathbb{E}_t\left[\hat \kappa_{c,\theta}(X_t,X_{t+1},Y_{t+1})|X_t=x\right]$, $\mathbf{I}$ is the identify matrix, and $\mathbf{P}$ is the transition matrix of $\{X_t\}$. We identify $\hat v_{c,\theta}$ to be the one that is orthogonal to the kernel of $\mathbf{I}-\mathbf{P}$, which must be uniquely determined. Moreover, $\hat v_{c,\theta}$ is linear in  $\hat w_{c,\theta}-\eta_{c,\theta}\mathbf 1$, so there exists a constant $L>0$ such that $\|\hat v_{c,\theta}\|\le L\|\hat w_{c,\theta}-\eta_{c,\theta}\mathbf 1\|$ for all $(c,\theta)$, where $\|\cdot\|$ stands for the Euclidean norm. }

Now, we prove the boundedness of $\{c_nf_{c_n,\theta_n}\}_{n\ge 1}$ for the case in which $\rho<1$ and $\gamma\neq 1$. Recalling that $f_{c_n,\theta_n}$ is the fixed point of $\mathbb{T}_{c_n,\theta_n}$, we conclude from \eqref{eq:ChangeOfMeasure2} that
\begin{align*}
  f_{c_n,\theta_n}(x)v_{c_n,\theta_n}(x)^{-\frac{1}{1-\gamma}} &=  H\bigg(v_{c_n,\theta_n}(x)^{-\frac{1}{1-\gamma}},\varpi_{c_n,\theta_n}(x)v_{c_n,
  \theta_n}(x)^{-\frac{1}{1-\gamma}}\\
  &\quad +\delta_{c_n,\theta_n}u^{-1}\left(\tilde \expect_t\left[u\big(f_{c_n,\theta_n}(X_{t+1})v_{c_n,\theta_n}(X_{t+1})^{-\frac{1}{1-\gamma}}\big)|X_t=x\right]\right)\bigg)\\
  &\le H\bigg(\|v_{c_n,\theta_n}^{-\frac{1}{1-\gamma}}\|_\infty,\|\varpi_{c_n,\theta_n}^+v_{c_n,
  \theta_n}^{-\frac{1}{1-\gamma}}\|_\infty+\delta_{c_n,\theta_n}\|f_{c_n,\theta_n}v_{c_n,\theta_n}^{-\frac{1}{1-\gamma}}\|_\infty\bigg),
\end{align*}
where $\|\cdot\|_\infty$ stands for the $L^\infty$ norms of functions on $\mathbb{X}$.
 Consequently, {\add recalling the relation $\hat v_{c,\theta}=c^{1-\gamma}v_{c,\theta}$, we have
\begin{align*}
  \|c_nf_{c_n,\theta_n}\hat v_{c_n,\theta_n}^{-\frac{1}{1-\gamma}}\|_\infty\le H\bigg(\|c_n\hat v_{c_n,\theta_n}^{-\frac{1}{1-\gamma}}\|_\infty,\|\hat \varpi_{c_n,\theta_n}^+\hat v_{c_n,
  \theta_n}^{-\frac{1}{1-\gamma}}\|_\infty+\delta_{c_n,\theta_n}\|c_nf_{c_n,\theta_n}\hat v_{c_n,\theta_n}^{-\frac{1}{1-\gamma}}\|_\infty\bigg),
\end{align*}
where $\hat \varpi_{c,\theta}:=c\varpi_{c,\theta}$.
With $\rho<1$, we have $(y_1+y_2)^{1-\rho}\le y_1^{1-\rho}+y_2^{1-\rho}$ for any $y_1,y_2\ge 0$, so
\begin{align*}
  \|c_nf_{c_n,\theta_n}\hat v_{c_n,\theta_n}^{-\frac{1}{1-\gamma}}\|_\infty^{1-\rho}&\le (1-\beta) \|c_n\hat v_{c_n,\theta_n}^{-\frac{1}{1-\gamma}}\|_\infty^{1-\rho}+\beta \left( \|\hat \varpi_{c_n,\theta_n}^+\hat v_{c_n,
  \theta_n}^{-\frac{1}{1-\gamma}}\|_\infty+\delta_{c_n,\theta_n}\|f_{c_n,\theta_n}v_{c_n,\theta_n}^{-\frac{1}{1-\gamma}}\|_\infty\right)^{1-\rho}\\
  &\le (1-\beta) \|c_n\hat v_{c_n,\theta_n}^{-\frac{1}{1-\gamma}}\|_\infty^{1-\rho}+\beta \|\hat \varpi_{c_n,\theta_n}^+\hat v_{c_n,
  \theta_n}^{-\frac{1}{1-\gamma}}\|_\infty^{1-\rho}+\beta \delta_{c_n,\theta_n}^{1-\rho}\|c_nf_{c_n,\theta_n}\hat v_{c_n,\theta_n}^{-\frac{1}{1-\gamma}}\|_\infty^{1-\rho}.
\end{align*}
As a result,
\begin{align}\label{eq:Bound}
   \|c_nf_{c_n,\theta_n}\hat v_{c_n,\theta_n}^{-\frac{1}{1-\gamma}}\|_\infty^{1-\rho}\le \frac{(1-\beta) \|c_n\hat v_{c_n,\theta_n}^{-\frac{1}{1-\gamma}}\|_\infty^{1-\rho}+\beta \|\hat \varpi_{c_n,\theta_n}^+\hat v_{c_n,
  \theta_n}^{-\frac{1}{1-\gamma}}\|_\infty^{1-\rho}}{1-\beta \delta_{c_n,\theta_n}^{1-\rho}}.
\end{align}
We first recall that $\sup_{(c,\theta)\in{\cal A}}\beta \delta_{c,\theta}^{1-\rho}<1$. In addition, $\sup_{(c,\theta)\in {\cal A}}\|\hat \varpi_{c_n,\theta_n}^+\|_\infty<+\infty$ because $c(x)\in(0,1)$ and $J_x$ is compact for any $x\in\mathbb{X}$. Thus, by \eqref{eq:Bound}, to prove that $\{c_nf_{c_n,\theta_n}\}_{n\ge 1}$ is bounded, we only need to show that $\{\hat v_{c_n,\theta_n}\}_{n\ge }$ is uniformly bounded from above by a constant and from below by another positive constant. Because $\hat v_{c_n,\theta_n}$ is the Perron-Frobenius eigenvector of $\hat{\mathbb{U}}_{c_n,\theta_n}$ with unitary $L^1$ norm, its entries must be bounded by 1. Thus, we only need to prove that $\{\hat v_{c_n,\theta_n}\}_{n\ge 1}$ is uniformly bounded from below by a positive constant. Equivalently, denoting $L_{c_n,\theta_n}$ as the ratio of the maximum entry of $\hat v_{c_n,\theta_n}$ divided by the minimum entry of $\hat v_{c_n,\theta_n}$, we only need to prove that $\{L_{c_n,\theta_n}\}_{n\ge 1}$ is bounded.

Recall that $\max_{\bar \theta\in J_x}D_{\Phi_{n}}(x,\bar \theta)>0,\forall x\in\mathbb{X}$ for all $n\ge 1$ and that $\mathbb{X}$ is of finite elements, so there exists $\epsilon>0$ such that $\max_{\bar \theta\in J_x}D_{\Phi_{n}}(x,\bar \theta)\ge \epsilon$, for all $x\in\mathbb{X}$ and all $n\ge 1$. Recall the definition of $(c_n,\theta_n)$. Then, the bound in \eqref{eq:OptimalCBound} yields that there exists $\tau\in (0,1)$ such that $c_n(x)\le  \tau$, for all $x\in\mathbb{X}$ and all $n\ge 1$. Recall $\hat \kappa_{c_n,\theta_n}$ and $\hat{\mathbb{U}}_{c_n,\theta_n}$, and denote the matrix representing $\hat{\mathbb{U}}_{c_n,\theta_n}$ as $\hat{\mathbf{P}}^{c_n,\theta_n}$. Then, we have
\begin{align*}
  \hat{\mathbf{P}}^{c_n,\theta_n}_{x,y} &= \mathbf P_{x,y} \mathbb{E}_t\left[u\left(e^{\hat\kappa_{c_n,\theta_n}(X_t,X_{t+1},Y_{t+1})}\right)|X_t=x,X_{t+1}=y\right] = \mathbf P_{x,y} u(1-c_n(x))\mathbf{Q}^{\theta_n}_{x,y},
\end{align*}
where $\mathbf{Q}^{\theta}_{x,y}:=\mathbb{E}_t\left[u\left(R_\theta(X_t,X_{t+1},Y_{t+1})\right)|X_t=x,X_{t+1}=y\right]$. Because $\{X_t\}$ is irreducible, there exists $k\ge 1$ such that $\mathbf{P}^k$ is positive. Consequently, because $c_n\in(0,1)$ and $R_{\theta_n}>0$, we conclude that $(\hat{\mathbf{P}}^{c_n,\theta_n})^k$ is positive. Note that $\eta_{c_n,\theta_n}^k$ and $\hat v_{c_n,\theta_n}$ are, respectively, the Perron-Frobenius eigenvalue and eigenvector of $(\hat{\mathbf{P}}^{c_n,\theta_n})^k$, so\footnote{For a positive matrix $A$, denote its Perron-Frobenius eigenvalue and eigenvector as $\eta$ and $v$, respectively, and denote the maximum and lowest entries of $v$ as $v_{i^*}$ and $v_{j^*}$, respectively. Then, the Collatz-Wielandt formula shows that the eigenvalue $\eta\le \max_{i,j}A_{i,j}$. On the other hand, from the identity $Av=\delta v$, we conclude that $\delta v_{j^*} = \sum_{j}A_{j^*,j}v_j\ge A_{j^*,i^*}v_{i^*}\ge (\min_{i,j}A_{i,j})v_{i^*}$. Thus, we conclude $v_{i^*}/v_{j^*}\le \max_{i,j}A_{i,j}/\min_{i,j}A_{i,j}$.}
\begin{align*}
  L_{c_n,\theta_n}\le \max_{x,y}(\hat{\mathbf{P}}^{c_n,\theta_n})^k_{xy}/\min_{x,y}(\hat{\mathbf{P}}^{c_n,\theta_n})^k_{xy}.
\end{align*}
Because $c_n(x)\le  \tau$ and thus $1-c_n(x)\in (1-\tau,1)$ for all $x\in\mathbb{X}$ and all $n\ge 1$ and because $J_x$ is compact for each $x\in\mathbb{X}$, it is straightforward to see that $a_{\mathrm{min}}:=\inf_{n\ge 1}\min_{x,y}u(1-c_n(x))\mathbf{Q}^{\theta_n}_{x,y}>0$ and $a_{\mathrm{max}}:=\sup_{n\ge 1}\max_{x,y}u(1-c_n(x))\mathbf{Q}^{\theta_n}_{x,y}<+\infty$. Consequently, we have $a_{\mathrm{min}}^k\mathbf{P}^k_{xy}\le (\hat{\mathbf{P}}^{c_n,\theta_n})^k_{xy}\le a_{\mathrm{max}}^k\mathbf{P}^k_{xy}$, and thus
\begin{align*}
  \sup_{n\ge 1}L_{c_n,\theta_n}\le \max_{x,y}[a_{\mathrm{max}}^k\mathbf{P}^k_{xy}]/\min_{x,y}[a_{\mathrm{min}^k}\mathbf{P}^k_{xy}] = \big(a_{\mathrm{max}}^k/a_{\mathrm{min}}^k\big)\big(\max_{x,y}\mathbf{P}^k_{xy}/\min_{x,y}\mathbf{P}^k_{xy}\big)<\infty.
\end{align*}

Next, we consider the case in which $\rho<1$ and $\gamma=1$. Following the same calculation as above, we conclude that
\begin{align*}
   \|c_nf_{c_n,\theta_n}e^{-\hat v_{c_n,\theta_n}}\|_\infty^{1-\rho}\le \frac{(1-\beta) \|c_ne^{-\hat v_{c_n,\theta_n}}\|_\infty^{1-\rho}+\beta \|\hat \varpi_{c_n,\theta_n}^+e^{-\hat v_{c_n,\theta_n}}\|_\infty^{1-\rho}}{1-\beta \delta_{c_n,\theta_n}^{1-\rho}},
\end{align*}
and we only need to show that $\{\|v_{c_n,\theta_n}\|\}_{n\ge 1}$ is uniformly bounded. Recalling that $\|\hat v_{c,\theta}\|\le L\|\hat w_{c,\theta}-\eta_{c,\theta}\mathbf 1\|$ for some $L>0$ and all $(c,\theta)$, that $\eta_{c,\theta}$ is a probability distribution on $\mathbb{X}$ and thus the sum of its entries is equal to 1, that $J_x$ is compact for all $x\in\mathbb{X}$, and that $1-c_n(x)\in (1-\tau,1)$ for some $\tau\in(0,1)$ and all $x\in\mathbb{X}$ and $n\ge 2$, we immediately conclude that $\{\|v_{c_n,\theta_n}\|\}_{n\ge 1}$ is uniformly bounded.

Next, we consider the case in which $\rho=1$. The same calculation as above shows that
\begin{align*}
  \left\|\frac{c_nf_{c_n,\theta_n}}{u^{-1}(\hat v_{c_n,\theta_n})}\right\|_\infty\le H\bigg(\left\|\frac{c_n}{u^{-1}(\hat v_{c_n,\theta_n})}\right\|_\infty,\left\|\frac{\hat \varpi_{c_n,\theta_n}^+}{u^{-1}(\hat v_{c_n,\theta_n})}\right\|_\infty+\delta_{c_n,\theta_n}\left\|\frac{c_nf_{c_n,\theta_n}}{u^{-1}(\hat v_{c_n,\theta_n})}\right\|_\infty\bigg),
\end{align*}
$\hat \varpi_{c_n,\theta_n}^+$ is uniformly bounded from above by a positive constant $K$, and
$\big(u^{-1}(\hat v_{c_n,\theta_n})\big)^{-1}$ is uniformly bounded from below by a positive constant $b_{\mathrm{min}}$ and bounded from above by a positive constant $b_{\mathrm{max}}$. When $\gamma\neq 1$, recall that $\eta_{c_n,\theta_n}^k$ is the Perron-Frobenius eigenvalue $(\hat{\mathbf{P}}^{c_n,\theta_n})^k$, and that we already proved that $ a_{\mathrm{min}}^k\mathbf{P}^k_{xy}\le (\hat{\mathbf{P}}^{c_n,\theta_n})^k_{xy}\le a_{\mathrm{max}}^k\mathbf{P}^k_{xy}$ for some positive constants $a_{\mathrm{min}}$ and $a_{\mathrm{max}}$. Consequently,
\begin{align*}
  a_{\mathrm{min}}^k\min_{x,y}\mathbf{P}^k_{xy}\le \min_{x,y}(\hat{\mathbf{P}}^{c_n,\theta_n})^k_{xy}\le \delta_{c_n,\theta_n}^k\le \max_{x,y}(\hat{\mathbf{P}}^{c_n,\theta_n})^k_{xy}\le a_{\mathrm{max}}^k\max_{x,y}\mathbf{P}^k_{xy}.
\end{align*}
Therefore, there exists $\bar \delta>0$ such that $\sup_{n\ge 1}\delta_{c_n,\theta_n}=\sup_{n\ge 1}u^{-1}(\eta_{c_n,\theta_n})\le \bar \delta$. When $\gamma=1$, $\delta_{c_n,\theta_n}=e^{\eta_{c_n,\theta_n}}$, where $\eta_{c_n,\theta_n} = \mathbb{E}\left[\kappa_{c_n,\theta_n}(X_t,X_{t+1},Y_{t+1})\right]$, where the expectation is taken under the stationary probability of $\{X_t\}$. Recalling \eqref{eq:kappa}, we immediately conclude that
\begin{align*}
  \eta_{c_n,\theta_n} = \mathbb{E}[\ln(1-c_n(X_t))] + \mathbb{E}[\ln R_{\theta_n}(X_t,X_{t+1},Y_{t+1})].
\end{align*}
Because there exists $\tau\in(0,1)$ such that $c_n\le \tau$ for all $n\ge 1$ and because $J_x$ is compact for all $x\in\mathbb{X}$, we immediately conclude that $\sup_{n\ge 1}|\eta_{c_n,\theta_n}|<+\infty$ and thus that $\sup_{n\ge 1}\delta_{c_n,\theta_n}\le \bar \delta$ for certain constant $\bar \delta$. Then, for any $\gamma>0$, we have
\begin{align*}
  \|c_nf_{c_n,\theta_n}/u^{-1}(\hat v_{c_n,\theta_n})\|_\infty\le H\bigg(b_{\mathrm{max}},Kb_{\mathrm{max}}+\bar \delta\|c_nf_{c_n,\theta_n}/u^{-1}(\hat v_{c_n,\theta_n})\|_\infty\bigg).
\end{align*}
Recalling that $H(c,z) = e^{(1-\beta)\ln c + \beta\ln z}$, we immediately conclude that $\|c_nf_{c_n,\theta_n}u^{-1}(\hat v_{c_n,\theta_n})\|_\infty$, and thus $\|c_nf_{c_n,\theta_n}\|_\infty$ are uniformly bounded in $n\ge 1$.

}

We have proved that the limit of $\{\Phi_n\}$ exists and must be in ${\cal X}_{++}$. Then, by the continuity of $\mathbb{W}$, the limit must be a fixed point of $\mathbb{W}$ in ${\cal X}_{++}$.

Next, we consider the case in which $\varpi_{c,\theta}\ge 0$ for any $(c,\theta)\in {\cal A}$. In this case, for any $(c_0,\theta_0)\in {\cal A}$, $\mathbb{W}F_{c_0,\theta_0}$ is well defined. Moreover, we have $\mathbb{W}F_{c_0,\theta_0}\ge c_0 \mathbb{T}_{c_0,\theta_0}(F_{c_0,\theta_0}/c_0)=F_{c_0,\theta_0}$, so $\{\mathbb{W}^nF_{c_0,\theta_0}\}_{n\ge 0}$ is an increasing sequence. On the one hand, the sequence $\{\mathbb{W}^nF_{c_0,\theta_0}\}_{n\ge 0}$ is in ${\cal X}_{++}$ because $F_{c_0,\theta_0}\in{\cal X}_{++}$. On the other hand, following the same proof as in the previous case, we can show that this sequence is bounded from above. As a result, this sequence converges and the convergent point is a fixed point of $\mathbb{W}$ in ${\cal X}_{++}$.

\medskip
\noindent {\em Part Two: uniqueness of the fixed point}
\medskip

{\add  Consider any fixed point $\Phi$ of $\mathbb{W}$ in ${\cal X}_{++}$. If $\varpi_{c,\theta}\ge 0$ for any $(c,\theta)\in {\cal A}$, then we have $\max_{\bar \theta\in J_x}D_\Phi(x,\bar \theta)>0,\forall x\in\mathbb{X}$, so there exists $(c^*,\theta^*)\in{\cal A}$ such that $(c^*(x),\theta^*(x))$ solves the maximization problem in \eqref{eq:WOperator} for each $x\in\mathbb{X}$. If $\varpi_{c,\theta}(x)< 0$ for some $x\in\mathbb{X}$ and some $(c,\theta)\in {\cal A}$, we have $\Phi=\mathbb{W}\Phi\ge \Phi_0$ where $\Phi_0$ is as defined in \eqref{eq:Phi0}. Consequently, there exists $m\ge 1$ such that
\begin{align*}
  \max_{\bar \theta\in J_x}D_{\Phi}(x,\bar \theta) = \max_{\bar \theta\in J_x}D_{\mathbb{W}\Phi^{m}}(x,\bar \theta) \ge \max_{\bar \theta\in J_x}D_{\mathbb{W}^m\Phi_0}(x,\bar \theta)>0,\; x\in\mathbb{X}.
\end{align*}
Therefore, there also exists $(c^*,\theta^*)\in{\cal A}$ such that $(c^*(x),\theta^*(x))$ solves the maximization problem in \eqref{eq:WOperator} for each $x\in\mathbb{X}$. Now, Proposition \ref{prop:DynamicProgramming} yields that $\Phi$ must be the optimal value of \eqref{eq:PortfolioSelectionMarkovian}, and thus the fixed point of $\mathbb{W}$ in ${\cal X}_{++}$ is unique.

}


\medskip
\noindent {\em Part Three: computing the fixed point}
\medskip


Denote the unique fixed point of $\mathbb{W}$ as $\Phi^*$. We first consider the case in which $\varpi_{c,\theta}\ge 0$ for any $(c,\theta)\in{\cal A}$. Note that for any $\Phi\in{\cal X}_{++}$, because $\mathbb{X}$ is finite, there exists $r>1$ such that $(1/r)\Phi^*(x)\le \Phi(x)\le r\Phi^*(x),x\in\mathbb{X}$. Because $\varpi_{c,\theta}\ge 0$ for all $(c,\theta)\in{\cal A}$ and $\beta< 1$,
\begin{align*}
 \mathbb{W}(r\Phi^*)(x) = \max_{ (c,\theta)\in {\cal A}} c(x){\mathbb{T}}_{c,\theta}(r\Phi^*/c)(x) \le  \max_{ (c,\theta)\in {\cal A}} c(x)r{\mathbb{T}}_{c,\theta}(\Phi^*/c)(x) = r\mathbb{W}\Phi^*(x) = r\Phi^*(x).
\end{align*}
Therefore, $\{\mathbb{W}^n(r\Phi^*)\}_{n\ge 0}$ is a decreasing sequence. Similarly, $\{\mathbb{W}^n((1/r)\Phi^*)\}_{n\ge 0}$ is an increasing sequence. Moreover, $\mathbb{W}^n((1/r)\Phi^*)\le \mathbb{W}^n\Phi\le \mathbb{W}^n(r\Phi^*)$ because $\mathbb{W}$ is increasing. As a result, both $\{\mathbb{W}^n(r\Phi^*)\}_{n\ge 0}$ and $\{\mathbb{W}^n((1/r)\Phi^*)\}_{n\ge 0}$ converge in ${\cal X}_{++}$ and the convergent points are fixed points of $\mathbb{W}$ in ${\cal X}_{++}$. Because the fixed point of $\mathbb{W}$ in ${\cal X}_{++}$ is unique, both $\{\mathbb{W}^n(r\Phi^*)\}_{n\ge 0}$ and $\{\mathbb{W}^n((1/r)\Phi^*)\}_{n\ge 0}$ converge to this fixed point, i.e., to $\Phi^*$. By the squeeze theorem, $\{\mathbb{W}^n\Phi\}_{n\ge 0}$ converges to $\Phi^*$ as well.

Next, we consider the case in which $\varpi_{c,\theta}(x)<0$ for some $x\in\mathbb{X}$ and some $(c,\theta)\in{\cal A}$. In this case, for any $\Phi$ {\add in the domain of $\mathbb{W}$}, we have
\begin{align*}
  \mathbb{W}\Phi = \max_{ (c,\theta)\in {\cal A}} c(x){\mathbb{T}}_{c,\theta}(\Phi/c)\ge \max_{ (c,\theta)\in {\cal A}} c(x)f_{0,c,\theta}=\Phi_0.
\end{align*}
As a result, $\mathbb{W}^n\Phi\ge \mathbb{W}^{n-1}\Phi_0$, $n\ge 1$. On the other hand, consider the following operator:
\begin{align*}
  \mathbb{W}_+\Phi:=\max_{ (c,\theta)\in {\cal A}} c(x){\mathbb{T}}_{+,c,\theta}(\Phi/c),
\end{align*}
where ${\mathbb{T}}_{+,c,\theta}$ is defined by replacing $\varpi_{c,\theta}$ in $\mathbb{T}_{c,\theta}$ with $\varpi^+_{c,\theta}$. We already showed that $\mathbb{W}_+\Phi$ has a unique fixed point in ${\cal X}_{++}$, and we denote this fixed point as $\Phi^*_+$. Because $\mathbb{X}$ is finite, there exists $r\ge 1$ such that $\Phi\le r\Phi^*_+$. Then,
\begin{align*}
  \mathbb{W}(\Phi)\le \mathbb{W}(r\Phi^*_+) \le \mathbb{W}_+(r\Phi^*_+)\le r\mathbb{W}_+\Phi^*_+ = r\Phi^*_+,
\end{align*}
where the first inequality is the case because $\mathbb{W}$ is increasing, the second inequality is the case because $\mathbb{W}_+$ dominates $\mathbb{W}$, the third inequality is the case because $\beta < 1$ and $\varpi_{c,\theta}^+\ge 0$ for all $(c,\theta)\in {\cal A}$, and the equality is the case because $\Phi^*_+$ is the fixed point of $\mathbb{W}_+$. As a result, $\{\mathbb{W}^n(r\Phi^*_+)\}_{n\ge 1}$ is a decreasing sequence and dominates $\{\mathbb{W}^n(\Phi)\}_{n\ge 1}$, and thus it dominates $\{\mathbb{W}^{n-1}(\Phi_0)\}_{n\ge 1}$ as well. We already showed that $\{\mathbb{W}^{n-1}(\Phi_0)\}_{n\ge 1}$ converges to $\Phi^*$, so $\{\mathbb{W}^n(r\Phi^*_+)\}_{n\ge 1}$ must converge in ${\cal X}_{++}$, and the convergent point is a fixed point of $\mathbb{W}$ in ${\cal X}_{++}$. Because the fixed point of $\mathbb{W}$ in ${\cal X}_{++}$ is unique, we conclude that $\{\mathbb{W}^n(r\Phi^*_+)\}_{n\ge 1}$ converges to $\Phi^*$ as well. By the squeeze theorem, we conclude that
$\{\mathbb{W}^n\Phi\}_{n\ge 0}$ converges to $\Phi^*$.\halmos
\proofend

{\add

\proof{Proof of Proposition \ref{prop:SufficientCond}}
Suppose $\rho\ge 1$. In the proof of Theorem \ref{prop:DynamicProgramingSolution}, we already showed that $\mathbb{W}^n\Phi_0\in{\cal X}_{++}$ for sufficiently large $n$. Then, we must have $\max_{\bar \theta\in J_x}D_{\mathbb{W}^n\Phi_0}(x,\bar \theta)>0,\forall x\in\mathbb{X}$ because
\begin{align*}
  0<\mathbb{W}^{n+1}\Phi_0(x) =\max_{\bar c\in I_x}H(\bar c,(1-\bar c)\max_{\bar \theta\in J_x}D_{\mathbb{W}^n\Phi_0}(x,\bar \theta)),\; \forall x\in\mathbb{X}
\end{align*}
and because $H(c,0)=0$ in the case $\rho\ge 1$.

In the following, we assume $\rho<1$, denote $D_{\mathbb{W}^{-1}\Phi_0}(x,\bar \theta):=\big(\sum_{i=1}^n\bar \theta_ig_i(x)\big)^+$, and, for the sake of contradiction, we assume that for any $N$ there exists $n\ge N$ such that $\max_{\bar \theta\in J_x}D_{\mathbb{W}^n\Phi_0}(x,\bar \theta)=0$ for some $x\in\mathbb{X}$. Because $\mathbb{X}$ is of finite elements, there exists $x\in\mathbb{X}$ such that $\max_{\bar \theta\in J_x}D_{\mathbb{W}^n\Phi_0}(x,\bar \theta)=0$ is true for infinitely many $n$ and thus true for any $n\ge 0$ due to the monotonicity of $D_{\Phi}$ in $\Phi$. Denote the set of such $x$ as $\mathbb{X}_0$. For any $x\in\mathbb{X}_0$, we have
\begin{align*}
  \mathbb{W}^n\Phi_0(x) = \max_{\bar c\in I_x}H\left(\bar c,(1-\bar c)\max_{\bar \theta\in J_x}D_{\mathbb{W}^{n-1}\Phi_0}(x)\right) =\max_{\bar c\in I_x}H(\bar c,0)>0
\end{align*}
for any $n\ge 0$. Now, consider $y\in \mathbb{X}$ such that $y$ is reachable from certain $x\in\mathbb{X}_0$, i.e., $\prob(X_{t+1}=y\mid X_t=x)>0$. We claim that $\mathbb{W}^{n}\Phi_0(y)$ is a constant in $n\ge 0$. For the sake of contradiction, suppose $\mathbb{W}^{n_0}\Phi_0(y)>\mathbb{W}^{n_0-1}\Phi_0(y)$ for certain $n_0\ge 1$. Then, because $y$ is reachable from $x$ and because $\mathbb{W}^{n_0}\Phi_0\ge \mathbb{W}^{n_0-1}\Phi_0$, we conclude from the definition of $D_\Phi$ in \eqref{eq:DPhi} that $\max_{\bar \theta\in J_x}D_{\mathbb{W}^{n_0}\Phi_0}(x,\bar \theta)> \max_{\bar \theta\in J_x}D_{\mathbb{W}^{n_0-1}\Phi_0}(x,\bar \theta)$ and, consequently, $\mathbb{W}^{n_0+1}\Phi_0(x)>\mathbb{W}^{n_0}\Phi_0(x)$, which is a contradiction. Denote by $\mathbb{X}_1$ the union of the set of such $y$ as above and $\mathbb{X}_0$, and the above analysis shows that $\mathbb{W}^{n}\Phi_0(x)$ is a constant in $n\ge 0$ for any $x\in\mathbb{X}_1$. The same argument as above then shows that for any $y$ that is reachable from certain $x\in\mathbb{X}_1$, $\mathbb{W}^{n}\Phi_0(y)$ is also constant in $n\ge 0$. Because $\{X_t\}$ is irreducible, we can eventually show that $\mathbb{W}^{n}\Phi_0(x)$ is constant in $n\ge 0$ for any $x\in \mathbb{X}$. This implies that
\begin{align}\label{eq:AuxEq1}
  \max_{\bar \theta\in J_x}D_{\Phi_0}(x,\bar \theta)= \max_{\bar \theta\in J_x}\big(\sum_{i=1}^n\bar \theta_ib_ig_i(x)\big)^+,\; \forall x\in\mathbb{X}.
\end{align}
We further claim that
\begin{align}\label{eq:AuxEq3}
  \max_{\bar \theta\in J_x}\sum_{i=1}^n\bar \theta_ib_ig_i(x)<0,\; \forall x\in\mathbb{X}.
\end{align}
 For the sake of contradiction, suppose there exists $x_0\in\mathbb{X}$ such that $\max_{\bar \theta\in J_{x_0}}\sum_{i=1}^n\bar \theta_ib_ig_i(x_0)\ge 0$. Then, because $\Phi_0$ is positive, we have
\begin{align}\label{eq:AuxEq2}
   \max_{\bar \theta\in J_{x_0}}D_{\Phi_0}(x_0,\bar \theta) >\max_{\bar \theta\in J_{x_0}}\sum_{i=1}^n\bar \theta_ib_ig_i(x_0).
\end{align}
Combining \eqref{eq:AuxEq1} and \eqref{eq:AuxEq2} and noting that $\max_{\bar \theta\in J_{x_0}}\big(\sum_{i=1}^n\bar \theta_ib_ig_i(x_0)\big)^+=\max_{\bar \theta\in J_{x_0}}\sum_{i=1}^n\bar \theta_ib_ig_i(x_0)$ because the latter value is nonnegative, we arrive at a contradiction.

Now, if $\max_{\bar \theta\in J_x}\sum_{i=1}^n\bar \theta_ib_ig_i(x)\ge 0$ for certain $x\in\mathbb{X}$, then \eqref{eq:AuxEq3} cannot hold, so we must have $\max_{\bar \theta\in J_x}D_{\mathbb{W}^n\Phi_0}(x,\bar \theta)>0,\forall x\in\mathbb{X}$ for sufficiently large $n$. On the other hand, suppose there exists $a\in(0,1)$ such that $J_x\subset (0,a]$ for any $x\in \mathbb{X}$. Then, because $I_x$ is closed relative to $(0,1)$, there exists $c^*(x)\in I_x$ such that $H(c^*(x),0)=\max_{\bar c\in I_x}H(\bar c,0)$. Consequently, \eqref{eq:AuxEq3} implies that for a given feasible $\theta$,
\begin{align*}
  \Phi_0(x) = \max_{\bar c\in I_x}H(\bar c,0) = H(c^*(x),0) = H\left(c^*(x),\left(\sum_{i=1}^n \theta_i(x)b_ig_i(x)\right)^+\right) = c^*(x)f_{c^*,\theta}(x),\; \forall x\in\mathbb{X}.
\end{align*}
By the definition of $\mathbb{W}$, we have $\mathbb{W}^n\Phi_0\ge  \mathbb{V}_{c^*,\theta}(\mathbb{W}^{n-1}\Phi_0)=c^*\mathbb{T}_{c^*,\theta}(\mathbb{W}^{n-1}\Phi_0/c^*)$ for any $n\ge 1$, so we have $\mathbb{W}^n\Phi_0\ge c^*\mathbb{T}^n_{c^*,\theta}(\Phi_0/c^*)= c^*\mathbb{T}_{c^*,\theta}^{n}f_{c^*,\theta} $ for any $n\ge 1$. Because of \eqref{eq:AuxEq3}, we conclude $\varpi_{c^*,\theta}(x)<0$ for any $x\in\mathbb{X}$, so Assumption \ref{as:Feasibility} implies that $\mathbb{T}_{c^*,\theta}^{m}f_{c^*,\theta}>f_{c^*,\theta}$ for certain $m\ge 1$. Consequently, we have
\begin{align*}
  \mathbb{W}^m\Phi_0\ge c_0^*\mathbb{T}_{c^*,\theta}^{m}f_{c^*,\theta}>c^*f_{c^*,\theta} = \Phi_0,
\end{align*}
which is a contradiction. \halmos
\proofend

\proof{Proof of Proposition \ref{prop:VerifyCon}}
For each fixed $(c,\theta)\in{\cal A}$, Proposition \ref{prop:Perron}-(iii) and \eqref{eq:kappa} yields that
\begin{align}
  \delta_{c,\theta} &= \max_{f\in{\cal X}_{++}} \min_{x\in\mathbb{X}} \frac{u^{-1}\Big(\expect_t\left[u\big((1-c(X_t))c(X_t)^{-1}R_{\theta}(X_t,X_{t+1},Y_{t+1})C(X_{t+1})f(X_{t+1})\big)|X_t=x\right]\Big)}{f(x)}.\notag \\
  & = \max_{\tilde f\in{\cal X}_{++}} \min_{x\in\mathbb{X}} \left\{(1-c(x)) \frac{u^{-1}\Big(\expect_t\left[u\big(R_{\theta}(X_t,X_{t+1},Y_{t+1})\tilde f(X_{t+1})\big)|X_t=x\right]\Big)}{\tilde f(x)}\right\}.\label{eq:deltaRepAlt}
\end{align}
By considering $\tilde f\equiv 1$, we immediately conclude that
\begin{align}
  \delta_{c,\theta}\ge \min_{x\in\mathbb{X}} \left\{(1-c(x)) u^{-1}\Big(\expect_t\left[u\big(R_{\theta}(X_t,X_{t+1},Y_{t+1})\big)|X_t=x\right]\Big)\right\}.\label{eq:deltaLB}
\end{align}
On the other hand, for any $\tilde f\in {\cal X}_{++}$, there exists $x_{\tilde f}\in \mathbb{X}$ such that $f(x_{\tilde f})=\max_{x\in\mathbb{X}}\tilde f(x)$. Consequently,
\begin{align*}
  &\min_{x\in\mathbb{X}} \left\{(1-c(x)) \frac{u^{-1}\Big(\expect_t\left[u\big(R_{\theta}(X_t,X_{t+1},Y_{t+1})\tilde f(X_{t+1})\big)|X_t=x\right]\Big)}{\tilde f(x)}\right\}\\
  &\le (1-c(x_{\tilde f})) \frac{u^{-1}\Big(\expect_t\left[u\big(R_{\theta}(X_t,X_{t+1},Y_{t+1})\tilde f(X_{t+1})\big)|X_t=x_{\tilde f}\right]\Big)}{\tilde f(x_{\tilde f})}\\
  &\le (1-c(x_{\tilde f})) u^{-1}\Big(\expect_t\left[u\big(R_{\theta}(X_t,X_{t+1},Y_{t+1})\big)|X_t=x_{\tilde f}\right]\Big)\\
  &\le \max_{x\in\mathbb{X}} \left\{(1-c(x)) u^{-1}\Big(\expect_t\left[u\big(R_{\theta}(X_t,X_{t+1},Y_{t+1})|X_t=x\right]\Big)\right\},
\end{align*}
where the second inequality is the case because $\tilde f(y)/\tilde f(x_{\tilde f})\le 1$ for any $y\in\mathbb{X}$. Consequently, we conclude
\begin{align}
  \delta_{c,\theta}\le \max_{x\in\mathbb{X}} \left\{(1-c(x)) u^{-1}\Big(\expect_t\left[u\big(R_{\theta}(X_t,X_{t+1},Y_{t+1})\big)|X_t=x\right]\Big)\right\}.\label{eq:deltaUB}
\end{align}
The conclusion of the proposition then follows from \eqref{eq:deltaLB} and \eqref{eq:deltaUB}.\halmos
\proofend

\proof{Proof of Proposition \ref{prop:VerifyCon2}}
For fixed $(c,\theta)\in {\cal A}^-$, the monotonicity of $H(c,z)$ in $z$ yields that $\mathbb{T}_{c,\theta}f_{0,c,\theta}>f_{0,c,\theta}$ if and only if
\begin{align*}
  \varpi_{c,\theta}(x) + u^{-1}\Big(\expect_t\left[u\big(e^{\kappa_{c,\theta}(X_t,X_{t+1},Y_{t+1})}f_{c,\theta}(X_{t+1})\big)|X_t=x\right]\Big)>\varpi_{c,\theta}^+(x),\; \forall x\in\mathbb{X}.
\end{align*}
Straightforward calculation shows that the above is equivalent to
\begin{align*}
  u^{-1}\left(\expect_t\left[u\left(R_\theta(X_t,X_{t+1},Y_{t+1})H\left(c(X_{t+1}),(1-c(X_{t+1}))\left(\sum_{i=1}^n \theta_i(X_{t+1})b_ig_i(X_{t+1}) \right)^+\right)\right)|X_t=x\right]\right)\\
  >\left(\sum_{i=1}^n \theta_i(X_{t+1})b_ig_i(X_{t+1}) \right)^-,\; x\in\mathbb{X}.
\end{align*}
Because $H(\bar c,(1-\bar c)z)$ is concave in $\bar c$ for any $z\ge 0$, one can see that, for the above to hold for any $(c,\theta)\in {\cal A}^-$, it is sufficient to have \eqref{eq:VerifyCon2}.

Next, suppose $\rho<1$. Note that
\begin{align*}
  \min_{\bar c\in\{\underline{i},\bar{i}\}}H\left(\bar c,(1-\bar c)\left(\sum_{i=1}^nb_i\theta_i(X_{t+1})g_i(X_{t+1})\right)^+\right)\ge H(\underline{i},0),
\end{align*}
so one can see that a sufficient condition for \eqref{eq:VerifyCon2} is given by \eqref{eq:VerifyCon2Suff}.\halmos
\proofend

}

\proof{Proof of Proposition \ref{th:ExistenceGeneral}.}
Following the proof of Theorem \ref{th:ExistUniqueFiniteStatePositive}, we can show that $f$ is a fixed point of $\mathbb{T}$ in ${\cal X}_{++}$ if and only if $g(x):=\big(f(x)/u^{-1}(v(x))\big)^{1-\rho},x\in\mathbb{X}$ is a fixed point of $\mathbb{S}$ in ${\cal X}_{++}$. In addition, when $\alpha\ge 1$, inequality \eqref{eq:ContractionMapping} holds for any $g_1,g_2\in \tilde L^\alpha_+(\mathbb X)$. As a result, because $\mathbb{S}0\in \tilde L^\alpha_+(\mathbb X)$, this inequality implies that $\mathbb{S}g\in \tilde L^\alpha_+(\mathbb X)$ for any $g\in\tilde L^\alpha_+(\mathbb X)$. Moreover, because $\beta \delta^{1-\rho}<1$, $\mathbb{S}$ is a contraction mapping on $\tilde L^\alpha_+(\mathbb X)$ and thus admits a unique fixed point. In particular, $\{\mathbb{S}^n0\}_{n\ge 0}$ is an increasing sequence converging to the fixed point. Because $\mathbb{S}0(x)\ge (1-\beta)/\tilde u^{-1}(v(x))>0,x\in\mathbb{X}$, we conclude that the fixed point must be positive.

When $\alpha<1$, define
\begin{align*}
\begin{split}
  \tilde {\mathbb{S}}g(x):=\frac{1-\beta}{\tilde u^{-1}(v(x))} + \beta \delta^{1-\rho}\left\{\left(\tilde \expect_t\left[g(X_{t+1})^{\alpha'}|X_t=x\right]\right)^{\frac{1}{\alpha'(1-\rho)}}+
\frac{\delta^{-1}\varpi(x)}{ u^{-1}\big(v(x)\big)}\right\}^{1-\rho}.
\end{split}
\end{align*}
Because $\tilde {\mathbb{S}}0 = \mathbb{S}0\in \tilde L^{\alpha'}_+(\mathbb X)$ and $\beta \delta^{1-\rho}<1$, we can show that $\tilde{\mathbb{S}}$ is a contraction mapping on $L^{\alpha'}_+(\mathbb X)$. In particular, $\{\tilde{\mathbb{S}}^n0\}_{n\ge 0}$ is an increasing sequence and converges to the unique fixed point of $\tilde{\mathbb{S}}$. Noting that $\{\mathbb{S}^n0\}_{n\ge 0}$ is an increasing sequence and is dominated by $\{\tilde{\mathbb{S}}^n0\}_{n\ge 0}$ due to the inequality $\alpha<1$, we conclude that the limit of $\{\mathbb{S}^n0\}_{n\ge 0}$ exists and belongs to $\tilde L^{\alpha'}_+(\mathbb X)$. Moreover, by the monotone convergence theorem and the observation that $\mathbb{S}0(x)>0,x\in\mathbb{X}$, the limit, denoted as $\mathbb{S}^\infty0$, must be a fixed point of $\mathbb{S}$ in ${\cal X}_{++}$. Finally, for any other fixed point $g$ of $\mathbb{S}$, we have $g\ge 0$, so $g=\mathbb{S}^ng\ge \mathbb{S}^n0$ for any $n\ge 0$. As a result, $g\ge \mathbb{S}^\infty0$.

Finally, we prove part (iii) of the proposition, where we assume $\alpha\in(0,1)$. It is straightforward to see that $f$ is a fixed point of $\mathbb{T}$ in ${\cal X}_{++}$ if and only if $h(x)=u(f(x))/v(x),x\in\mathbb{X}$ is a fixed point of $\bar{\mathbb{S}}$ in ${\cal X}_{++}$. Note also that $\bar{\mathbb{S}}$ is an increasing mapping. Consider function $\varphi(z):=\tilde u\big(a + \beta\delta^{1-\rho}\tilde u^{-1}(z)\big), z\ge 0$ for certain $a>0$. Then, because
\begin{align*}
\varphi'(z) = \left(\beta \delta^{1-\rho}z^{\frac{1}{\alpha}}/(\beta \delta^{1-\rho}z^{\frac{1}{\alpha}} + a)\right)^{1-\alpha}(\beta \delta^{1-\rho})^\alpha< (\beta \delta^{1-\rho})^\alpha,
\end{align*}
we conclude that $|\varphi(z_1)-\varphi(z_2)|< (\beta \delta^{1-\rho})^\alpha|z_1-z_2|$ for any $z_1,z_2\ge 0$. As a result, for any $h_1$ and $h_2$ in $\tilde L^{1}_+(\mathbb X)$, we have
\begin{align*}
\begin{split}
  |\bar{\mathbb{S}} h_1 (x)-\bar{\mathbb{S}} h_2(x)|&\le  (\beta \delta^{1-\rho})^\alpha\left|\tilde \expect_t\left[h_1(X_{t+1}) |X_t=x\right])-\tilde \expect_t\left[h_2(X_{t+1}) |X_t=x\right]\right|\\
  &\le (\beta \delta^{1-\rho})^\alpha\tilde \expect_t\left[\left|h_1(X_{t+1})-h_2(X_{t+1})\right| |X_t=x\right].
\end{split}
\end{align*}
Taking expectation on both sides of the inequality under the stationary distribution of $X_t$, we conclude that
\begin{align*}
  \tilde \expect\left[|\bar{\mathbb{S}} h_1 (X_t)-\bar{\mathbb{S}} h_2(X_t)|\right]\le (\beta \delta^{1-\rho})^\alpha\tilde \expect_t\left[\left|h_1(X_{t+1})-h_2(X_{t+1})\right| \right].
\end{align*}
This inequality implies (i) $\bar{\mathbb{S}}$ is a mapping from $\tilde L^{1}_+(\mathbb X)$ into $\tilde L^{1}_+(\mathbb X)$ because $\bar{\mathbb{S}} 0 \in \tilde L^{1}_+(\mathbb X)$ and (ii) $\bar{\mathbb{S}}$ is a contraction mapping because $(\beta \delta^{1-\rho})^\alpha<1$. Consequently, $\bar{\mathbb{S}}$ admits a unique fixed point in $\tilde L^{1}_+(\mathbb X)$, and the fixed point must be positive because $\bar{\mathbb{S}} h\ge \bar{\mathbb{S}}0\in {\cal X}_{++}$ for any $h\in {\cal X}_+$. \halmos
\proofend

\proof{Proof of Proposition \ref{prop:ExistUniqueGeneralStateMyopic}}
Consider function $\varphi(z):=\ln(ae^{z}+b)$ for some $a>0$ and $b\ge 0$. Straightforward computation yields $\varphi'(z) = ae^{z}/(ae^{z}+b)\le 1$, so $|\varphi(z_1)-\varphi(z_2)|\le |z_1-z_2|$. Consequently, recalling the definition of $\mathbb{S}$, we conclude that, for any functions $g_1$ and $g_2$ on $\mathbb{X}$,
\begin{align*}
|\mathbb{S}g_1(x)-\mathbb{S}g_2(x)|\le \beta \expect_t\left[|g_1(X_{t+1})-g_2(X_{t+1})|\mid X_t=x\right],\; x\in \mathbb{X}.
\end{align*}
Replacing $x$ with $X_t$ in the above inequality and taking expectation on both sides under the stationary distribution of $\{X_t\}$, we immediately conclude that
\begin{align*}
\expect\left[|\mathbb{S}g_1(X_t)-\mathbb{S}g_2(X_t)|\right]\le \beta \expect\left[|g_1(X_{t+1})-g_2(X_{t+1})|\right] = \beta \expect\left[|g_1(X_t)-g_2(X_t)|\right].
\end{align*}
This inequality implies that (i) $\mathbb{S}$ is a mapping from $L^1(\mathbb{X})$ into $L^1(\mathbb{X})$ because $\mathbb{S}g\in L^1(\mathbb{X})$ for some $g\in L^1(\mathbb{X})$ and (ii) $\mathbb{S}$ is a contraction mapping. \halmos
\proofend

\proof{Proof of Proposition \ref{prop:ExistUniqueNonMarkovMyopic}}
Using the same proof as that of Proposition \ref{prop:ExistUniqueGeneralStateMyopic}, we can show that for any two adapted processes $\{Z_{i,t}\}$, $i=1,2$,
\begin{align*}
|({\cal S}Z_1)_t - ({\cal S}Z_2)_t |\le \beta \expect_t\left(|Z_{1,t+1}-Z_{2,t+1}|\right) = (\beta/\alpha)\cdot \alpha\expect_t\left(|Z_{1,t+1}-Z_{2,t+1}|\right),\; t=0,1,\dots.
\end{align*}
In consequence,
\begin{align*}
\expect[|({\cal S}Z_1)_t - ({\cal S}Z_2)_t |]\le  (\beta/\alpha)\cdot \alpha\expect\left[|Z_{1,t+1}-Z_{2,t+1}|\right],
\end{align*}
which yields $||{\cal S}Z_1-{\cal S}Z_2||\le (\beta/\alpha)||Z_1-Z_2||$. Because ${\cal S}Z\in {\cal L}^{1,\alpha}$ for some $Z\in {\cal L}^{1,\alpha}$ and ${\cal L}^{1,\alpha}$ is a complete normed space, ${\cal S}$ is a contraction mapping and thus admits a unique fixed point on this space.\halmos
\proofend

\bibliography{LongTitles,BibFile}

\end{document}